\documentclass[preprint,times, 12pt]{elsarticle}

\usepackage{epsfig}

\usepackage{amssymb}
\usepackage{bm}
\usepackage{subfig}
\usepackage{graphicx}

\usepackage{csquotes}

\usepackage{amsmath,mathtools}

\usepackage{comment}
\usepackage{braket}

\newcommand{\be}{\begin{equation}}
\newcommand{\ee}{\end{equation}}

\newcommand{\tr}{\mathrm{Tr}}
\newcommand{\Tr}{\mathrm{Tr}}

\newcommand{\cH}{\mathcal{H}}

\newcommand{\ra}{\rangle}

\journal{Physics Reports}

\begin{document}

\begin{frontmatter}

\title{Quantum entanglement in cosmology}

\author[siena]{Alessio Belfiglio}
\ead{alessio.belfiglio@unisi.it}
\affiliation[siena]{DSFTA, University of Siena, Via Roma 56, 53100 Siena, Italy.}

\author[infn,camerino,inaf,kazak]{Orlando Luongo\corref{cor1}\fnref{orl}}
\ead{orlando.luongo@unicam.it}
\cortext[cor1]{Corresponding author at: School of Science and Technology, University of Camerino, Via Madonna delle Carceri, 62032, Italy. }
\affiliation[infn]{Istituto Nazionale di Fisica Nucleare (INFN), Sezione di Perugia, Perugia, 06123, Italy.}
\affiliation[camerino]{School of Science and Technology, University of Camerino, Via Madonna delle Carceri, 62032, Italy.}
\affiliation[inaf]{Istituto Nazionale di Astrofisica (INAF), Osservatorio Astronomico di Brera, Milano, Italy.}
\affiliation[kazak]{Al-Farabi Kazakh National University, Almaty, 050040, Kazakhstan.}

\author[infn,camerino]{Stefano Mancini}
\ead{stefano.mancini@unicam.it}

\begin{abstract}
We discuss recent progress in the study of entanglement within cosmological frameworks, focusing on both momentum and position-space approaches and also reviewing the possibility to directly extract entanglement from quantum fields. Entanglement generation in expanding spacetimes can be traced back to the phenomenon of gravitational particle production, according to which the background gravitational field may transfer energy and momentum to quantum fields. The corresponding entanglement amount and its mode dependence are both sensitive to the field statistics and to the details of spacetime expansion, thus encoding information about the background. Gravitational production processes also play a key role in addressing the quantum-to-classical transition of cosmological perturbations. 
In order to directly extract entanglement from quantum fields, local interactions with additional quantum systems, working as detectors, have been suggested, leading to the formulation of the entanglement harvesting protocol. Despite harvesting procedures are currently unfeasible from an experimental point of view, various proposals for implementation exist and a proper modeling of detectors and local interactions is crucial to address entanglement extraction via realistic setups.
In the final part of the work, we address entanglement characterization in position space, primarily focusing on black hole spacetimes. We first investigate a possible interpretation of Bekenstein-Hawking black hole entropy in terms of the entanglement entropy arising in discrete quantum field theories, on account of the area law. Then, we discuss the resolution of the black hole information paradox via the gravitational fine-grained entropy formula, which provides a new way to compute the entropy of Hawking radiation and allows to preserve unitarity in black hole evaporation processes.
\end{abstract}


\end{frontmatter}

\tableofcontents



\section{Introduction} \label{secINTRO}

In recent years, quantum information theory (QIT) has attracted the interest of several, apparently uncorrelated, branches of physics. Among them, tools and techniques from QIT are giving new insights into relativistic scenarios described by quantum field theory (QFT) and cosmology, inspiring novel research avenues that may lead to significant developments, both from a theoretical and experimental point of view. 

Within this picture, \emph{quantum entanglement} \cite{Schrödinger_1935,RevModPhys.81.865} emerges as a fundamental resource. Entangled states exhibit correlations with no classical analogue, and the appearance of such correlations within relativistic quantum field frameworks may have relevant implications on the latest technologies \cite{Erhard2020} and on our current understanding of universe evolution. Quantum technologies indeed promise to revolutionize computation \cite{Nielsen:2012yss,10167529}, communication \cite{RevModPhys.82.665}, metrology \cite{Giovannetti2011,RevModPhys.89.035002} and imaging \cite{Defienne2024}, and the possibility to extract (\emph{harvest}) entanglement \cite{Reznik:2002fz,PhysRevD.92.064042} directly from quantum fields, despite not yet experimentally achieved, has been widely investigated in recent years by probing such fields with simpler localized quantum systems, usually in the form of Unruh-DeWitt detectors \cite{PhysRevD.14.870,PhysRevD.29.1047}. Despite a fully relativistic approach to harvesting procedures is still under investigation \cite{PhysRevD.109.045018}, the protocol has been refined and studied exhaustively in several different scenarios, showing how particle detectors may encode information about the universe dynamics.

Field entanglement is currently expected to provide relevant information about spacetime expansion and the history of our universe. In particular, gravitational particle production (GPP) due to the universe evolution \cite{PhysRev.183.1057,PhysRevD.3.346,PhysRevD.17.964} has been shown to produce entanglement in the final state of quantum fields, and the corresponding entanglement spectrum exhibits peculiar features according to the field statistics and the universe background dynamics \cite{Ball:2005xa,Fuentes:2010dt}. Such a phenomenon may play a crucial role in the early phases of universe evolution, where a short stage of strongly accelerated expansion, known as \emph{inflation} \cite{Starobinsky:1980te,Guth:1980zm,RevModPhys.78.537}, has been theorized in order to address several puzzles of the hot big-bang cosmological model. The development of cosmic inflation represents one of the major achievements of modern cosmology, despite the nature of the theorized inflaton field(s) is still to be determined, due to the significant degeneracy of theoretical models with current observational data \cite{Planck:2018jri}. Furthermore, we would like to understand the back-reaction \cite{Schander:2021pgt} and decoherence \cite{Lombardo:2005iz,Franco:2011fg,Liu:2016aaf,Martin:2018lin} mechanisms by which initial quantum fluctuations evolved into classical perturbations in the post-inflationary universe. Such \emph{quantum-to-classical transition} \cite{Kiefer:1998qe, Burgess:2006jn,PhysRevD.102.023512} has been recently addressed using the tools of quantum information theory \cite{PhysRevD.102.043529}, and momentum-space entanglement \cite{Martin-Martinez:2012chf,PhysRevD.86.045014, Brahma:2023hki} of fluctuation modes is expected to play a key role in the characterization of primordial perturbations. In particular, we expect that GPP associated with inhomogeneous inflaton fluctuations \cite{Frieman:1985fr,Cespedes:1989kh} may generate entanglement entropy through the Hubble horizon \cite{Belfiglio:2022cnd,Belfiglio:2022yvs,Belfiglio:2023moe}, thus suggesting that a proper characterization of such entanglement dynamics may reveal how quantum fluctuations become classical at late times. Moreover, the quantum implications of primordial particle creation processes may provide additional insights into the longstanding \emph{cosmological constant problem} \cite{RevModPhys.61.1,Martin:2012bt}, thus enriching our understanding of vacuum energy throughout the universe evolution, and particularly its dynamics during the inflationary phase. Additionally, purely gravitational mechanisms of particle production are among the most promising candidates to explain the origin of dark matter (DM) and other cosmological relics \cite{RevModPhys.96.045005}, and a proper characterization of their quantum features would be crucial in the attempt to theorize possible detection setups.

At the same time, entanglement is expected to play a crucial role in addressing the origin of black hole entropy \cite{Bekenstein:1972tm,Bekenstein:1973ur,Bekenstein:1974ax,Hawking:1974rv,Hawking:1975vcx}, whose microscopic nature is still unknown \cite{Wald:1999vt,Page:2004xp}. The entropy-area proportionality emerging for black holes suggests significant differences from the usual thermodynamic entropy (e.g. the entropy of a thermal gas in a box), which typically scales with the total volume of the system. As we know, the entropy of ordinary matter is expected to arise from the number of quantum states accessible to the matter at given values of the energy and other state parameters. Accordingly, one would like to obtain a similar understanding in the case of black holes, by identifying and counting the corresponding quantum degrees of freedom. Among the plethora of theoretical proposals, quantum entanglement is recently gaining considerable attention \cite{2008arXiv0806.0402D,Solodukhin:2011gn}. Within this picture, black hole entropy is traced back to the position-space entanglement entropy of quantum fields between the interior and the exterior of the black hole and, more recently, a one-to-one correspondence between entanglement mechanics and black hole thermodynamics has been proposed in static scenarios \cite{PhysRevD.102.125025}, suggesting in particular that entanglement characterization close to black hole horizons may capture relevant information about their thermodynamic structure. Despite promising, such an entanglement-based approach has still to face several open issues. In particular, the need for a ultraviolet cutoff in the computation of entanglement entropy for discrete quantum fields does not find a counterpart in the case of Bekenstein-Hawking entropy, which is a finite quantity. Such divergences may be healed by a renormalization of Newton's constant \cite{PhysRevD.87.084047,Cooperman:2013iqr}, despite the validity of this approach for high-spin fields is still under investigation \cite{PhysRevX.14.011024}.  At the same time, the presence of nonminimal coupling to the spacetime curvature may lead to significant deviations from an area law behaviour \cite{Belfiglio:2023sru}. Even more importantly, when taking into account Hawking radiation due to quantum fields, its thermal behaviour should imply that the entanglement entropy outside the black hole increases monotonically as the black hole evaporates \cite{Hawking:1975vcx, RevModPhys.93.035002}. However, quantum mechanics would require this entropy to be zero at the end of evaporation, in order to recover the initial pure state, thus leading to the well-known black hole \emph{information paradox} \cite{PhysRevD.14.2460}. The time evolution of the generalized black hole and radiation entanglement entropy can be described by the so-called Page curve \cite{Page:1993wv,Page:2013dx} and, accordingly, the information paradox can be rephrased in terms of if and how the Page curve can be reproduced without resorting to quantum gravity. Recent findings have suggested that this curve may emerge from the effect of disconnected regions known as islands \cite{Penington:2019npb,Almheiri:2019psf,Almheiri:2019yqk,Almheiri:2019hni}, highlighting how the Anti-de Sitter/conformal field theory (AdS/CFT) correspondence \cite{Maldacena:1997re,Ryu:2006bv} may shed further light on this paradox.

In the following, we delve into the above-presented topics, in order to convince the reader that entanglement is establishing itself as a main character in the latest developments across cosmology and towards a full quantum theory of gravity. The plan of the paper is the following. In Sec. \ref{secENTA}, we present the notion of entanglement for pure and mixed quantum states, introducing some widely-employed measures to quantify it in both scenarios. In Sec. \ref{secQFT}, we briefly review some aspects of field quantization in curved spacetimes, then introducing the GPP mechanism in expanding spacetimes and some notions on black hole spacetimes. In Sec. \ref{secMOMENT}, we discuss momentum-space entanglement generation from GPP, showing how the background spacetime evolution affects the amount and mode dependence of the produced entanglement. In the final part of the section, we apply momentum-space techniques to the study of primordial perturbations, focusing on the entanglement entropy associated with inflationary fluctuations. In Sec. \ref{secHARV}, we present the entanglement harvesting protocol, discussing the main differences between relativistic and non-relativistic models of particle detectors.  In Sec. \ref{secPOS}, we first discuss entanglement in discrete field theories, showing how entanglement entropy represents a plausible candidate for the origin of black hole entropy. Then, we derive the Page curve for the entanglement entropy of Hawking radiation during black hole evaporation.  Finally, in Sec. \ref{secCONCL} we draw our conclusions and present some future perspectives. We use units in which $\hbar=c=k_B=1$ throughout the review.

\section{Entanglement in a nutshell} \label{secENTA}

We provide a concise overview of the fundamental concepts regarding the entanglement of pure and mixed states, directing the reader to reviews on quantum entanglement for an in-depth exploration \cite{RevModPhys.81.865,Plenio:2007zz}.

Let us start considering a bipartite system with associated Hilbert space $\cH_A\otimes\cH_B$.
Any bipartite pure state $|\Psi_{AB}\ra\in\cH_A\otimes\cH_B$ is termed \emph{separable}
if there exist  $\ket{\psi_A}\in\cH_A, \ket{\psi_B}\in\cH_B$ such that  
\begin{equation}
\ket{\Psi_{AB}}=\ket{\psi_A}\ket{\psi_B},
\end{equation} 
or, in other words, if it can be written as a tensor product of vectors belonging to the Hilbert spaces of the subsystems
(for  pure states the notion of separability coincides with that of factorability).
On the other hand, if it does not exist any $\ket{\psi_A}\in\cH_A, \ket{\psi_B}\in\cH_B$ of such kind,
the state $|\Psi_{AB}\ra$ is termed \emph{entangled}.

Given orthonormal bases $\{\ket{e_A^i}\}_i$ for $\cH_A$ and $\{\ket{e_B^j}\}_j$ for $\cH_B$, 
we can expand $\ket{\psi_{AB}}$ as
\begin{equation}
\label{PsiAB}
\ket{\psi_{AB}}=\sum_{i=1}^{d_A}\sum_{j=1}^{d_B}\Psi_{ij} \ket{e_A^i}\ket{e_B^j},
\end{equation}
where $d_A=\dim\cH_A$ and $d_B=\dim\cH_B$. 
Then, by means of singular value decomposition of the matrix $\Psi$ of coefficients $\Psi_{ij}$, Eq.\eqref{PsiAB} can be taken into the form
\begin{equation}
\label{Schmidtdec}
\ket{\psi_{AB}}=\sum_{i=1}^{\min\{d_A,d_B\}} \lambda_i \ket{\tilde{e}_A^i}\ket{\tilde{e}_B^i},
\end{equation}
where $\lambda_i$ are non-zero singular 
eigenvalues of $\Psi$, and
 $\{\ket{\tilde{e}_A^i}\ket{\tilde{e}_B^i} \}_{i}$ is a  bi-orthonormal basis for $\cH_A\otimes\cH_B$.
Eq.\eqref{Schmidtdec} is known as \emph{Schmidt decomposition} and the quantities $\lambda_i$ as \emph{Schmidt coefficients}.

It turns out that the state $\ket{\psi_{AB}}$ is separable if and only if $\Psi$ has rank one, namely if and only if there is a unique $\lambda_i$ different from zero (and equal to one). In contrast, $\ket{\psi_{AB}}$ is entangled if and only if more than one $\lambda_i$ is different from zero and it will be maximally entangled if and only of $\lambda_i=\frac{1}{\sqrt{d}}$, with $d=\min\{d_A,d_B\}$.

Entanglement remains invariant under local unitary operations, denoted as $U_A \otimes U_B$. The coefficients $\lambda_i$ are the sole parameters invariant under such transformations, thus fully determining bipartite entanglement.

A quantitative measure of entanglement must satisfy at least two fundamental properties:
\begin{itemize}
 \item[i)] it cannot increase under local operations and classical communication;
\item[ii)] it must vanish for separable states.
\end{itemize}
Additionally, normalization is often required, ensuring the entanglement measure equals $\log d$ when  $\ket{\psi_{AB}}=\sum_{i=1}^d  \ket{\tilde{e}_A^i}\ket{\tilde{e}_B^i}/\sqrt{d}$, i.e., for a maximally entangled state.

The \emph{entropy of entanglement}, satisfies these conditions and will be considered throughout this paper (it can be also proved that in the asymptotic regime all pure states entanglement measures correspond to this one). 

It is defined as the (von Neumann) entropy of subsystem (either $A$ or $B$)
\begin{equation}
\label{ententdef}
{\cal S}_A=-\tr\left( \rho_A\log\rho_A\right)={\cal S}_B=-\tr\left(\rho_B\log\rho_B\right),
\end{equation}
where $\rho_A=\tr_B(|\Psi_{AB}\rangle\langle\Psi_{AB}|)$ and $\rho_B=\tr_A(|\Psi_{AB}\rangle\langle\Psi_{AB}|)$ are the reduced density operators\footnote{Despite, by convention, the von Neumann entropy associated with discrete systems should contain $\log_2$ instead of $\log$, we will employ Eq. \eqref{ententdef} in both scenarios, for consistency.}.

Moving on to mixed states, any bipartite density operator $\rho_{AB}$ defined on Hilbert space ${\cal H}_A\otimes {\cal H}_B$ 
is separable if and only if it can be expressed in  the following form
\begin{equation}
\rho_{AB}=\sum_i p_i \rho_A^{(i)} \otimes \rho_B^{(i)},
\end{equation}
where $\rho_A^{(i)}$ and $\rho_B^{(i)}$
are defined on local Hilbert spaces ${\cal H}_A$,
${\cal H}_B$ and $p_i$ is a probability mass function. 

A criterion called positive partial transpose (PPT) states that 
if $\rho_{AB}$ is separable, then the new matrix $\rho_{AB}^{\Gamma}$
with entries defined in some fixed product basis as
\begin{equation}
\langle k_B | \langle m_A | \rho_{AB}^{\Gamma} | n_A \rangle |_B l\rangle \equiv 
\langle l_B | \langle m_A | \rho_{AB}  | n_A \rangle | k_B \rangle,
\end{equation}
is a valid density operator (i.e. has nonnegative spectrum),
which means automatically that $ \rho_{AB}^{\Gamma}$
is also a quantum state. 
The operation $\Gamma$, known as partial transpose (the symbol $\Gamma$ is a part of the letter $T$ typically denoting transposition), 
corresponds to transposing the indices associated with  the second subsystem and has interpretation
as a partial time reversal. Notably, the same conclusion holds if the transposition is applied to the indices of the first subsystem instead.

A fundamental fact is that the PPT condition serves as both a necessary and sufficient criterion for the separability of 
$2\otimes 2$ and $2\otimes 3$ systems. 
While it does not provide a complete characterization of separability in general quantum systems, the PPT criterion has inspired a simple and computationally accessible measure of entanglement, known as  \emph{Negativity} 
\begin{equation}
{\cal N}=\sum_{\lambda <0}\lambda
\end{equation}
where $\lambda$ are eigenvalues of $\rho^\Gamma$. 
A version of the measure called \emph{logarithmic negativity}, satisfying conditions i) and ii), is given by
\begin{equation}
E_{\cal N}=\log \frac{2{\cal N}(\rho)+1}{2}.
\end{equation}


The extension of the above arguments to infinite-dimensional Hilbert spaces requires careful consideration. Systems with an associated Hilbert space isomorphic to $L^2(\mathbb{R})$ (or equivalently $\ell^2(\mathbb{C})$) are typically referred to as continuous variable (CV) systems.

For a bipartite CV pure state in $\ell^2(\mathbb{C})_A \otimes \ell^2(\mathbb{C})_B$, where Fock bases are conventionally employed, separability occurs if and only if the state has a Schmidt rank of one. This means that one Schmidt coefficient equals one while all others are zero. However, maximally entangled states cannot exist in this context, as a state with all Schmidt coefficients being equal would result in an infinite norm.

The entropy of entanglement remains a valid measure for pure CV states. Nevertheless, it is unbounded and can diverge to infinity for specific states. To address this issue, appropriate constraints must be introduced, which can be implemented more effectively within the framework of Gaussian states \cite{Serafini:2023rrn}.

A Gaussian state $\rho_{AB}$ of two bosonic modes is characterized by the first two moments.
Consider $\rho_{AB}$ on the Hilbert space $L^2(\mathbb{R})_A\otimes L^2(\mathbb{R})_B$ of functions of 
position variables $(q_A,q_B)$, we have the displacement vector (first moments)
\begin{equation}
 d_i=\tr(\rho_{AB} R_i),
 \end{equation}
 and the covariance matrix (second moments) 
 \begin{equation}
 \sigma_{ij}=\tr\left\{\rho_{AB}[(R_i-d_i)(R_j-d_j)+(R_j-d_j)(R_i-d_i)]\right\},
 \end{equation}
 where $R=(Q_A,P_A,Q_B,P_B)$ and $Q$, $P$ are mode position and momentum observables respectively.
These operators satisfy the canonical commutation relations $[R_k,R_{k'}]= 2 i \Sigma_{kk'}$ where 
\be
\Sigma=\bigoplus_{i=1}^2 
\left(
\begin{array}{cc}
0 & 1 \\
-1 & 0
\end{array}
\right),
\ee
is the symplectic form (we adopt the convention that the covariance matrix of the vacuum state is $I$).

Since the displacement $d$ can be eliminated through local unitary operations, the covariance matrix $\sigma$ alone determines the entanglement properties. It is important to note that the Heisenberg uncertainty imposes constraints on $\sigma$
\begin{equation}
\label{Hrel}
\sigma+i \Sigma \ge 0.
\end{equation}
The covariance matrix $\sigma$ describes a pure state if and only if $(\sigma \Sigma)^2=-I$.

To prevent divergences in physical quantities, it is customary within the Gaussian state manifold to restrict the mean value of the 'energy' (assuming free, non-interacting oscillators) in each subsystem. This requires fixing the following values
\begin{align}
E_A&=\Tr\{\rho_{AB}(a^\dag a+a a^\dag)\}, \label{Vener}\\
E_B&=\Tr\{\rho_{AB}(b^\dag b+b b^\dag)\},
\end{align}
where $a,a^\dag$ (resp. $b,b^\dag$) are ladder operators, with real and hermitian part given by
$Q_A,P_A$ (resp. $Q_B,P_B$), poviding link to $\ell^2(\mathbb{C})_A\otimes \ell^2(\mathbb{C})_B$.  

Let $\sigma$ represent the covariance matrix of a bipartite Gaussian pure state $\rho_{AB}$ with $1 + 1$ modes. 
The reduced density operator $\rho_A$ remains Gaussian and is described by a covariance matrix $\sigma_A$. 
The matrix $\sigma_A$ can always be diagonalized using a symplectic matrix $S_A$, 
yielding $S_A \sigma_A S_A^T = \mathrm{diag}(\nu, \nu)$, where $\nu \in[1,\infty)$ is the so called symplectic eigenvalue of $\sigma_A$
(for a pure Gaussian state $\rho_{AB}$, the symplectic eigenvalues of $\sigma_A$ are equal to those of $\sigma_B$). 

Similarly, in the case of a bipartite Gaussian state with $n_A + n_B$ modes, symplectic diagonalization yields a set of $n_A$ symplectic eigenvalues, $\nu_1, \nu_2, \dots, \nu_{n_A}$ (assuming $n_A \leq n_B$). It can be shown that any entanglement measure depends solely on the symplectic eigenvalues. Specifically, the entropy of entanglement, given in equation (\ref{ententdef}), can be expressed as a function of these eigenvalues \cite{PhysRevA.63.032312},
\begin{equation}
{\cal S}_A=\sum_{i=1}^{n_A} h(\nu_i),
\end{equation}
where 
\begin{equation}
h(x)=\frac{x+1}{2}\log\left(\frac{x+1}{2}\right)
-\frac{x-1}{2}\log\left(\frac{x-1}{2}\right).
\end{equation}

The logarithmic negativity of a bipartite Gaussian state with $n_A + n_B$ modes turns out to be
\begin{equation}
E_{\cal N}= \sum_{i=1}^{n_A+n_B} \max\{0, -\log \tilde{\nu}_i\},
\end{equation}
where $\tilde{\nu}_i$ are the symplectic eigenvalues of the partially
transposed covariance matrix $\tilde{\sigma}=\Gamma\sigma\Gamma$,
obtainable through $\Gamma=\left(\oplus_{i=1}^{n_A} I_2\right) \oplus \left(\oplus_{i=1}^{n_B}\sigma_z\right)$,
being $I_2$ and $\sigma_z$ the identity matrix and the $z$-Pauli matrix respectively.

\section{Quantum fields in curved spacetimes} \label{secQFT}

In this Section, we present the basic notions of QFT in curved spacetime, highlighting how particle (and vacuum) states can be properly defined in the presence of a gravitational field. We then focus on the phenomenon of GPP due to spacetime expansion and we also discuss possible corrections arising from inhomogeneities, at a perturbative level. Finally, a short introduction to black hole spacetimes is provided. For a broader introduction to QFT in curved spacetime, the reader may consult Refs. \cite{Birrell:1982ix,Parker:2009uva,Hollands:2014eia}.

\subsection{Field quantization on curved backgrounds} \label{secQFT.1}

QFT in curved spacetime aims to study the dynamics of quantum fields propagating in a classical, curved spacetime, which is described according to the laws of general relativity. Due to its classical treatment of the background, QFT in curved spacetime cannot represent a fundamental theory. However, it is expected to provide an accurate description of quantum phenomena well below Planck energy scales ($M_{\rm pl}=1.22 \times 10^{19}$ GeV), where quantum gravity effects should be negligible.

The basic generalization of the particle concept from flat to curved spacetimes can be readily addressed \cite{Birrell:1982ix}. However, the physical interpretation of field excitations becomes much more difficult in a curved background. The presence of a gravitational field indeed breaks Poincar\'e invariance \cite{wald2010general}, on which Minkowskian QFT significantly relies, having as a main the consequence the fact that a unique vacuum state cannot be defined in this case. Accordingly, the concept of particle becomes generally ambiguous, apart from such regions where the spacetime expansion (or contraction) is sufficiently slow.

The simplest scenario arises in the presence of a real scalar field, whose Lagrangian density reads
\be \label{scaldens}
\mathcal{L}_S=\frac{1}{2}  g^{\mu \nu} \phi_{,\,\mu} \phi_{,\,\nu}  - V(\phi),
\ee
for a generic potential $V(\phi)$ and a given metric tensor $g^{\mu \nu}$, with $\phi_{,\,\mu} \equiv \partial \phi/\partial x^\mu$. 

Let $\{ f_j \}$ be a complete set of positive norm solutions of the Klein-Gordon (KG) equation
\be \label{kg_gen}
\square \phi \equiv \frac{1}{{\sqrt{-g}}} \partial_{\nu} \left( \sqrt{-g} g^{\mu \nu} \partial_{\nu} \phi \right)= V_{,\, \phi},
\ee
with $V_{,\phi} \equiv \partial V/\partial \phi$ , namely
\be \label{kgproduct}
\left( f_j,f_j \right)= -i \int \left( f_j^* \partial_\mu f_j-f_j \partial_\mu f_j^* \right) d\Sigma^\mu > 0,\ \ \ \forall j,
\ee
where $\left( f_j,f_j \right)$ is the \emph{KG scalar product} and $d\Sigma^\mu=d\Sigma n^\mu$, with $d\Sigma$ denoting the volume element in a given spacelike hypersurface and $n^\mu$ the timelike unit vector normal to this hypersurface \cite{Ford:2021syk}. Accordingly, the scalar field can be quantized as
\be \label{KG_quant}
\hat{\phi}= \sum_j
( \hat{a}_j f_j+ \hat{a}_j^\dagger f_j^*  ),
\ee
where the ladder operators satisfy the canonical commutation relations $[a_j,a_{j^\prime}^\dagger]=\delta_{j,j^\prime}$. This field expansion then defines a vacuum state $\ket{0}$, satisfying $a_j \ket{0}=0\ \ \forall j$. In flat spacetime, we take positive norm solutions to be positive frequency solutions, namely $f_j \propto e^{-i \omega_k t}$, with $\omega_k=\sqrt{k^2+m^2}$ in the simplest scenario of a scalar field with mass $m$ and corresponding potential $V(\phi)=m^2 \phi^2/2$. Independently of the Lorentz frame in which $t$ is the time coordinate, this procedure allows to unambiguously identify the unique Minkowski vacuum state.

On the other hand, in a generic curved spacetime there is no unique choice of the $\{ f_j\}$ and thus no unique notion of vacuum state. A possible solution to this issue consists in resorting to some quantities different from the particle content to label quantum states, e.g. local expectation values \cite{Wald:1995yp}. However, the notion of particle (and vacuum) can be preserved in the case of asymptotically flat spacetimes.

In particular, let us consider a spacetime which is asymptotically flat in the past and in the future, but is non-flat in the intermediate region. Such a toy model may simulate an early phase of fast inflationary expansion, followed by radiation and matter-dominated epochs, during which the universe expansion is much slower. We take $\{ f_j \}$ as positive frequency solutions in the past (\emph{in} region) and we denote by $\{ F_j \}$ the positive frequency solutions in the future (\emph{out} region), also assuming these sets of solutions to be orthonormal
\be
\begin{aligned} \label{orthof_cond}
& (f_j,f_{j^\prime})= (F_j,F_{j^\prime})= \delta_{j j^\prime}, \\
& (f_j^*,f_{j^\prime}^*)= (F_j^*,F_{j^\prime}^*)= -\delta_{j j^\prime}, \\ 
& (f_j,f_{j^\prime}^*)= (F_j,F_{j^\prime}^*)= 0.
\end{aligned}
\ee
Since both sets of functions are solutions of the scalar field equation of motion, we may expand, for example, the \emph{in} modes in terms of the \emph{out} modes
\be \label{bogo_exp}
f_j = \sum_k (\alpha_{jk} F_k+ \beta_{jk} F_k^*),
\ee
where $\alpha_{jk}$ and $\beta_{jk}$ are called \emph{Bogoliubov coefficients} and, exploiting the orthonormality conditions in Eq. \eqref{orthof_cond}, they satisfy 
\be \label{bogo_orthno}
\sum_k (\alpha_{jk} \alpha^*_{j^\prime k}-\beta_{jk} \beta^*_{j^\prime k})= \delta_{j j^\prime}.
\ee
The field operator $\hat{\phi}$ may be then equivalently expanded as
\be \label{scalainout_exp}
\hat{\phi}= \sum_j ( \hat{a}_{j,{\rm in}} f_j+\hat{a}^\dagger_{j,{\rm in}} f^*_j ) = \sum_j (\hat{a}_{j,{\rm out}} F_j + \hat{a}_{j,{\rm out}}^\dagger F_j^*),
\ee
where now the $\hat{a}_{j,{\rm in}}$ and $a_{j,{\rm in}}^\dagger$ are annihilation and creation operators, respectively, for the \emph{in} region, whereas the $\hat{a}_{j,{\rm out}}$ and $\hat{a}_{j,{\rm out}}^\dagger$ are the corresponding operators for the \emph{out} region. Accordingly, the \emph{in} vacuum is defined from $\hat{a}_{j,{\rm in}} \ket{0}_{\rm in}=0\ \ \forall j$ and, similarly, $\hat{a}_{j,{\rm out}} \ket{0}_{\rm out}=0\ \ \forall j $.

Noting that $\hat{a}_{j,{\rm in}}=(\hat{\phi},f_j)$ and $\hat{a}_{j,{\rm out}}=(\hat{\phi},F_j)$, we can expand the two sets of ladder operators in terms of one another:
\begin{align}
    &\hat{a}_{j,{\rm in}}= \sum_k (\alpha_{jk}^* \hat{a}_{k,{\rm out}}- \beta_{jk}^* \hat{a}^\dagger_{k,{\rm out}} ), \label{inout_bogo}\\
    &\hat{a}_{k,{\rm out}}= \sum_j (\alpha_{jk} \hat{a}_{j,{\rm in}}+ \beta_{jk}^* \hat{a}^\dagger_{j,{\rm in}} ) \label{outin_bogo},
\end{align}
which will be useful in the following to compute expectation values of the particle number operator.

In addition to particle states and Bogoliubov coefficients, it turns useful to introduce Green functions, which allow to identify vacuum expectation values of various products of free field operators. In analogy with its flat-space counterpart, we define
\be \label{pau_jor}
iG(x_1,x_2)= \bra{0} \left[ \phi(x_1) \phi(x_2)  \right]  \ket{0},
\ee
where $\left[ \phi(x_1), \phi(x_2)\right] \equiv \phi(x_1) \phi(x_2)-\phi(x_2) \phi(x_1) $ is the field commutator and $G$ is known as Pauli-Jordan or Schwinger function. This Green function can be split into its positive and negative frequency parts
\be \label{green_poneg}
iG(x_1,x_2)= W^+(x_1,x_2)-W^-(x_1, x_2),
\ee
where $G^{\pm}$ are known as Wightman functions, defined by 
\begin{align}
    &W^+ = \bra{0} \phi(x_1) \phi(x_2) \ket{0}, \label{green_pl} \\
    &W^- = \bra{0} \phi(x_2) \phi(x_1) \ket{0}. \label{green_min}
\end{align}
Accordingly, we can introduce the Feynman propagator $G_F$, obtained from the time-ordered product of fields
\begin{align} 
 iG_F(x_1,x_2) &=  \bra{0} \mathcal{T} \left(\phi(x_1) \phi(x_2) \right) \ket{0} \notag \\
 & = \theta(t_1-t_2) W^+ (x_1, x_2) + \theta(t_2 - t_1 ) W^-(x_1, x_2),
\end{align}
where $\mathcal{T}$ is the time-ordering operator and $\theta(x)$ the Heaviside step function. We underline again that, in curved spacetime, the vacuum state $\ket{0}$ must be selected with due care and, more generally, the specification of boundary conditions is not so simple, depending on the global features of the spacetime under investigation \cite{Birrell:1982ix}.

\subsection{Gravitational particle production in expanding spacetimes} \label{secQFT.2}

As we have seen, an initial vacuum state in an expanding spacetime is no longer seen as a vacuum by an observer which lives in the \emph{out} region. Accordingly, the gravitational field may transfer part of its energy to quantum fields, resulting in the phenomenon of GPP, which we first present in the context of homogeneous and isotropic expanding spacetimes for real scalar and Dirac fields. See also Ref. \cite{RevModPhys.96.045005} for an introduction to GPP in case of higher-spin fields.

\subsubsection{Scalar field} \label{secQFT.2.1}

Working within the Heisenberg picture, the mean number of particles in the \emph{out} region for a given mode $k$ can be computed from Eqs. \eqref{inout_bogo}-\eqref{outin_bogo}, giving 
\be \label{meanpart}
 N_k^{(0)}  = \prescript{}{\rm in}\langle 0 \lvert a_{k,\rm{out}}^\dagger a_{k, \rm{out}} \rvert 0 \rangle_{\rm in} = \sum_j \lvert \beta_{jk} \rvert^2,
\ee
where the superscript denotes the perturbative order, thus being zero when inhomogeneities are not taken into account.
Eq. \eqref{meanpart} implies that, if at least one of the $\beta_{jk}$ coefficients is nonzero, some mixing of positive and negative frequency solutions occurs, leading to particle production by the gravitational field.

In order to apply this formalism to relevant cosmological frameworks, we now specialize to the case of a spatially flat Friedmann-Robertson-Walker (FRW) spacetime, described by the line element 
\be \label{FRWline}
ds^2=g_{\mu \nu} dx^\mu dx^\nu= a^2(\tau) \left[dt^2-dr^2 -r^2\left( d\theta^2+ \sin^2 \theta d\phi^2 \right)\right],
\ee
where $a(\tau)$ is the scale factor and $\tau$ the conformal time, related to the usual cosmic time via $\tau=\int dt/a(t)$. Accordingly, asymptotic flatness is now defined by the condition
\be \label{asyFRW}
a(\tau) \rightarrow
\begin{cases}
    a_i,\ \ \ \ \ \tau \rightarrow -\infty,\\
    a_f,\ \ \ \ \ \tau \rightarrow +\infty,
\end{cases}
\ee
where $a_i$, $a_f$ are constants and $a_f > a_i$ for an expanding spacetime. Within this framework, the positive norm solutions of the KG equation, Eq. \eqref{kg_gen}, may be taken to be
\be \label{modsol_FRW}
f_{\bf k}({\bf x},\tau)= \frac{e^{i {\bf k} \cdot {\bf x}}}{(2 \pi)^{{3/2}}a(\tau) } \chi_k(\tau),
\ee
where the $\chi_k(\tau)$ satisfy
\be \label{timemodsol_FRW}
\chi^{\prime \prime}_k(\tau)+ \left( k^2- \frac{a^{\prime \prime}}{a} \right) \chi_k(\tau)+ a^2(\tau) \frac{\partial V(\phi)}{\partial \phi} \bigg \rvert_{\phi=\chi_k}=0.
\ee
and $k= \lvert {\bf k} \rvert$. Defining now $\chi_k^{(\rm in)}(\tau)$ as pure positive frequency solutions of Eq. \eqref{timemodsol_FRW} in the limit $\tau \rightarrow -\infty$ and $\chi_k^{(\rm out)}(\tau)$ as the corresponding solutions for $\tau \rightarrow +\infty$, we can write
\be \label{bogo_expmodFRW}
\chi_k^{(\rm in)}(\tau)= \frac{a_i}{a_f} [\alpha_k \chi_k^{(\rm out)}(\tau)+ \beta_k \chi_k^{(\rm out)}(\tau)^*],
\ee
leading to the relation
\be \label{bogo_expFRW}
f_{\bf k}({\bf x}, \tau)= \alpha_k F_{\bf k}({\bf x}, \tau)+ \beta_k F_{\bf k}^*({\bf x}, \tau),
\ee
which is the analogue of Eq. \eqref{bogo_exp} in the case of a FRW spacetime, with corresponding Bogoliubov coefficients 
\begin{align}
    & \alpha_{{\bf k}{\bf k^\prime}}= \alpha_k \delta_{{\bf k}{\bf k^\prime}}, \label{alpha_bogoFRW} \\
    & \beta_{{\bf k} {\bf k^\prime}} = \beta_k \delta_{{\bf k},-{\bf k^\prime}}. \label{beta_bogoFRW}
\end{align}
Eqs. \eqref{alpha_bogoFRW}-\eqref{beta_bogoFRW} show that an unperturbed FRW spacetime only mixes modes with equal and opposite momenta via the coefficients $\beta_k$, i.e., it creates particle-antiparticle pairs with corresponding spectrum
\be \label{meanpart_FRW}
 N_k^{(0)} = \lvert \beta_k \rvert^2.
\ee
Moving then to the continuum limit
\be \label{contlim}
\sum_{\bf k} \rightarrow \frac{V}{(2 \pi)^3} \int d^3k,
\ee
where $V$ is the total volume of the system, the number of particles created per unit volume reads
\be \label{totnumb_FRW}
n^{(0)} \equiv \frac{N^{(0)}}{V} = \frac{1}{(2 \pi a_f)^3} \int d^3k\  \lvert \beta_k \rvert^2.
\ee
As we will see, GPP also generates entanglement in the final state of the scalar field.

\subsubsection{Dirac field} \label{secQFT.2.2}

The Lagrangian density of a Dirac field $\psi$ with mass $m$ in curved spacetime can be written in the form \cite{RevModPhys.48.393}
\be \label{dirac_ld}
\mathcal{L}_D = \frac{1}{2}\big[\bar{\psi} \tilde{\gamma}^{\mu} D_{\mu} \psi - (D_{\mu} \bar{\psi}) \tilde{\gamma}^{\mu} \psi \big]- m \bar{\psi} \psi,
\ee
where $\tilde{\gamma}^\mu$ are the curved Dirac-Pauli matrices, related to the usual flat-space Dirac matrices by
\be \label{dir_curvfla}
\tilde{\gamma}^{\mu}= e^{\mu}_\alpha \gamma^\alpha,
\ee
with $e^{\mu}_\alpha$ denoting the \emph{tetrad} (or \emph{vierbein}) field \cite{PhysRevD.10.411,PhysRevD.14.2505}, satisfying
\be \label{vierbein}
\text{g}^{\mu \nu} e^\alpha_{\mu} e^\beta_{\nu}= \eta^{\alpha \beta}.
\ee
and $\eta^{\alpha \beta}$ is the Minkowski metric tensor.

The covariant derivative $D_\mu$ of the Dirac field can be expressed in the form \cite{Shapiro:2001rz}
\be \label{spin_covar}
D_{\mu} \psi = (\partial_{\mu}-\Omega_{\mu}) \psi,
\ee
where $\Omega_\mu$ is the Levi-Civita \emph{spin connection}, namely \cite{Benisty:2019jqz}
\be \label{LC_spin}
\Omega_\mu=  -\frac{1}{4} g_{\beta \nu} [\Gamma^{\nu}_{\ \alpha \mu}- e^{\nu}_j \partial_{\mu} e^j_{\alpha}] \tilde{\gamma}^\beta \tilde{\gamma}^\alpha
\ee
and the $\Gamma^{\nu}_{\ \alpha \mu}$ are the usual Christoffel symbols of GR \cite{wald2010general}. On a FRW background, the corresponding Dirac equation for the spinor mode functions $U_r$, $V_r$ reads
\begin{subequations}
    \begin{align}
& (i \gamma^0 \partial_0 - {\boldsymbol \gamma} \cdot {\bf k} - M) U_r({\bf k}, \tau) = 0,   \label{udir} \\[2pt]
& (i \gamma^0 \partial_0 - {\boldsymbol \gamma} \cdot {\bf k} - M) V_r(-{\bf k}, \tau) = 0   ,\label{vdir}
    \end{align}
\end{subequations}
where $M=ma(\tau)$, while $r$ labels the spin of particle and antiparticles, and ${\boldsymbol  \gamma}=(\gamma^1,\gamma^2,\gamma^3)$. These equations can be solved via the ansatz 
\begin{subequations}
    \begin{align}
    & U_r({\bf k}, \tau) =  (i \gamma^0 \partial_0 - {\boldsymbol \gamma} \cdot {\bf k} +  ma) \phi_k^-(\tau) u_r, \label{ansmod_fer1}\\
    & V_r({\bf k}, \tau) = (i \gamma^0 \partial_0 - {\boldsymbol \gamma} \cdot {\bf k} +  ma) \phi^+_k(\tau) v_r, \label{ansmod_fer2}
    \end{align}
\end{subequations}
where the ``minus" denotes particle solutions and the ``plus" their corresponding antiparticle ones. Furthermore, $u_r= \begin{pmatrix}  \xi_r \\[3.5pt] 0    \end{pmatrix}$, $v_r= \begin{pmatrix}  0 \\[3.5pt] \xi_r    \end{pmatrix}$, and the two-component spinors $\xi_r$ are chosen to be helicity eigenstates, i.e., $
{\boldsymbol \sigma} \cdot {\bf k} = r k \xi_r$ with $ r= \pm 1$ and ${\boldsymbol \sigma}=(\sigma^1,\sigma^2,\sigma^3)$ is a vector of Pauli matrices. 

From Eqs. \eqref{udir}-\eqref{vdir}, the mode functions $\phi_k^\pm$ satisfy \cite{Fuentes:2010dt}
\be \label{dmodes_time}
(\partial_0^2+M^2 \pm  i M^\prime+k^2) \phi_k^{(\pm)}=0,
\ee
As in the scalar field scenario, in order to quantize the field we need to identify positive and negative frequency modes. Again, this cannot be done globally. However, for an asymptotically flat spacetime we can identify positive and negative frequency modes in the far past and future, then writing 
\be \label{asymod_dir}
\phi_{k, \text{in}}^{(\pm)}= A_k^{(\pm)} \phi_{k, \text{out}}^{(\pm)} + B_k^{(\pm)} \phi_{k, \text{out}}^{(\mp)*}.
\ee
We can then write the following Bogoliubov transformations for ladder operators:
\begin{align}
& \hat{a}_{\text{in}}({\bf k},d)= \alpha^*(k)\  \hat{a}_{\text{out}}({\bf k},d)- \sum_{s} \beta_{ds}^*(-k)\  \hat{b}_{\text{out}}^\dagger(-{\bf k},s), \label{abogo_dir} \\
& \hat{b}^\dagger_{\text{in}}(-{\bf k},d)=  \sum_s \beta_{sd}(-k)\ \hat{a}_{\text{out}}({\bf k},s)+ \alpha(k)\  \hat{b}^\dagger_{\text{out}}(-{\bf k},d), \label{bbogo_dir}
\end{align}
where $\hat{a}$ operators refer to particles and $\hat{b}$ to antiparticles, while the $\alpha(k)$ and $\beta(k)$ Bogoliubov coefficients ($d,s=\uparrow,\downarrow$) can be derived from the coefficients in Eq. \eqref{asymod_dir}, see e.g \cite{Pierini:2015jma}. Eqs. \eqref{abogo_dir}-\eqref{bbogo_dir} can be simplified resorting to charge and angular momentum conservation. The number density of particle-antiparticle pairs can be then computed in analogy with Sec. \ref{secQFT.2.1}.


\subsection{Gravitational particle production from inhomogeneities} \label{secQFT.3}

Let us now introduce perturbations within the above-presented FRW background, according to the general prescription
\be \label{pert_FRW}
g_{\mu \nu}=a^2(\tau) (\eta_{\mu \nu}+h_{\mu \nu}).
\ee
In Eq. \eqref{pert_FRW}, $h_{\mu \nu} \equiv h_{\mu \nu}(x)$ describes spacetime perturbations, with $\lvert h_{\mu \nu} \rvert \ll 1$ and we require $h_{\mu \nu} \to 0$ in the asymptotic limit $\tau \to \pm \infty$.

The presence of inhomogeneities is able to enhance the mechanism of gravitational production from vacuum. In order to show that, let us consider the first-order interaction Lagrangian density \cite{Frieman:1985fr}
\be \label{intlag}
    \mathcal{L}_{int}=-\frac{1}{2}\sqrt{-g_{(0)}}H^{\mu\nu}T^{\left(0\right)}_{\mu\nu},
\ee
where $g_{(0)}$ is the determinant of the background unperturbed metric tensor, $H_{\mu \nu}= a^2(\tau) h_{\mu \nu}$ and $T_{\mu \nu}^{(0)}$ is the zero-order energy-momentum tensor for a given quantum field.

In all our cases of interest, it can be shown that the equality $\mathcal{L}_{int}=-\mathcal{H}_{int}$  holds \cite{Cespedes:1989kh}. This implies that we can write the $S$ matrix at first-order in Dyson expansion as 
\be \label{smatr_fir}
\hat{S} \simeq 1 + i \mathcal{T} \int d^4x \mathcal{L}_{int},
\ee
If the interaction Lagrangian is quadratic in the field and its derivatives\footnote{Generalization to higher order terms is straightforward, despite it may require renormalization in some cases, see e.g. Ref. \cite{Belfiglio:2023moe}.}, an initial vacuum state $\ket{0}_{\rm in}$ evolves within the interaction picture as 
\be \label{intevol_sc}
\ket{\phi}= \hat{S} \ket{0}_{\rm in}= N\left( \ket{0}_{\rm in}+ \frac{1}{2} \int d^3 k_1 d^3 k_2\  \prescript{}{\rm in}\langle {\bf k}_1, {\bf k}_2 \lvert \hat{S} \rvert 0 \rangle_{\rm in}\ \ket{{\bf k}_1, {\bf k}_2}_{\rm in}\right),
\ee
where $N$ is a normalization factor. Accordingly, we can compute the the final particle density starting from the number density operator in the \emph{out} region 
\be \label{numbdensop}
\hat{n}= \frac{1}{( 2 \pi a_f )^3} \int d^3q\  \hat{a}^\dagger_{{\bf q},{\rm out}} \hat{a}_{{\bf q},{\rm out}}
\ee
where ladder operators evolve with the unperturbed field Hamiltonian.
In the simplest case of a scalar field, from Eqs. \eqref{outin_bogo} and \eqref{beta_bogoFRW}, we have
\be \label{outin_bogoFRW}
\hat{a}_{\bf k,{\rm out}}= \alpha_k \hat{a}_{{\bf k},{\rm in}}+ \beta_k^* \hat{a}_{-{\bf k},{\rm in}},
\ee
obtaining 
\be \label{numbdensop_exp}
\bra{\phi} \hat{n} \ket{\phi} = n^{(0)}+ n^{(1)}+ n^{(2)},
\ee
where $n^{(0)}$ is the non-perturbative contribution derived in Eq. \eqref{totnumb_FRW} and 
\begin{align} 
    & n^{(1)}= \frac{\lvert N \rvert^2}{( 2 \pi a_f )^3} \int d^3 k_1 d^3 k_2\ \delta({\bf k}_1+ {\bf k}_2)\ \text{Re}\left[ \mathcal{C}_{{\bf k}_1,{\bf k}_2} (\alpha_{k_1} \beta_{k_1}+\alpha_{k_2} \beta_{k_2})\right], \label{numbdens1}\\
    & n^{(2)}= \frac{\lvert N \rvert^2}{( 2 \pi a_f )^3} \int d^3 k_1 d^3 k_2\ \lvert \mathcal{C}_{{\bf k}_1,{\bf k}_2} \rvert^2 (1+ \lvert \beta_{k_1} \rvert^2+ \lvert \beta_{k_2} \rvert^2) \label{numbdens2},
\end{align}
with $\mathcal{C}_{{\bf k}_1,{\bf k}_2} \equiv \langle {\bf k}_1,{\bf k}_2 \lvert \hat S \rvert 0 \rangle$ denoting the probability amplitude. Accordingly, we observe that:
\begin{itemize}
    \item[-]  At first perturbative order, particles are still produced as particle-antiparticle pairs, due to the combined effects of expansion and inhomogeneities.
    \item[-] At second perturbative order, a nonzero contribution emerges even in the case of negligible Bogoliubov coefficients. This contribution is also present for fields with nonzero spin (see e.g. Ref. \cite{Bassett:2001jg}) and it may play a fundamental role in describing gravitational production processes involving conformally coupled scalar fields or minimally coupled fermionic ones, particularly in the limit of small field mass.
\end{itemize}
In the next section, we will discuss how gravitational production processes are able to generate entanglement in the final state of quantum fields and we will quantify the entanglement entropy associated with primordial inflationary perturbations. 

\subsection{Modeling black hole spacetimes}

We end this section by briefly reviewing some basic concepts related to black holes, which will be useful in Sec. \ref{secPOS}.

We will primarily deal with static and spherically symmetric black hole configurations, which can be described by a line element of the form
\be \label{linel_stsph}
ds^2=  f(r) dt^2- \frac{dr^2}{f(r)}- r^2(d\theta^2+\sin^2 \theta d\phi^2),
\ee
where spherical coordinates have been employed and $f(r)$ is known as the lapse function of the black hole.

The simplest black hole configuration is described by the Schwarzschild geometry \cite{Schwarzschild:1916uq}, which represents the unique source-free solution of Einstein equations with spherical symmetry\footnote{This result is known as \emph{Birkhoff's theorem} \cite{GDB1927}, which more generally states that all spherically symmetric spacetimes with $R_{\mu \nu}=0$ are static. See, e.g. Ref. \cite{wald2010general} for more details and a rigorous definition of static spacetimes.} and approaches flat Minkowski spacetime at large distances. The corresponding lapse function takes the form
\be \label{schwalap}
f(r)= 1-\frac{2 G M}{r}, \ \ \ \ 0<r \leq \infty.
\ee
where $G$ is the Newton gravitational constant and $M$ is the total mass, describing a point-like contribution at $r=0$ and giving rise to a physical singularity. Such pathology can be described in a coordinate invariant way as the divergence of the fully contracted Riemann tensor $R_{\alpha \beta \gamma \delta} R^{\alpha \beta \gamma \delta}$ (\emph{Kretschmann scalar}), leading to formally infinite tidal forces. On the other hand, the additional singularity corresponding to $r=r_s$, where $r_S \equiv 2GM$ is known as Schwarzschild radius, can be removed by an appropriate change of coordinates and provides the black hole horizon.

Black hole singularities are a peculiar feature of classical general relativity, which should be resolved within a complete theory of quantum gravity. However, even from a classical and semiclassical perspective, some black hole solutions avoiding true singularities have been proposed in the last decades, collectively denoted as \emph{regular black hole} models \cite{Lan:2023cvz}.

The first regular black hole solution was implemented by Bardeen \cite{1968qtr..conf...87B} and lately interpreted in terms of a nonzero magnetic monopole source \cite{Ayon-Beato:2000mjt}. Starting from Eq. \eqref{linel_stsph}, we can write the corresponding lapse function for \emph{Bardeen black hole} in the form
\be \label{bardlaps}
f(r)= 1- \frac{2 G M r^2}{\left(r^2+q^2 \right)^{3/2}},
\ee
where $q$ is the magnetic charge of the monopole and $M$ the Arnowitt-Deser-Misner (ADM) mass \cite{PhysRev.116.1322,Manko:2016ixk} of the Bardeen solution, respectively. The limiting case of Schwarzschild black hole is here recovered when $q = 0$.
\begin{figure}[t]
    \centering
    \includegraphics[scale=0.45]{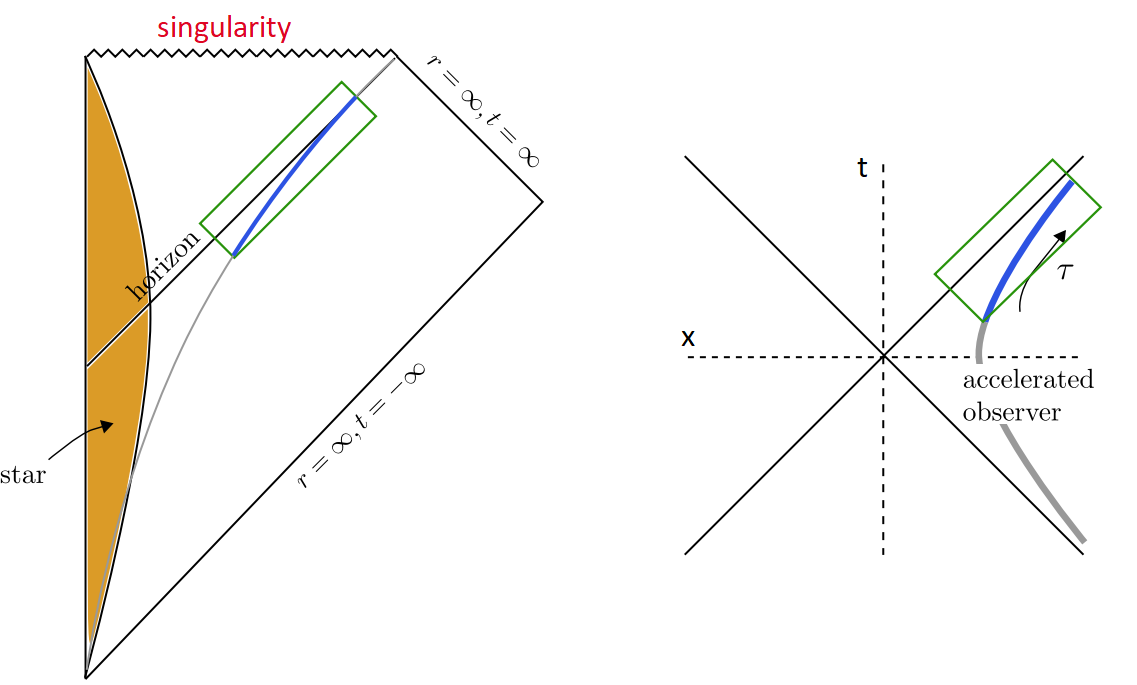}
    \caption{Penrose diagram of a black hole formed by gravitational collapse. On the right, the near-horizon region has been enlarged, showing the trajectory of a uniformly accelerated observer. Figure adapted from \cite{RevModPhys.93.035002}.}
    \label{fig_pnr_diag}
\end{figure}

An alternative solution was proposed more recently by Hayward \cite{PhysRevLett.96.031103}, considering a vacuum energy term which ensures regularity as $r \rightarrow 0$. The \emph{Hayward solution} is described by a lapse function of the form
\be \label{haylaps}
f(r)= 1- \frac{2 G M r^2}{r^3+2 G M b^2},
\ee
where $M$ is the usual point-like mass of the black hole and $b$ is related to the central vacuum energy density, which is assumed to be positive. Similarly to the Bardeen solution\footnote{It has been recently shown that Bardeen and Hayward spacetimes are special cases of a more general two-parameter class of black hole solutions, as discussed in Ref. \cite{Neves:2014aba}.}, Eq. \eqref{haylaps} reduces to Schwarzschild if $l=0$ and it is flat for $M=0$.

In order to illustrate the global causal structure of black hole spacetimes, Penrose diagrams are widely employed \cite{PhysRevLett.10.66}. Such diagrams allow to reproduce four-dimensional spacetimes in terms of finite-coordinate diagrams, via a conformal transformation which also allows to handle points at infinity. In particular, when dealing with spherically symmetric spacetimes, each point on the Penrose diagram corresponds to a 2-dimensional sphere. In a Penrose diagram, light rays move at 45 degrees, thus clarifying the causal structure at each point of the spacetime. Accordingly, it is much easier to determine if, for instance, an event might be able to influence another, i.e., if there exists a causal future-pointing curve connecting the first to the latter. 

In Fig. \ref{fig_pnr_diag}, we show the Penrose diagram associated with an astrophysical black hole, i.e., a black hole that has formed by gravitational collapse. We can see that whenever a particle reaches a point inside the event horizon, its entire causal future lies inside the horizon, so that all of its possible future world-lines finally intersect the singularity. The horizon is a lightlike surface, in agreement with the fact that it represents the boundary of the set of points from which a light ray could reach $(r=+\infty, t=+\infty)$.
 .


\section{Momentum-space entanglement techniques} \label{secMOMENT}

As we have seen, GPP processes rely on the mixing between quantum field modes with different momenta, due to the background gravitational field. More generally, when dealing with interacting QFTs, we can typically access only a limited subset of the total degrees of freedom, which however may be entangled with the inaccessible, higher-energy modes. Accordingly, momentum-space entanglement techniques \cite{PhysRevD.86.045014,PhysRevD.95.065023} are usually more suitable in describing QFT processes, with respect to position-space ones, which may result in unphysical divergences. The study of entanglement generation associated with GPP processes may enable the possibility to characterize the produced particles and to deduce cosmological parameters related to the background dynamics. At the same time, such techniques are gaining considerable attention in the characterization of primordial perturbations, which represent the seeds for structure formation in the universe and are typically probed in momentum space \cite{Planck:2018jri}.


\subsection{Entanglement from spacetime expansion} \label{sec4.1}

The GPP mechanism presented in Sec. \ref{secQFT.2} has been shown to generate entanglement in the final state of the quantum fields involved. The corresponding entanglement entropy exhibits different features according to the field statistics.


\subsubsection{Scalar field} \label{sec4.1.1}

Let us write the initial particle-antiparticle vacuum state as a Schmidt decomposition of \emph{out} states \cite{Ball:2005xa}
\be \label{schm_out}
\ket{0_k; 0_{-k}}_{\text{in}}=  \sum_{n=0}^{\infty} c_n \ket{n_k; n_{-k}}_{\text{out}},
\ee
where $c_n$ are the Schmidt coefficients and $n_k$ labels the number of excitations in the field mode $k$, as seen by an inertial observer in the \textit{out} region. The presence of nonzero $c_n$ then implies  that particle-antiparticle pairs are entangled in the \emph{out} region.

An explicit expression for the Schmidt coefficients can be found by inverting Eq. \eqref{outin_bogoFRW}, leading to
\be 
0 = \hat{a}_{{\bf k},{\rm in}} \lvert 0_k; 0_{-k} \rangle_{\text{in}}= \left( \alpha_k^* \hat{a}_{{\bf k},{\rm out}} -\beta_k^* \hat{a}^\dagger_{-{\bf k},{\rm out}} \right) \sum_{n=0}^{\infty} c_n \lvert n_k; n_{-k} \rangle_{\text{out}},
\ee
and thus giving the following recurrence relation
\be \label{recurr_schm}
c_n= \left( \frac{\beta_k^*}{\alpha_k^*} \right)^n  c_0,\ \ \ c_0=\sqrt{1-\lvert \beta_k/\alpha_k \rvert^2}.
\ee

In order to quantify the entanglement entropy in the \emph{out} region, we start by introducing the density operator 
\be \label{densop_out}
\rho_{k,-k}^{(\rm in)}=\ket{ 0_k; 0_{-k}}_{\rm in} \bra{ 0_k;0_{-k} },
\ee
which can be rewritten in terms of the \emph{out} state via Eq. \eqref{schm_out}. Such density operator describes a pure state, so its corresponding von Neumann entropy $S\left(\rho_{k,-k}^{(\rm in)/(\rm out)}\right)$ is zero by definition. Accordingly, the reduced density operator\footnote{We can equivalently take the partial trace over either particle or antiparticle states, since the total density operator describes a pure state and thus the von Neumann entropy of the two subsystems will be the same.},
\be \label{redop_part}
\rho_k^{(\rm out)}=\tr_{-k} \left(\rho_{k,-k}^{(\rm out)}\right)=  \sum_{m=0}^{\infty} \langle m_{-k} \rvert \rho_{k,-k}^{(\rm out)} \rvert m_{-k} \rangle,
\ee
allows to properly quantify the particle-antiparticle entanglement in the \emph{out} region
\be \label{subentrop}
S_{KG}\left(\rho_k^{(\rm out)}\right)= -\tr\left(\rho_k^{(\rm out)} \log \rho_k^{(\rm out)}\right)= \log \frac{\gamma^{\gamma/(\gamma-1)}}{1-\gamma},
\ee
where $\gamma= \left \lvert \beta_k/\alpha_k \right \rvert^2$.
\begin{figure}
    \centering
    \includegraphics[scale=0.34]{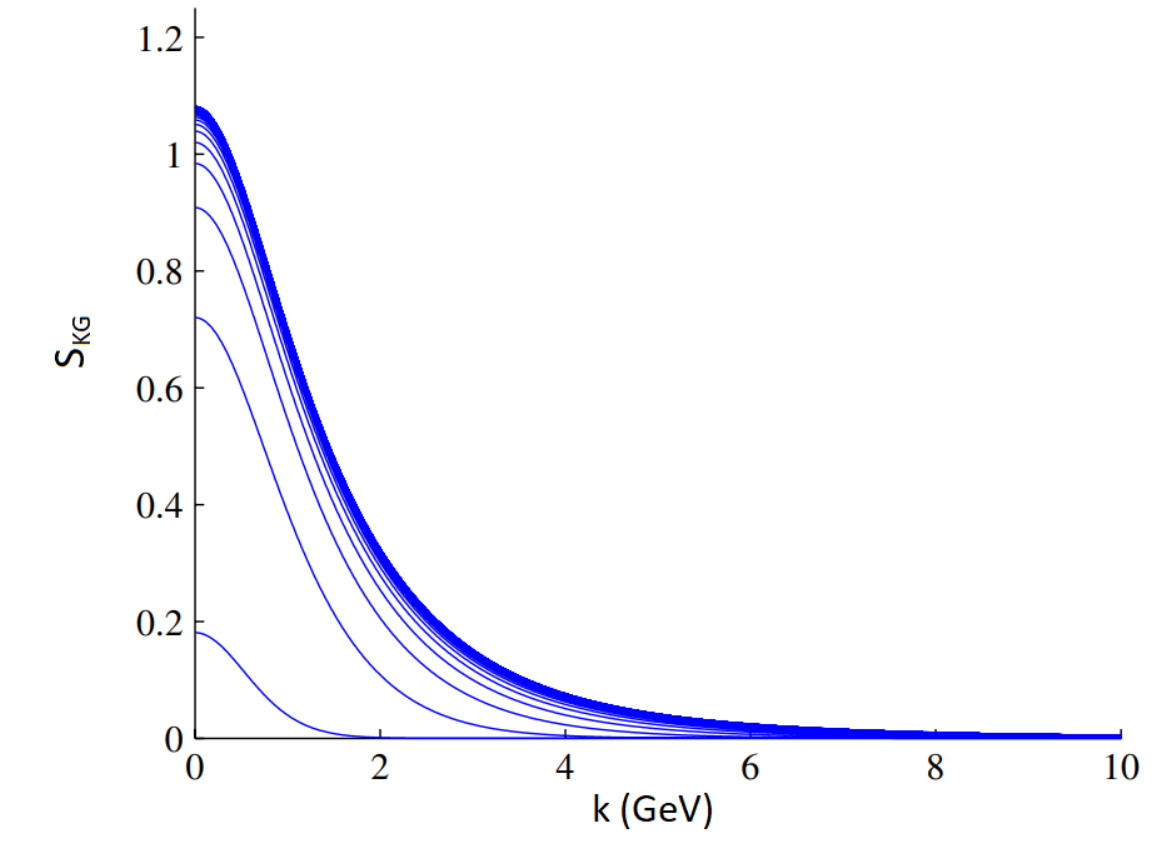}
    \includegraphics[scale=0.34]{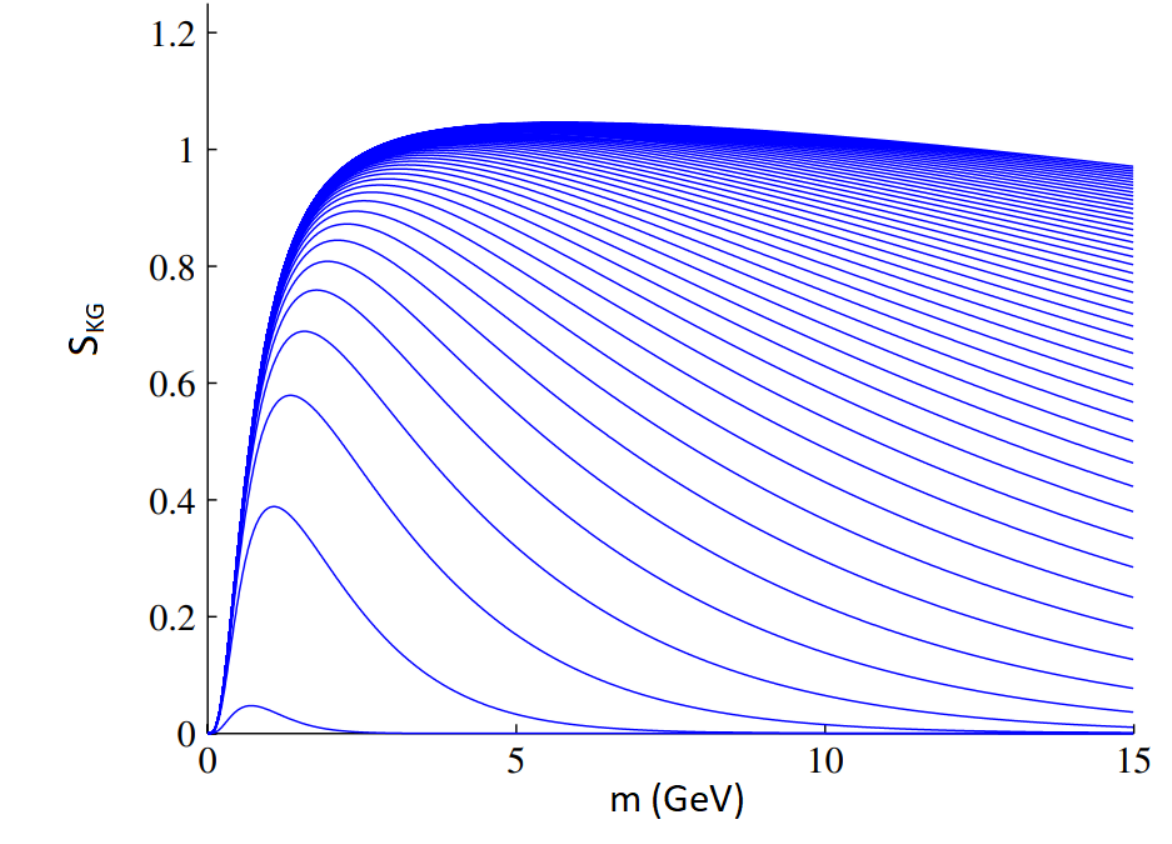}
    \caption{Entanglement entropy for the KG field in a $(1+1)$-dimensional FRW spacetime described by Eq. \eqref{dunctoy_scal}, with $\epsilon_S=1$ and $\sigma=1, \dots, 100$ GeV. On the left, the entropy is plotted for $m=1$ GeV as function of the momentum $k$, while on the right it is plotted for $k=1$ GeV and varying mass. Figure adapted from \cite{Fuentes:2010dt}.}
    \label{fig_KGent}
\end{figure}

The coefficients $\gamma$ can be computed analytically for some particular toy models of spacetime evolution. In particular, assuming a massive scalar field in a $(1+1)$-dimensional FRW spacetime, with corresponding potential
\be \label{quadpot}
V(\phi)= \frac{1}{2}m^2 \phi^2,
\ee
and selecting \cite{PhysRevD.17.964}
\be \label{dunctoy_scal}
a^2(\tau)= 1+ \epsilon_S(1+\tanh \sigma \tau),
\ee
where $\epsilon_S$ and $\sigma$ are positive real parameters.

The corresponding entanglement entropy is plotted in Fig. \ref{fig_KGent} as function of the field momentum (on the left) and the field mass (on the right), showing that larger entanglement is obtained for particle-antiparticle pairs with lower momentum, in agreement with the fact that it is typically easier to entangle modes with lower energy. It can be also shown that, in the case of a massless field ($m=0$), we obtain $\gamma=0$, thus implying vanishing entropy. This fact is related to the conformal invariance of massless scalar field theories in $(1+1)$-dimensional spacetimes, which gives $\beta_k=0$ for a conformally flat background \cite{PhysRevD.17.964,Ball:2005xa}. More recently, the entanglement entropy of gravitationally produced ultralight scalar particles has been studied in a more realistic scenario involving an instantaneous transition from inflation to the radiation-dominated era of universe evolution \cite{PhysRevD.102.063532}, confirming that the entanglement spectrum peaks at low momentum $\propto 1/k^3$.


\subsubsection{Dirac field} \label{sec4.1.2}

Here, analytical solutions can be found by selecting the scale factor \cite{PhysRevD.17.964}
\be \label{dunctoy_dir}
a(\tau)= 1+ \epsilon_S(1+\tanh \sigma \tau).
\ee

Assuming now charge and angular momentum conservation, Eqs. \eqref{abogo_dir}-\eqref{bbogo_dir} become \cite{Pierini:2015jma}
\begin{align}
& \hat{a}_{\text{out}}(d)= \alpha\  \hat{a}_{\text{in}}(d)- \beta_{d,-d}^*\  \hat{b}_{\text{in}}^\dagger(-d), \label{abogo_dir_cm} \\
& \hat{b}^\dagger_{\text{out}}(d)=  \beta_{-d,d}\ \hat{a}_{\text{in}}(-d)+ \alpha\  \hat{b}^\dagger_{\text{in}}(d), \label{bbogo_dir_cm}
\end{align}
where we have omitted the momentum dependence to simplify the notation. 
By means of the Schmidt decomposition, we can now write
\begin{align} \label{dir_invac}
\ket{ 0_k; 0_{-k}}_{\text{in}}= \alpha^2 \bigg( &\ket{0_k; 0_{-k}}_{\text{out}} - \frac{\beta^*_{\uparrow \downarrow}}{\alpha} \ket{ \uparrow_k ; \downarrow_{-k} }_{\text{out}} \notag \\    
 &- \frac{\beta_{\downarrow \uparrow}^*}{\alpha} \ket{ \downarrow_k ; \uparrow_{-k} }_{\text{out}}+ \frac{\beta^*_{\uparrow \downarrow} \beta^*_{\downarrow \uparrow}}{\alpha^2} \ket{ \uparrow \downarrow_k ; \uparrow \downarrow_{-k} }_{\text{out}} \bigg),
\end{align}
\begin{figure}
    \centering
    \includegraphics[scale=0.331]{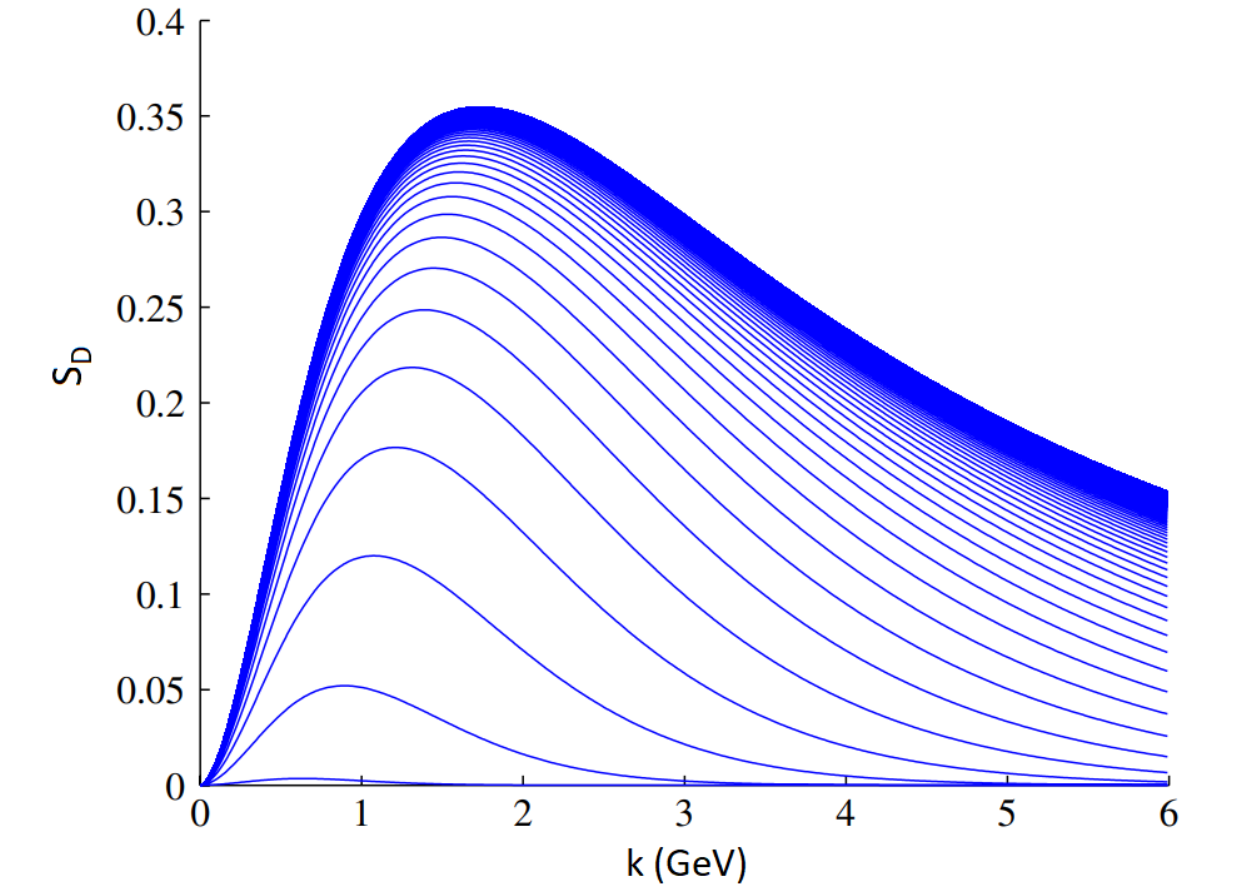}
    \includegraphics[scale=0.331]{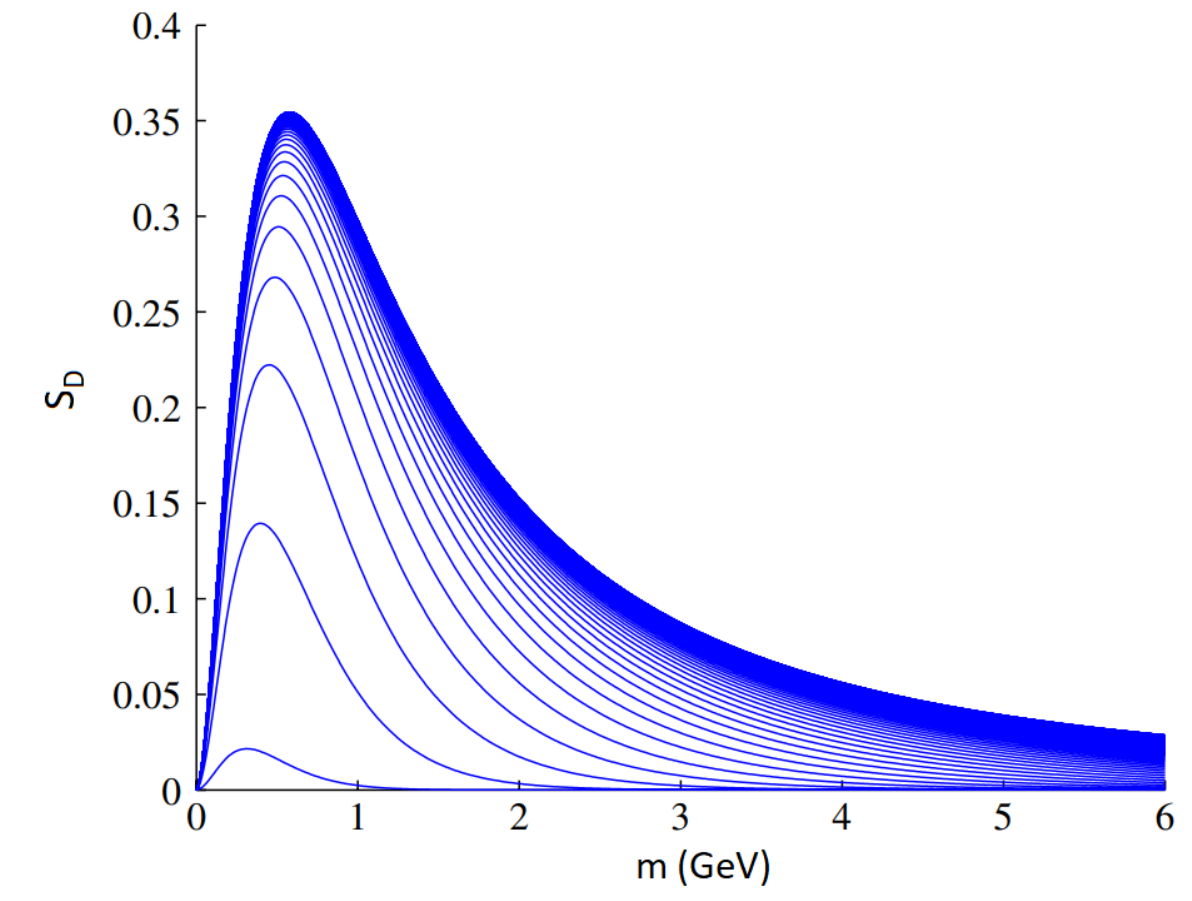}
    \caption{Entanglement entropy for the Dirac field in a (1+1)-dimensional FRW spacetime described by Eq. \eqref{dunctoy_dir}, with $\epsilon_S=1$ and $\sigma=1, \dots, 100$ GeV. On the left, the entropy is plotted for $m=1$ GeV as function of the momentum $k$, while on the right it is plotted for $k=1$ GeV and varying mass. Figure adapted from \cite{Fuentes:2010dt}.}
    \label{fig_dirent}
\end{figure}
The von Neumann entropy associated with the density operator $\rho_k^{(\text{out})}= \tr_{-k} \left( \rho_{k, -k}^{(\text{out})} \right)$ obtained from Eq. \eqref{dir_invac} is then
\begin{align}
S_D\left(\rho_k^{(\text{out})}\right)= & -\alpha^4 \log (\alpha^4)- \alpha^2\lvert \beta_{\uparrow \downarrow} \rvert^2 \log(\alpha^2 \lvert \beta_{\uparrow \downarrow} \rvert^2)- \alpha^2 \lvert \beta_{\downarrow \uparrow} \rvert^2 \log(\alpha^2 \lvert \beta_{\downarrow \uparrow} \rvert^2) \notag \\
&- \lvert \beta_{\uparrow \downarrow} \rvert^2 \lvert \beta_{\downarrow \uparrow} \rvert^2 \log (\lvert \beta_{\uparrow \downarrow} \rvert^2 \lvert \beta_{\downarrow \uparrow} \rvert^2), \label{subentrop_dir}
\end{align}
which is plotted in Fig. \ref{fig_dirent} as function of the field momentum (on the left) and the field mass (on the right), in the limit of a $(1+1)$-dimensional spacetime. We observe that, differently from the KG case, the fermionic entanglement entropy peaks at a certain momentum $k \neq 0$, which depends on the spacetime parameters in Eq. \eqref{dunctoy_dir} and on the field mass $m$. Intuitively, this can be explained in terms of the Pauli exclusion principle \cite{Fuentes:2010dt}, which does not allow to excite several low-frequency modes at the same time. On the other hand, for bosons this constraint does not exist, so it is energetically ``cheaper" to excite $k \rightarrow 0$ modes. Accordingly, these results suggest that field statistics plays a key role in determining how the universe expansion entangles quantum field modes. This outcome has been recently confirmed in Ref. \cite{PhysRevD.102.063532} within a more realistic cosmological scenario, focusing on gravitational DM production during the early phases of universe evolution.


\subsection{Entanglement in anisotropic spacetimes} \label{sec4.2}

The presence of anisotropies is able to affect entanglement generation in cosmological settings. This has been studied in the case of a Bianchi I type metric with weak anisotropy \cite{Pierini:2016rkk,Pierini:2018wki},
\be \label{bianchi_met}
ds^2=a^2(\tau) \left\{ d\tau^2-[1+h_i(\tau)] (dx^i)^2  \right\},\ \ \ \ \ \ \ i=1,2,3,
\ee
where $\max_\tau \lvert h_i(\tau) \rvert \ll 1$ is assumed. Explicit expressions for the Bogoliubov coefficients can be found by setting $\sum_{i=1}^3 h_i(\tau)=0$. A convenient choice is represented by
\be \label{anisans} 
h_i(\tau)=e^{-\rho \tau^2} g_i(\tau),
\ee
where $g_i(\tau)=\cos(\epsilon \tau^2+\delta_i)$ and $\delta_i=\pi/2,7\pi/6,11\pi/6\dots$, describing attenuation of anisotropies in a FRW spacetime. Further choosing
\be \label{scalef_anis}
a(\tau)=1-A e^{-\rho^2 \tau^2},
\ee
where $A$ and $\rho$ are positive constants, it can be shown that:
\begin{itemize}
    \item[-] In the case of a scalar field:
    \begin{align}
       & \alpha_{\bf k}= 1+ \alpha_{\bf k}^{(\text{iso})}+\alpha_{\bf k}^{(\text{aniso})},\label{alpha_an_kg}\\
       & \beta_{\bf k}= \beta_{\bf k}^{(\text{iso})}+\beta_{\bf k}^{(\text{aniso})}.
    \end{align}
    \item[-] In the case of a Dirac field:
    \begin{align}
       & \alpha_{\bf k}= 1+ \alpha_{\bf k}^{(\text{iso})}+\alpha_{\bf k}^{(\text{aniso})}, \label{alpha_an_d} \\
       & \beta_{{\bf k} \uparrow \uparrow} =  \beta^{(\text{iso})}_{{\bf k} \uparrow \uparrow} + \beta^{(\text{aniso})}_{{\bf k} \uparrow \uparrow},\\
       & \beta_{{\bf k} \uparrow \downarrow} =  \beta^{(\text{iso})}_{{\bf k} \uparrow \downarrow} + \beta^{(\text{aniso})}_{{\bf k} \uparrow \downarrow},
    \end{align}
    with $\beta_{{\bf k} \uparrow \uparrow}=-\beta_{{\bf k} \downarrow \downarrow}$ and $\beta_{{\bf k} \uparrow \downarrow}=-\beta^*_{{\bf k} \downarrow \uparrow}$.
\end{itemize}
Accordingly, in both cases, the contribution associated with anisotropies can be separated from the isotropic one, which only depends on the field mass and eventual couplings to the spacetime curvature\footnote{See Ref. \cite{Pierini:2018wki} for a detailed derivation of the Bogoliubov coefficients for both fields.}.

\begin{figure}
    \centering
    \includegraphics[scale=0.43]{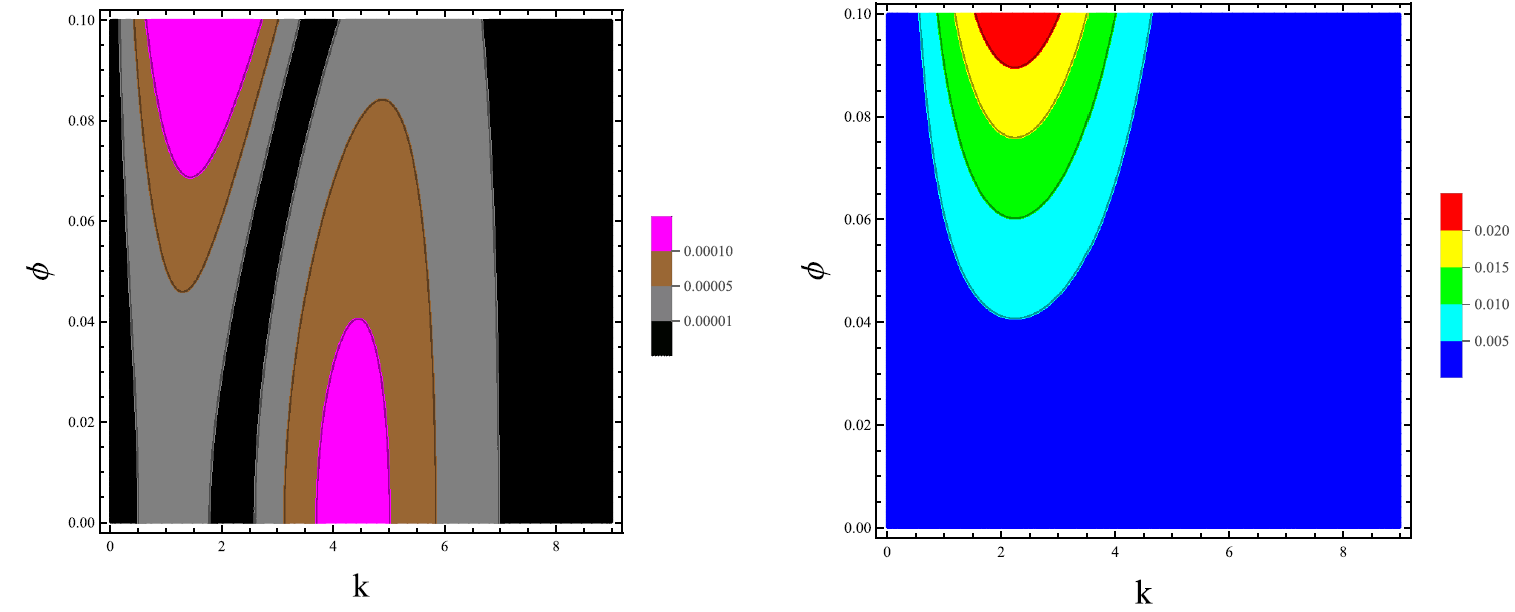}
    \caption{Density plot of the anisotropic subsystem entropy $S$ of a conformally coupled scalar field (left) and a minimally coupled Dirac field (right), as function of the field momentum $k \equiv \lvert {\bf k} \rvert$ and $\phi$. The other parameters are: $\rho=10$, $\epsilon=0.1$, $\theta=\pi/2$. Figure adapted from \cite{Pierini:2018wki}.}
    \label{fig_anis}
\end{figure}

Since the metric in Eq. \eqref{bianchi_met} is still homogeneous, only particle pairs with opposite momenta can get entangled, despite now the entanglement amount also depends on the direction of the momentum ${\bf k}$.  The corresponding particle entanglement entropy can be expressed in the compact form
\begin{align} \label{enta_anis}
S(\rho_{\bf k}^{(\text{out})})&=\tr \left( \rho_{\bf k}^{(\text{out})} \log \rho_{\bf k}^{(\text{out})}    \right) \notag \\[1pt]
&= (1-3\delta_F) \lvert \alpha_{\bf k} \rvert^2 \log \lvert \alpha_{\bf k} \rvert^2- (1-3\delta_F) (\lvert \alpha_{\bf k} \rvert^2-1) \log\left[ (1-2\delta_F) (\lvert \alpha_{\bf k} \rvert^2-1)  \right],
\end{align}
where $\delta_F$ is 0 for the scalar and 1 for the Dirac field.

From Eq. \eqref{alpha_an_kg} and Eq. \eqref{alpha_an_d}, we then observe that anisotropy may entangle particle-antiparticle pairs even when the background expansion does not contribute to entanglement generation, e.g. for conformally coupled massless scalar fields or minimally coupled massless Dirac fields. These two scenarios are depicted in Fig. \ref{fig_anis}, where the entanglement entropy $S$ is plotted as function of the momentum ${\bf k}= \left(k \sin \theta \cos \phi, k \sin \theta \sin \phi, k\cos \theta \right)$, by fixing the azimuthal angle $\theta$. The total entropy is larger in the Dirac case, suggesting that anisotropy may play an important role in entangling fermionic particles with negligible mass like neutrinos.

\subsection{Entanglement entropy corrections from inhomogeneities} \label{sec4.3}

The presence of inhomogeneities can be handled according to the perturbative techniques presented in Sec. \ref{secQFT.3}. For simplicity, we review only the scalar field case.

As shown by Eq. \eqref{numbdens1}, at first perturbative order particles are produced in pairs with opposite momenta. Accordingly, we can focus on a single pair and write the final state of the system from Eq. \eqref{intevol_sc} as
\be \label{final_partant}
\ket{\phi_{k,-k}}_{\rm out}= N \left(  \ket{0}_{\rm in}+ \frac{1}{2} \mathcal{C}_{k,-k} \ket{1_k; 1_{-k}}_{\rm in} \right).
\ee
This state can be rewritten in the \emph{out} basis by means of Bogoliubov transformations, recalling the Schmidt decomposition of Eq. \eqref{schm_out}. In particular,
 \begin{align} \label{B1}
\lvert 1_k;1_{-k} \rangle_{\text{in}} &= \big(\alpha_k \hat{b}_{\bf k}^\dagger -\beta_k \hat{b}_{-{\bf k}} \big) \big(\alpha_k \hat{b}_{-{\bf k}}^\dagger -\beta_k \hat{b}_{{\bf k}} \big)  \sum_{n=0}^{\infty} c_n \lvert n_k; n_{-k} \rangle_{\text{out}} \notag \\
&= \alpha_k^2 \sum_{n=0}^{\infty} (n+1) c_n\  \lvert n+1; n+1 \rangle_{\text{out}} - \alpha_k\beta_k \sum_{n=0}^{\infty} n c_n\  \lvert n_k; n_{-k} \rangle_{\text{out}} \notag \\
& \ \ \ - \alpha_k\beta_k \sum_{n=0}^{\infty} (n+1) c_n\  \lvert n_k; n_{-k} \rangle_{\text{out}} + \beta_k^2 \sum_{n=0}^{\infty}n c_n\ \lvert n-1;n-1 \rangle_{\text{out}}.
\end{align}
From Eq. \eqref{final_partant}, we then obtain the \emph{out} density operator $\rho^{\text{(out)}}_{k,-k}= \lvert \phi \rangle_{\text{out}} \langle \phi \rvert$ and, recalling the $\gamma$ coefficients introduced in Sec. \ref{sec4.1.1}, we arrive at the final expression for the normalized particle density operator, namely \cite{Belfiglio:2022cnd}
\be \label{redens_first}
\rho^{(\rm out)}_k=\frac{(1-\gamma)^2}{1-\gamma+(1+\gamma)\text{Re}\big( \mathcal{C}_{k,-k} \alpha_k \beta_k \big)}\notag \times \sum_{n=0}^\infty \gamma^n \bigg( 1+ \text{Re}\big( \mathcal{C}_{k,-k} \alpha_k \beta_k \big) (2n+1) \bigg) \ket{n_k} \bra{n_k}.
\ee
The corresponding von Neumann entropy is then
\be \label{voneu_first}
S\left(\rho_k^{(\rm out)}\right)= -\tr\left(\rho_k^{(\rm out)} \log \rho_k^{(\rm out)}\right)= - \sum_{n=0}^\infty \lambda_n \log \lambda_n,
\ee
where  $\lambda_n$ are the $\rho_k^{(\rm out)}$ eigenvalues, say
\be \label{eigenv_first}
\lambda_n= \frac{(1-\gamma)^2}{1-\gamma+(1+\gamma)\text{Re}\big( \mathcal{C}_{k,-k} \alpha_k \beta_k \big)} \times \gamma^n\left( 1+ \text{Re}\big( \mathcal{C}_{k,-k} \alpha_k \beta_k \big) (2n+1) \right).
\ee
We notice that, in the limit of vanishing perturbations ($\mathcal{C}_{k,-k}=0$), the well-known zero order eigenvalues are recovered, $\lambda_n^{(0)}= (1-\gamma)\gamma^n$, leading to the usual expression for the entropy, namely Eq. \eqref{subentrop}.

Moving to second order in perturbations, the generic state in the interaction picture takes the form of Eq. \eqref{intevol_sc}, due to mode mixing. Focusing on a single pair, we can write
\be \label{second_sta}
\ket{\phi_{k_1,k_2}}= N \left( \ket{ 0_{k_1} ; 0_{k_2} }_{\rm in} + \frac{1}{2} \mathcal{C}_{{\bf k}_1,{\bf k}_2} \ket{ 1_{k_1}; 1_{k_2}}_{\rm in} \right).
\ee
In analogy to the anisotropic scenario, let us now assume that the background expansion does not generate entanglement. Accordingly, we set $\beta_{k_1}=\beta_{k_2}=0$, implying that the \emph{in} and \emph{out} vacuum states coincide, i.e., $\ket{0}_{\rm in} \equiv \ket{0}_{\rm out}$.

Eq. \eqref{second_sta} shows up a bipartite pure state: in order to compute the corresponding entanglement entropy, we can trace out the ``$k_2$" (or ``$k_1$") contribution, obtaining 
\be \label{reduc_secord}
\rho_{k_1}= \lvert N \rvert^2 \left( \ket{0_{k_1}} \bra{0_{k_1}}+ \frac{1}{4} \lvert \mathcal{C}_{{\bf k}_1,{\bf k}_2} \rvert^2 \ket{1_{k_1}} \bra{1_{k_1}}  \right),
\ee
where $\lvert \mathcal{C}_{{\bf k}_1,{\bf k}_2} \rvert^2$ denotes the probability of pair production.

The entanglement entropy associated with this state has been first computed in Ref. \cite{Belfiglio:2022cnd}, by considering a FRW background expansion with 
\be \label{KG_ans_scal}
a^2(\tau)=A+B \tanh \sigma\tau
\ee
and a nearly Newtonian perturbation source, $h_{\mu \nu}=\text{diag}\left( 2\Psi, 2\Psi, 2\Psi, 2\Psi \right)$, with $\Psi(r)=- M/r$, where $M$ is the mass which generates the perturbation, and inhomogeneities are present only in a limited time interval $[\tau_i,\tau_f]$.

Moving to the synchronous gauge\footnote{For a systematic treatment of the linear theory of scalar gravitational perturbations in the synchronous and the longitudinal gauge, and the corresponding transformation laws, the reader may consult Ref. \cite{Ma:1995ey}.} ($h_{0 \nu}=0$), the pair production probability assuming slow expansion rate and conformal coupling can be computed analytically \cite{Cespedes:1989kh}, depending on the Fourier transforms of both the Weyl tensor and the scalar curvature\footnote{See, for example, Ref. \cite{Frieman:1985fr} for a detailed derivation of these quantities in linear perturbation theory.}, namely $C_{\mu \nu \rho \sigma}(q)$ and $R(q)$, with $q=(q^0, {\bf q} )=(k_1^0+k_2^0, \  {\bf k}_1+{\bf k}_2)$.

Accordingly, pair production probability can be written in terms of local geometric quantities, in the limit of slow background expansion.
\begin{figure} [t!]
    \centering
    \includegraphics[scale=0.48]{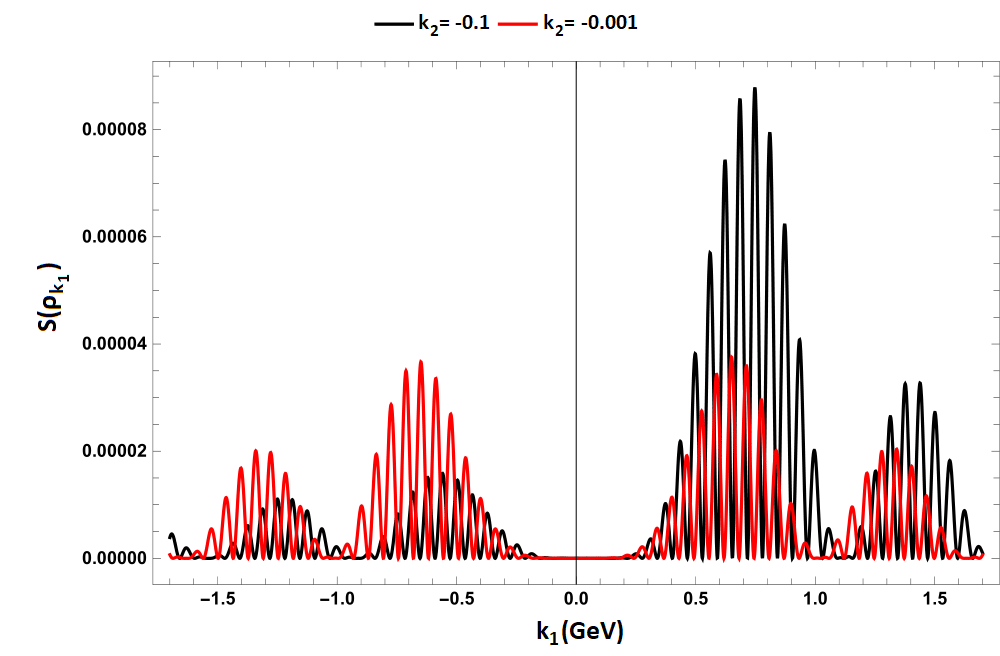}
    \hskip 0.3in
    \includegraphics[scale=0.48]{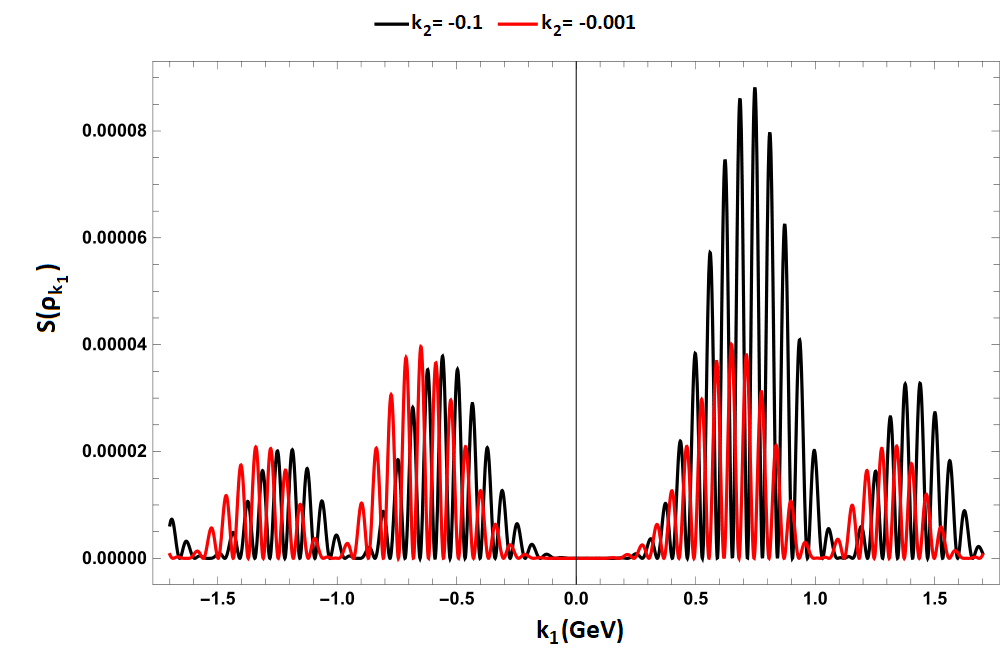}
    \caption{Entanglement entropy $S\left( \rho_{k_1} \right)$ of a KG field, as function of the particle momentum $k_1$, with $A=3$, $B=2$, $\sigma=1$ GeV, $M=10^{-5}$ GeV, $r_{\text{min}}=5$ GeV$^{-1}$, $\tau_i=-10^4$ GeV$^{-1}$ and $\Delta \tau \equiv \tau_f- \tau_i=100$ GeV$^{-1}$. On the left, the entropy is plotted for a massive field with $m=0.01$ GeV, while on the right a massless field is considered. Figure adapted from \cite{Belfiglio:2022cnd}.}
    \label{fig_entrogeo2}
\end{figure}
In order to compute the Fourier transform of $h_{\mu \nu}(x)$, one may assume that the particle momenta are along the $z$ direction, without losing generality, having
\be \label{zetapert}
\tilde{h}_{\mu \nu} (q)= \int d\tau\  e^{iq^0\tau} \int d^3r\  e^{i q_z  r \cos \theta} e^{-r}\  h_{\mu \nu}(x),
\ee
where $q_z$ is the total momentum and $e^{-r}$ a regularization factor \cite{Belfiglio:2022cnd}.

The corresponding subsystem entanglement entropy is plotted in Fig. \ref{fig_entrogeo2} as function of the particle momentum. The entanglement spectrum shows relevant differences with respect to the previously reviewed homogeneous scenarios, depending on the dynamics of the potential $\Psi$. Perturbations played a key role in early universe scenarios, and we will further discuss the generation of entanglement entropy from primordial perturbations in Sec. \ref{sec4.5}.

\subsection{Entanglement in Einstein-Cartan theory} \label{sec4.4}

Modifications to Einstein's gravity may, as well, affect entanglement generation from gravitational production processes. This has been recently investigated \cite{PhysRevD.104.043523} in the context of Einstein-Cartan theory, according to which a nonzero torsion tensor $T^\alpha_{\mu \nu}$ should be included within the spacetime dynamics \cite{Kibble:1961ba,RevModPhys.36.463}, namely 
\be \label{tors_tens}
T^\alpha_{\mu \nu} \equiv \Gamma^\alpha_{[\mu \nu]}= \frac{1}{2} (\Gamma^\alpha_{\mu \nu}-\Gamma^\alpha_{\nu \mu}),
\ee
thus representing the antisymmetric part of the affine connection $\Gamma^\alpha_{\mu \nu}$. In the case of a FRW background, one may set 
\be \label{torsion_FRW}
T_{\alpha \mu \nu}= f(\tau) \epsilon_{\alpha \mu \nu},\ \ \ \ \ T^\alpha_{\hphantom{\alpha}\mu 0}= h(\tau) \delta^\alpha_\mu,
\ee
where $\alpha, \mu, \nu=1,2,3$ and $f(\tau),h(\tau)$ are arbitrary functions of conformal time, while $\epsilon_{\alpha \mu \nu}$ and $\delta^\alpha_\mu$ are the three-dimensional Levi-Civita and Kronecker symbols, respectively.

\subsubsection{Scalar field} \label{sec4.4.1}

The KG equation for a scalar field propagating in a conformally flat FRW spacetime with metric described by Eq. \eqref{FRWline} and the additional presence torsion reads \cite{Buchbinder:1985ux,Buchbinder:1990ku}
\be \label{kgeq_tor}
\frac{1}{a^2} \square \phi - \frac{2 \dot{a}}{a^3} \dot{\phi}- \left( m^2+ \sum_{i=1}^5 \xi_i P_i \right) \phi=0.
\ee
Here $P_1=\tilde{R}$ (Riemannian scalar curvature), $P_2=\nabla_{\alpha}T^{\alpha}$, $P_3=T_{\alpha}T^{\alpha}$, $P_4=S_{\alpha}S^{\alpha}$, where $S_{\alpha}$ is the axial vector $S^{\alpha}=\epsilon^{\beta \mu \nu \alpha}T_{\beta \mu \nu}$. Finally $P_5=q_{\alpha \beta \gamma} q^{\alpha \beta \gamma}$, where $q^{\alpha}_{\hphantom{\alpha} \beta \gamma}$ is a tensor that satisfies the conditions 
\be \label{p5}
q^{\alpha}_{\hphantom{\alpha} \beta \alpha}=0\ \ \  \text{and}\ \ \  \epsilon^{\alpha \beta \mu \nu}q_{\alpha \beta \mu}=0.
\ee
\begin{figure}
    \centering
    \includegraphics[scale=0.313]{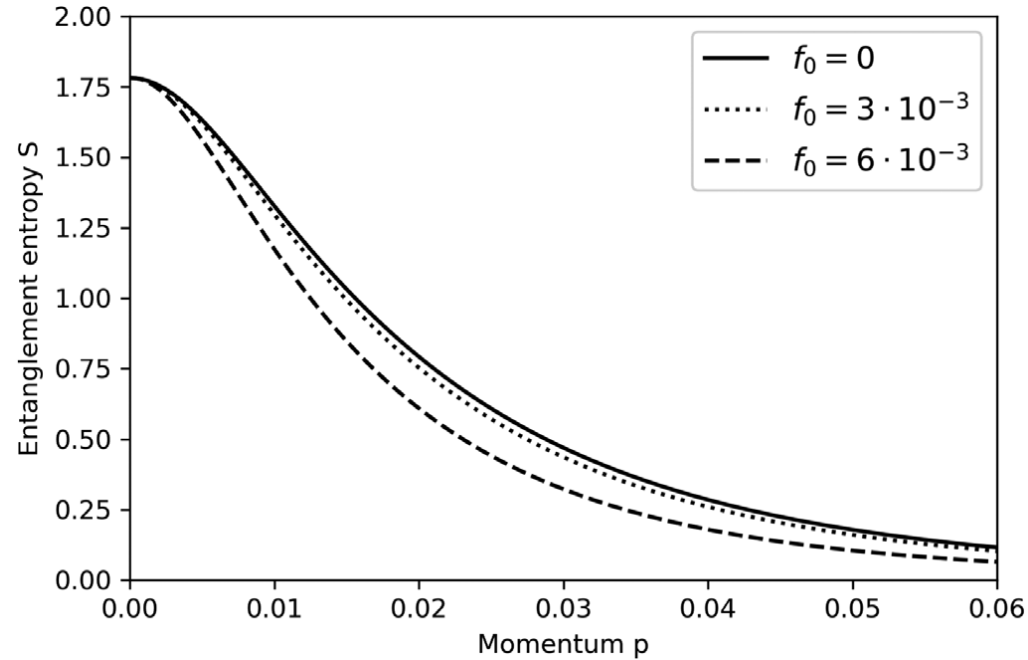}
    \includegraphics[scale=0.313]{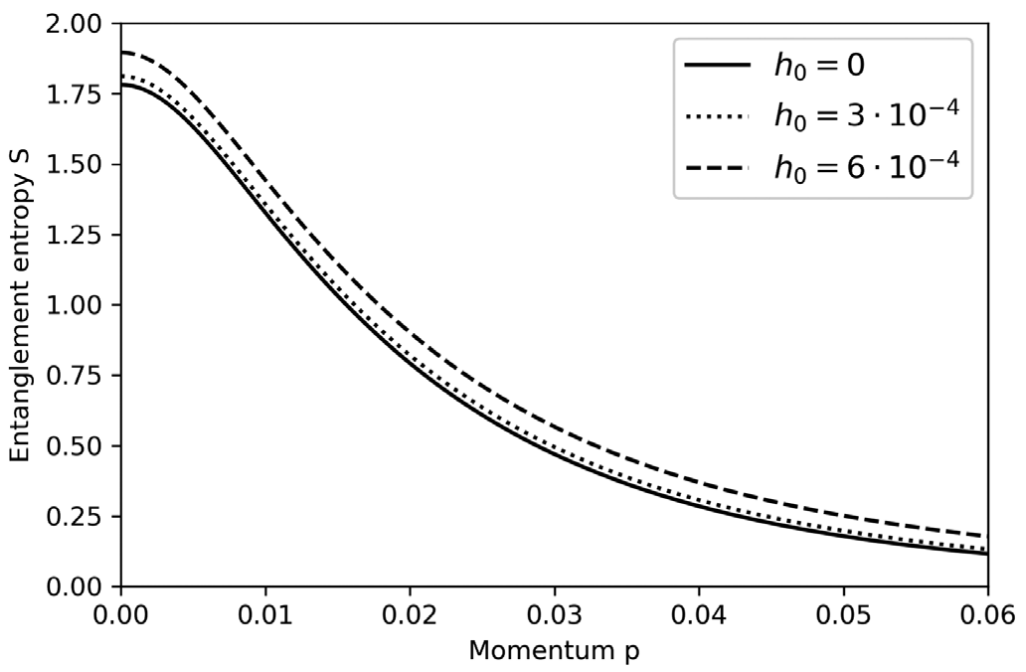}
    \caption{(Left) KG entanglement spectrum $S_{\rm KG}(\rho_k^{(\rm out)})$, for different values of the torsion parameter $f_0$. The other parameters are: $m=0.01$, $h_0=10^{-5}$, $A=3$, $B=2$, and $\rho=1$. (Right) The same entropy is plotted for varying $h_0$, with $f_0=10^{-5}$. Figure adapted from Ref. \cite{PhysRevD.104.043523}.}
    \label{fig_tor_KG}
\end{figure}

Eq. \eqref{kgeq_tor} can be simplified by assuming that the scalar field couples in the same way to all the components of the Riemann-Cartan scalar curvature\footnote{An alternative scenario is represented by a completely antisymmetric torsion \cite{Shapiro:2001rz,PhysRevD.95.095033}, which is also investigated in Ref. \cite{PhysRevD.104.043523}.}. In the case of conformal coupling, we would have $\xi_1= \dots = \xi_5=1/6$. Further recalling the ansatz of Eq. \eqref{KG_ans_scal}, analytical solutions for the KG equation can be found by setting
\be \label{ansa_KGtor}
f(\tau)=f_0a^3(\tau), \ \ \ \ \ \ \ h(\tau)=h_0a(\tau),
\ee
with $f_0,h_0$ constants. The evaluation of the entanglement entropy then follows the same steps of the torsionless scenario presented in Sec. \ref{sec4.1.1}. In Fig. \ref{fig_tor_KG} we show the particle entanglement spectrum, $S_{\rm KG}(\rho_k^{(\rm out)})$, for different values of the torsion parameters $f_0$ and $h_0$. We observe that a nonzero torsion typically modifies the amount and mode dependence of the entanglement entropy.

\subsubsection{Dirac field} \label{sec4.4.2}

\begin{figure}[tbp]
    \centering
    \includegraphics[scale=0.52]{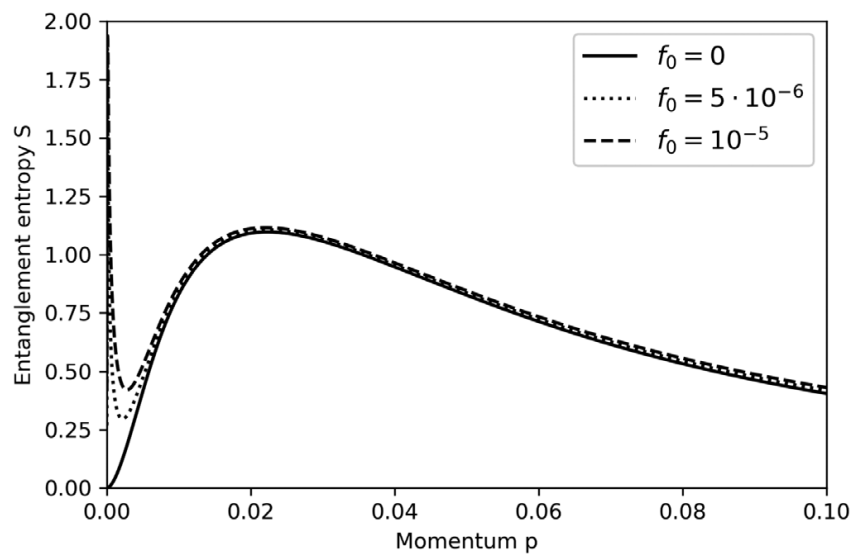}
    \caption{Dirac entanglement spectrum in presence of torsion. The values of the parameters are: $m=0.01$, $A=3$, $B=2$, $\rho=1$ and $r=6$. Figure adpated from Ref. \cite{PhysRevD.104.043523}.}
    \label{fig_dir_tor}
\end{figure}

In the presence of torsion, the Dirac Lagrangian of Eq. \eqref{dirac_ld} is modified according to \cite{Poplawski:2010kb,Cabral:2020mzw}
\begin{align}
    & D_\mu \psi \rightarrow D_\mu \psi - \frac{1}{4} K_{\alpha \beta \mu} \tilde{\gamma}^\alpha \tilde{\gamma}^\beta \psi, \\
    & D_\mu \psi \rightarrow D_\mu \bar{\psi} + \frac{1}{4} K_{\alpha \beta \mu} \bar{\psi} \tilde{\gamma}^\alpha \tilde{\gamma}^\beta,
\end{align}
where $K_{\alpha \beta \mu}$ is the contorsion tensor in its fully covariant form, namely
\be \label{contor_tens}
K_{\alpha \beta \mu} \equiv T_{\alpha \beta \mu} + 2 T_{(\beta \mu)\alpha}.
\ee
Recalling the ansatz \eqref{ansa_KGtor}, the Dirac equation in a FRW spacetime with torsion reads \cite{PhysRevD.104.043523}
\be \label{diraceq_tor}
\left[\frac{\gamma^{\mu}}{a} \left(\partial_{\mu}+ \frac{1}{4} \frac{\dot{a}}{a}[\gamma_{\mu}, \gamma^0] \right)+m  \right] \psi=- \frac{3i}{2} f(\tau)a \gamma^0 \gamma^5 \psi,
\ee
which can be solved via the standard ansatz of Eq. \eqref{dmodes_time}. The corresponding Dirac field modes can be derived by choosing $a(\tau)=A+B \tanh(\rho \tau)$ and
\be \label{anstor_dir}
f(\tau)= f_0\,a(\tau)^{-r},\ \ \ \ \ r \in \mathbb{N},\  r \geq 3,
\ee
where $f_0 \ll m$ to guarantee asymptotic flatness. At the same time, $t$ should be large enough to ensure that torsion quickly falls to zero when the universe starts its expansion.  

The particle entanglement entropy then takes the same form of Eq. \eqref{subentrop_dir} and it is plotted in Fig. \ref{fig_dir_tor}, showing that the torsion contribution becomes dominant at small momenta. In this limit, the entropy amount is negligible within torsionless scenarios, due to the Pauli exclusion principle. Accordingly, possible modifications to standard general relativity are expected to affect the entanglement profiles associated with GPP processes.


\subsection{Entanglement entropy of cosmological perturbations} \label{sec4.5}

Momentum-space entanglement techniques are recently gaining increasing attention in the characterization of cosmological perturbations \cite{Mukhanov:1990me,Riotto:2002yw,Brandenberger:2003vk}.

According to the standard inflationary picture \cite{Starobinsky:1980te,Guth:1980zm,RevModPhys.78.537}, quantum fluctuations associated with the inflaton field(s) are expected to generate the spacetime perturbations responsible for structure formation in our universe. 

Introducing a scalar inflaton field $\phi$, with corresponding Lagrangian density given by Eq. \eqref{scaldens}, we can expand it as
\be \label{infla_exp}
\phi({\bf x},\tau)= \phi_0({\bf x},\tau)+ \delta \phi({\bf x},\tau), 
\ee
thus separating the homogeneous background contribution $\phi_0$ from its unavoidable quantum fluctuations, described by $\delta \phi$. Such fluctuations then lead to the generation of inhomogeneities, resulting in the metric tensor \eqref{pert_FRW}. At linear perturbative order, a minimally coupled scalar field does not induce anisotropic stress \cite{Mukhanov:1990me,Brandenberger:2003vk}, so selecting the longitudinal gauge \cite{Bardeen:1980kt} we can describe the scalar degrees of freedom of inflationary perturbations in terms of a single potential, usually denoted by $\Phi$.

At first order, the potential $\Phi$ can be derived by perturbing Einstein equations, thus obtaining
\be \label{pertpot_1o}
\Phi^\prime+\mathcal{H} \Phi = \epsilon \mathcal{H}^2\frac{\delta \phi}{\phi^\prime},
\ee
where $\mathcal{H}=a^\prime/a$ and $\epsilon=  \left( V_{,\phi}/V \right)^2/16\pi G$
is the usual slow-roll parameter\footnote{In order to simplify the notation, we denote the background contribution $\phi_0$ by $\phi$ from now on.}. The slow-roll dynamics of the inflaton field is typically described via a quasi-de Sitter evolution \cite{Riotto:2002yw}, according to
\be \label{quasi_ds}
a(\tau)= - \frac{1}{H_I ( \tau -2\tau_f )^{1+\epsilon}},
\ee
where $H_I(\tau)$ is the inflationary Hubble rate and $\tau_f$ denotes the end of the slow-roll phase, while $\epsilon$ is assumed small and constant. Rescaling inflaton fluctuations by $\delta \chi= \delta \phi a$, we can describe their dynamics via 
\be \label{flures}
\delta \chi_k^{\prime \prime}+ \left[ k^2-\frac{1}{\eta^2} \left( 2+9\epsilon- \frac{ V_{,\phi \phi}}{H_I^2} \right)    \right] \delta \chi_k=0,
\ee
where $V_{,\phi \phi} \equiv \partial^2 V/\partial \phi^2$ and $\eta = \tau-2\tau_f$. The general solution of Eq. \eqref{flures} is usually derived by selecting the \emph{Bunch-Davies vacuum state} \cite{Bunch:1978yq,Danielsson:2003wb,Greene:2005wk} as the initial state for inflaton fluctuations, obtaining
\be \label{influct}
\delta \chi_k(\eta) = \frac{\sqrt{-\pi\eta}}{2} e^{i\left( \nu+ \frac{1}{2}\right) \frac{\pi}{2}}  H^{\left(1\right)}_{\nu}\left(-k\eta\right),
\ee
where $H_\nu^{(1)}$ is a Hankel function and $\nu= \sqrt{9/4+9\epsilon- V_{,\phi \phi}/H_I^2}$.

The inflaton fluctuations are stretched over cosmological distances during inflationary cosmic expansion, and entanglement can be generated between modes inside and outside the Hubble horizon. 

A proper characterization of entanglement generation processes associated with primordial fluctuations is then expected to shed further light on the inflationary epoch and on the properties of the cosmic microwave background (CMB) radiation.

\subsubsection{Squeezing entropy and quantum-to-classical transition} \label{sec4.5.1}

As we have seen, inflationary fluctuations are typically studied in Fourier space: to leading order, each Fourier mode evolves independently, obeying a harmonic oscillator equation with time-dependent mass (see Eq. \eqref{flures}). The comoving Hubble horizon $1/(aH_I)$ then emerges as a natural separation scale to describe the dynamics of inflaton fluctuations, which typically oscillate in time on sub-Hubble scales, while becoming frozen on super-Hubble ones \cite{Riotto:2002yw}.

From now on, we will consider the entanglement entropy associated with the space of super-Hubble modes
\be \label{suphub_space}
\mathcal{H}_A(t)= \prod_k \mathcal{H}_k,\ \ \ \ k< a H_I(\tau),
\ee
with respect to the ``bath" of sub-Hubble ones\footnote{Note that the boundary between the two Hilbert spaces $\mathcal{H}_A$ and $\mathcal{H}_B$ depends on time, and the dimension of $\mathcal{H}_A$ is increasing.}
\be \label{subhub_space}
\mathcal{H}_B(t)= \prod_k \mathcal{H}_k,\ \ \ \ k \geq a H_I(\tau),
\ee
where $\mathcal{H}_k$ is the harmonic oscillator Hilbert space of the $k$-th mode \cite{PhysRevD.102.043529}.

The mixing between sub and super-Hubble modes has been recently considered in several works (see e.g. Refs. \cite{Kiefer:1998qe,Martineau:2006ki,Burgess:2014eoa,Nelson:2016kjm}), and, more generally, decoherence processes \cite{Lombardo:2005iz,Franco:2011fg,Liu:2016aaf,Martin:2018lin} are expected to play a key role in explaining if and how cosmological perturbations become classical even though they have a quantum origin\footnote{See also Ref. \cite{Martin:2021znx} for a recent study of the relation between decoherence and quantum discord for primordial perturbations in a single-field inflationary model.}. 

A first evaluation of the entanglement entropy associated with scalar inflationary perturbations has been provided in Ref. \cite{PhysRevD.102.043529}, focusing on the role of cubic non-Gaussianities.
Working in the comoving gauge
\be \label{como_gaug}
ds^2= a^2(\tau) \left[ d\tau^2-(1+2\zeta) d{\bf x}^2  \right],
\ee
where $\zeta \equiv \zeta({\bf x},\tau)$ is the \emph{curvature perturbation}, we can consider the rescaled field \cite{PhysRevD.98.083535}
\be \label{resc_perturb}
s({\bf x},\tau)= z(\tau) \zeta({\bf x},\tau),
\ee
with 
\be \label{ztime} 
z^2(\tau)= 2 \epsilon a^2 M^2_{\rm pl} c_s^{-2}
\ee
and $c_s$ denoting the speed of sound of the matter source\footnote{In the following, we set $c^2_s=1$, which holds for single-field inflationary models with no derivative self-couplings.}, with $c^2_s= \partial P/\partial \rho$.
The metric and matter perturbations at first order can be canonically quantized as
\be \label{squeez_ham}
\hat{H}_2= \frac{1}{2} \int \frac{d^3k}{(2 \pi)^3} [c_s k ( \hat{c}_{\bf k} \hat{c}_{\bf k}^\dagger + \hat{c}_{-{\bf k}} \hat{c}_{-{\bf k}}^\dagger)]- \frac{1}{2} \int \frac{d^3k}{(2 \pi)^3} \left[i \left( \frac{z^\prime}{z} \right) (\hat{c}_{\bf k} \hat{c}_{-{\bf k}} - \hat{c}^\dagger_{\bf k} \hat{c}^\dagger_{-{\bf k}})    \right],
\ee
where, as usual, a prime denotes derivative with respect to conformal time. We notice that the second term in Eq. \eqref{squeez_ham}, describing the squeezing of perturbations, dominates on super-Hubble scales $k \ll aH_I$.

The Hamiltonian $\hat{H}_2$ leads to the following dynamics for ladder operators:
\be \label{solmot_ladd}
\hat{c}_{\bf k}(\tau)= e^{i \theta_k(\tau)} \cosh[r_k(\tau)] \hat{c}_{\bf k}(\tau_0) + e^{-i \theta_k(\tau)+2i\phi_k(\tau)} \sinh[r_k(\tau)] \hat{c}^\dagger_{-\bf k}(\tau_0),
\ee
where $\tau_0$ is a generic initial time and 
\begin{align}
    & r_k(\tau) = - \sinh^{-1} \left( \frac{1}{2 c_s k \tau} \right),\label{rsqueez}\\[3 pt]
    & \phi_k(\tau) = - \frac{\pi}{4}- \frac{1}{2} \tan^{-1}\left( \frac{1}{2 c_s k \tau} \right),\label{phisqueez} \\[3 pt]
    & \theta_k(\tau) = -k \tau - \tan^{-1}\left( \frac{1}{2 c_s k \tau} \right). \label{rotang}
\end{align}
From the Hamiltonian of Eq. \eqref{squeez_ham}, the time evolution operator acting on an initial vacuum state can be written as $\hat{U}_0=\hat{S}_k \hat{R}_k$, where $\hat{S}_k(r_k,\phi_k)$ and $\hat{R}_k(\theta_k)$ are the two-mode squeezing and rotation operators, respectively, defined by
\begin{align}
    & \hat{S}_k := \exp \left[ \frac{r_k}{2} (e^{-2i\phi_k} \hat{c}_{-{\bf k}} \hat{c}_{\bf k} -\text{h.c.}) \right], \\[3 pt]
    & \hat{R}_k := \exp [-i\theta_k(\hat{c}^\dagger_{\bf k} \hat{c}_{\bf k}+\hat{c}^\dagger_{-{\bf k}} \hat{c}_{-{\bf k}}+1)].
\end{align}
The effect of the rotation operator only consists in changing the global phase of the state, so it will be neglected from now on. The two-mode squeezing operator gives instead 
\begin{align} \label{squeezed_vac}
\ket{SQ(k,\tau)} & \equiv \hat{S}_k (r_k,\phi_k) \ket{0_k;0_{-k}} \notag \\[3pt]
&= \frac{1}{\cosh r_k} \sum_{n=0}^\infty e^{-2i n \phi_k} \tanh^n r_k \ket{n_k;n_{-k}},
\end{align}
where
\be \label{squevac}
\ket{n_k;n_{-k}} \equiv \left[ \frac{1}{n!} (\hat{c}^\dagger_{\bf k} \hat{c}^\dagger_{-{\bf k}})^n   \right]   \ket{0_k;0_{-k}}.
\ee
The squeezed vacuum corresponding to the complete set of modes is then simply given by the tensor product
\be \label{tens_modes} 
\ket{SQ(\tau)}= \prod_k \ket{SQ(k,\tau)},
\ee
leading to the density matrix
\begin{align} \label{squevac_dens}
\rho_{\rm sq}&= \ket{SQ(\tau)} \bra{SQ(\tau)}\notag \\
&= \prod_p \prod_k \sum_{n=0} \sum_{m=0} \frac{1}{\cosh r_k \cosh r_p} \notag \\[3pt]
& \ \ \ \ \ \ \times e^{-2i\phi_k(n-m)} \tanh^n r_k \tanh^m r_p \ket{n_k;n_{-k}} \bra{m_p;m_{-p}}.
\end{align}

If we now trace out either sub- or super-Hubble modes, the entanglement entropy corresponding to the reduced density operator is clearly zero, since the state displayed in Eq. \eqref{tens_modes} does not contain entanglement across the Hubble horizon.

A first attempt to get a non-vanishing von Neumann entropy from Eq. \eqref{squevac_dens} consists in neglecting its off-diagonal terms\footnote{Different arguments have been proposed to justify such procedure, e.g. the necessity to average over the squeezing angle \cite{PhysRevLett.69.3606,PhysRevD.48.2443} or to assume a distribution of coherent states as the initial state for the system, instead of the usual vacuum \cite{Gasperini:1993mq}.}, thus leading to the von Neumann entropy (per comoving volume)
\be \label{squered_entr}
S_{\rm sq}= \sum_k [(1+\sinh^2 r_k)\ln (1+\sinh^2 r_k)-\sinh^2 r_k \ln(\sinh^2 r_k)].
\ee
In the case of slow-roll inflation with an approximately constant Hubble parameter $H_I$, we can estimate the resulting entropy by focusing on super-Hubble modes which cross the horizon after the beginning of inflation. Setting then $a(\tau_0)=1$ when inflation starts, this condition is equivalent to integrate over momenta in the range $H_I < k < aH_I$, thus giving
\be \label{totentr_sque}
S_{\rm sq} \propto a^3 H_I^3,
\ee 
which provides the corresponding entropy density per unit volume 
\be \label{densentr_sque}
s_{\rm sq} \equiv \frac{S_{\rm sq}}{(2\pi a)^3} \propto H_I^3.
\ee
The classical and quantum features of inflationary squeezed states have been recently reviewed in Ref. \cite{PhysRevD.103.043521,Hsiang:2021kgh}, suggesting that primordial quantum perturbations may not decohere by themselves, even when highly squeezed. The above-presented squeezing procedure has also received some criticism \cite{Agullo:2022ttg}, which can be essentially traced back to the same ambiguities arising in GPP mechanisms when it is not possible to uniquely define particle states, e.g. during a phase of accelerated expansion.

Position-space entanglement characterization, by means of discretization procedures, may represent a viable alternative to probe the quantumness of primordial perturbations \cite{Maldacena:2012xp,Martin:2021qkg,Espinosa-Portales:2022yok, PhysRevD.109.023503} and we will come back to position-space techniques in Sec. \ref{secPOS}.

\subsubsection{Dynamics of non-Gaussianities} \label{sec4.5.2}

An additional contribution to the entanglement entropy of inflationary perturbations arises from cubic nonlinearities. Starting again from Eq. \eqref{como_gaug}, it can be shown that the dominant interaction term for perturbations at third order is given by the Hamiltonian \cite{Maldacena:2002vr,Chen:2006nt} 
\be \label{cubic_ham}
H_{int}= \frac{M^2_{\rm pl}}{2} \int d^3x \epsilon^2 a \zeta^2 (\partial^2 \zeta) ,
\ee
where, from now on, we set $c_s=1$. Accordingly, the total Hamiltonian for perturbations up to third order takes the form \cite{PhysRevD.102.043529}
\be \label{totham_per}
H_{\rm tot}= H_{\rm syst}+ H_{\rm bath}+ H_{int},
\ee
where $H_{\rm syst}$ and $H_{\rm bath}$ are given by Eq. \eqref{squeez_ham} restricted to super- and sub-Hubble modes, respectively. Starting from the vacuum state 
\be \label{totvac}
\ket{0,0}= \ket{0}_{k > aH} \otimes \ket{SQ(\tau)}_{k < aH},
\ee
where $\ket{SQ(\tau)}$ has been defined in Eq. \eqref{tens_modes}, we need to compute the probability amplitudes \cite{PhysRevD.86.045014}
\be \label{probamp}
\mathcal{C}_{n,N}= \bra{n,N} \left( -i \int_{\tau_0}^\tau d\tau^\prime H_{ int}(\tau^\prime) \right) \ket{0,0}+ \mathcal{O}(\lambda^2),
\ee
where $\lambda(\tau)=\sqrt{\epsilon}/(2\sqrt{2}a M_{\rm pl})$, with $\epsilon$ the usual slow-roll parameter, and $\ket{n}$ denotes a $n$-particle state for the system and $\ket{N}$ is the corresponding state for the bath. 

The quantization of the Hamiltonian in Eq. \eqref{cubic_ham} then leads to the following nonzero contributions:
\begin{itemize}
    \item[-] Terms of the form $\hat{c}^\dagger_{-{\bf k}} \hat{c}^\dagger_{-{\bf k}} \hat{c}^\dagger_{-{\bf k}}$. There can be either two system (super-Hubble) modes and one bath (sub-Hubble) mode or vice versa. 
    \item[-] Terms of the form $\hat{c}_{\bf k} \hat{c}^\dagger_{-{\bf k}} \hat{c}^\dagger_{-{\bf k}}$. There can be either two system modes and one bath mode or vice versa, but the annihilation operator must always correspond to the super-Hubble mode.
    \item[-] Terms of the form $\hat{c}_{\bf k} \hat{c}_{\bf k} \hat{c}^\dagger_{-{\bf k}}$. The two system modes must correspond to the two annihilation operators, so there is only one bath mode.
\end{itemize}
We also notice that terms of the form $\hat{c}_{\bf k} \hat{c}_{\bf k} \hat{c}_{\bf k}$ do not contribute to the probability amplitude in Eq. \eqref{probamp}, since the annihilation operator corresponding to any of the bath modes annihilates the Minkowski vacuum.

Moving to momentum space, we can perform the standard transformation \cite{PhysRevD.86.045014}
\be \label{postomom}
\sum_{n,N \neq 0} \rightarrow \sum_{\{ {\bf k}_j \} \gtrless \mu },
\ee
where $\mu$ will depend on the Hubble radius and $j=1,2,3$ for a cubic interaction.

The nonzero contributions to the probability amplitude $\mathcal{C}_{\{ {\bf k}_j \}}$ for the interaction Hamiltonian of Eq. \eqref{cubic_ham} have been computed in Ref. \cite{PhysRevD.102.043529}. The corresponding entanglement entropy takes the form, 
\be \label{entr_momsp}
S_{\rm cub}=- \int_{\{{\bf k}_j \} \gtrless \mu} \prod_{j}^3 d^3k_j \left[ \lvert \mathcal{C}_{\{{\bf k}_j \} \gtrless \mu} \rvert^2 \left( \ln\  \lvert \mathcal{C}_{\{{\bf k}_j \} \gtrless \mu} \rvert^2 -1\right)  \right] + \mathcal{O}(\lambda^3),
\ee
and one finds
\be \label{cubic_entro}
s_{\rm cub} \equiv \frac{S_{\rm cub}}{(2\pi a)^3} \simeq \mathcal{O}(1) \ln(\lambda^2) \epsilon H_I^2 M_{\rm pl} a^2.
\ee
Recalling now Eq. \eqref{densentr_sque} for the squeezing entropy, we can compute the ratio
\be \label{ratio_entr}
\frac{s_{\rm cub}}{s_{\rm sq}} \simeq \epsilon \left( \frac{M_{\rm pl}}{H_I} \right) a^2 \simeq 10^9 \left( \frac{H_I}{M_{\rm pl}} \right) e^{2N},
\ee
where in the last equation we assumed $\epsilon \simeq 10^9 (H_I/M_{\rm pl})^2$, as discussed in Ref. \cite{PhysRevD.102.043529}, and $N \equiv \ln\left[ a(\tau_f)/a(\tau_i) \right]$ is the number of inflationary e-foldings, with $\tau_i$ denoting the beginning of the inflationary phase.

Eq. \eqref{ratio_entr} implies that the entanglement entropy due to (cubic) gravitational nonlinearities is typically larger than the contribution associated with the squeezing part of the quadratic action, provided inflation lasts for a sufficiently long period of time, in agreement with observations \cite{Planck:2018jri}. The natural next step would be to compute the entanglement entropy corresponding to primordial gravitational waves, i.e., tensor modes of perturbations. Cubic tensor interactions have been recently considered in Ref. \cite{Brahma:2022yxu} by means of open-effective field theory (EFT) approaches. See also \cite{Salcedo:2024smn,Burgess:2022nwu,Banerjee:2021lqu} for applications of open EFT techniques to momentum-space inflationary perturbations.

\subsubsection{Entanglement from inhomogeneous particle production in inflation} \label{sec4.5.3}

Particle creation processes are expected to play a key role in entanglement generation and decoherence mechanisms during inflation. 

While the squeezing of perturbations presented in Sec. \ref{sec4.5.1} may be developed in the language of Bogoliubov coefficients (see e.g. Ref. \cite{Hsiang:2021qqo} for a discussion on this point), additional contributions may arise from perturbative production mechanisms due to inhomogeneities, which we have introduced in Sec. \ref{secQFT.3}. The presence of inhomogeneities indeed allows for mode mixing in particle creation processes and, accordingly, it may generate entanglement entropy across the Hubble horizon.

Superhorizon entanglement generation during inflation has been recently considered in Refs. \cite{Belfiglio:2022yvs,Belfiglio:2023moe} by coupling the inflaton quantum fluctuations to the scalar metric perturbations generated by the inflaton dynamics itself. 

Accordingly, one may assume an interaction Lagrangian density of the form of Eq. \eqref{intlag}, where 
\begin{align} \label{zerotens_flu}
T_{\mu \nu}^{(0)}=& \partial_{\mu} \delta \phi\  \partial_{\nu}\delta \phi-\frac{1}{2} g_{\mu \nu}^{(0)} \left[ g^{\rho \sigma}_{(0)}\  \partial_{\rho} \delta \phi\  \partial_{\sigma} \delta \phi - V(\delta \phi)  \right] \notag \\ 
&- \xi \left[ \nabla_{\mu} \partial_{\nu}- g_{\mu \nu}^{(0)} \nabla^\rho \nabla_\rho+R_{\mu \nu}^{(0)}-\frac{1}{2} R^{(0)} g_{\mu \nu}^{(0)}   \right] (\delta \phi)^2
\end{align}
is the energy-momentum tensor associated with the nonminimally coupled scalar inflaton fluctuations, $\delta \phi$, and the inflationary potential $V$ is left unspecified for the moment, while $\xi$ is the coupling constant to the spacetime scalar curvature. The time evolution of fluctuation modes is then described by Eq. \eqref{flures}, once the nonminimal coupling term is properly included in the inflaton dynamics.

Inflationary potentials of the form $V \propto \phi^{2n}$ ($n=1,2$) have been considered in Ref. \cite{Belfiglio:2023moe}, focusing on a quasi-de Sitter spacetime evolution, described by the scale factor \eqref{quasi_ds}.
    \begin{figure}
    \centering
    \includegraphics[scale=0.6]{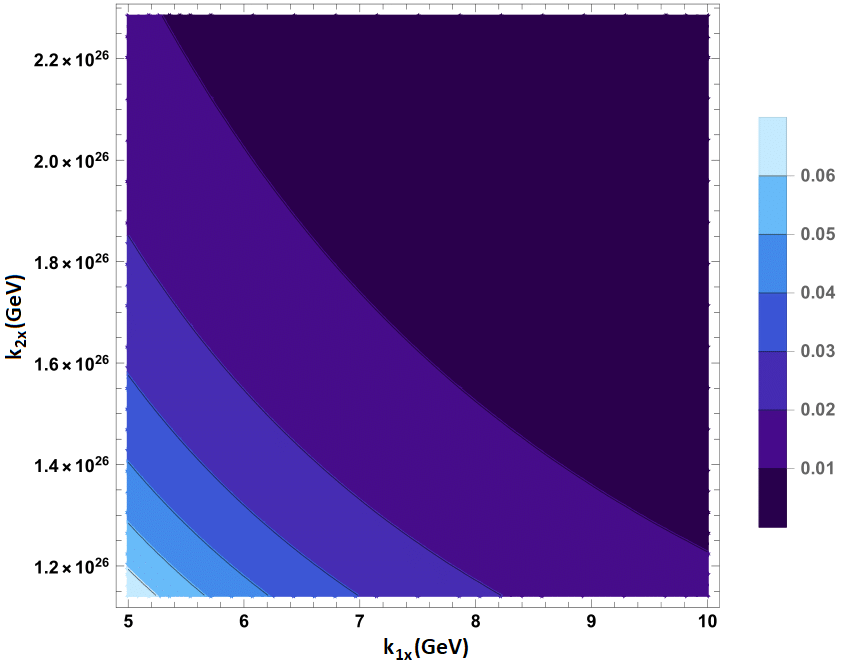}
    \includegraphics[scale=0.6]{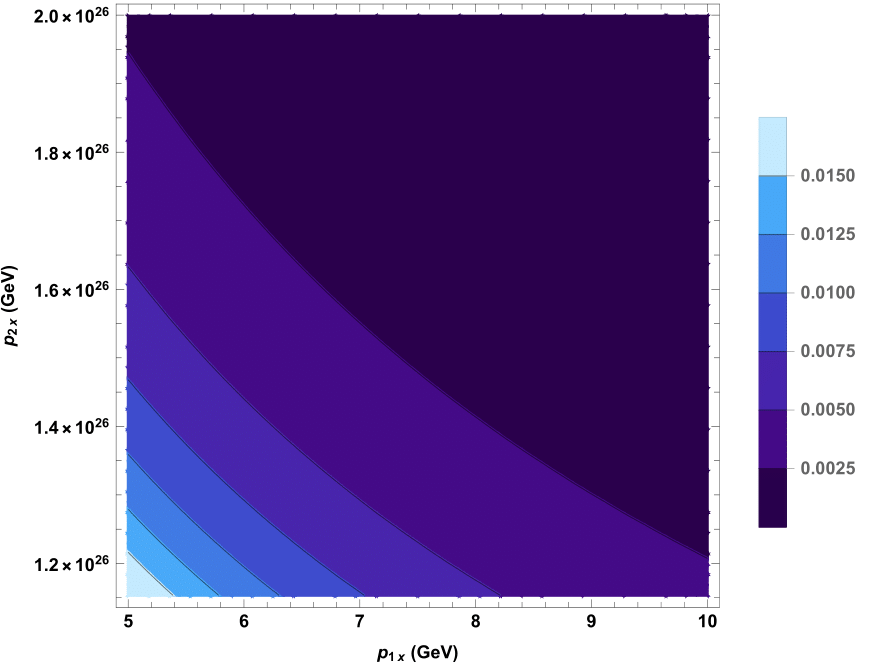}
    \caption{Superhorizon entanglement entropy $S_{\rm inh}$ at time $\tau_f$ from inhomogeneous particle production. The entropy is plotted as function of the super-Hubble mode $k_{1x}$ and the sub-Hubble mode $k_{2x}$, assuming a quartic self-coupling potential (left) and a quadratic one (right). The other momentum components are set to zero, for simplicity. We assume $\phi(\tau_i)=5$ $M_{\rm pl}$, $\lambda=10^{-15}$, $m=1.34\times 10^{12}$ GeV, $\epsilon=10^{-3}$, $\xi=10^{-4}$ and $\tau^*=\tau_0/1000=-10^{-3}$ GeV$^{-1}$. Figure adapted from \cite{Belfiglio:2023moe}.}
    \label{figquart_th}
    \end{figure}
Superhorizon pair production from inhomogeneities then implies
\be \label{setmom_flu}
\{ {\bf k}_j^{\rm inh} \}= \begin{cases}  a(\tau_i)H_I < \lvert {\bf k}_1 \rvert < a(\tau) H_I, \\ a(\tau) H_I < \lvert {\bf k}_2 \rvert < a(\tau) M_{\rm pl},   \end{cases}
\ee
where $M_{\rm pl}$ provides the natural ultraviolet cutoff and all modes of interest are in sub-Hubble form at the beginning of inflation, namely at $\tau_i$. Furthermore, a total number of e-foldings $N=60$ is imposed, where $\tau_f$ denotes the end of inflationary slow-roll.

The corresponding entanglement entropy can be expressed in the form
\be \label{entmom_fluctu}
S_{\rm inh}=- \int d^3k_1 d^3k_2 \   \lvert \mathcal{C}^{\rm inh}_{{\bf k}_1,{\bf k}_2} \rvert^2 \left( \ln\  \lvert \mathcal{C}^{\rm inh}_{{\bf k}_1,{\bf k}_2} \rvert^2 -1\right)   + \mathcal{O}(h^3),
\ee
where the probability amplitudes $\mathcal{C}^{\rm inh}_{{\bf k}_1,{\bf k}_2}$ are computed from Eqs. \eqref{pertpot_1o} and \eqref{influct}, following the prescriptions of Sec. \ref{secQFT.3}.

Within this picture, large-field inflationary models have been shown to produce an higher amount of entanglement entropy with respect to small-field ones. Below, we show the entanglement spectrum corresponding to $V=\lambda (\phi^2-v^2)^2/4$ (left) and $V=m^2\phi^2/2$ (right). Entanglement entropy is quantified in the range $\tau_* < \tau< \tau_f$, with $\tau_* > \tau_i$ to ensure the presence of super-Hubble modes. 

Here, particle production has been specified along the $x$ direction\footnote{The final result is independent of this choice, due to spherical symmetry.} and we notice that a significant amount of entanglement can be produced in such processes. We also observe that the entropy is larger at small momenta: in the case of super-Hubble modes, this implies that entanglement increases when approaching the IR cutoff and reflects the bosonic nature of inflationary fluctuations within this picture, in agreement with the non-perturbative results of Sec. \ref{sec4.1.1}.

\section{Entanglement harvesting} \label{secHARV}

We have seen that, when dealing with quantum fields in curved backgrounds, a significant amount of entanglement entropy can be found in Fourier space. However, this fact does not directly tell us how to reveal its presence experimentally and whether it can be employed for quantum information purposes. A natural possibility arises by locally coupling the quantum field to external detectors and by analyzing the emergence of quantum correlations between two (or more) of such spatially separated probes. 

One of the most striking implications of entanglement in QFT is the phenomenon known as \emph{entanglement harvesting} --the ability of two localized quantum systems (detectors) to become entangled via local interactions with a quantum field, even when they are separated by spacelike intervals \cite{Salton2015}. This possibility arises from the intrinsic entanglement structure of quantum field vacuum states, which allows for the presence of quantum correlations across spacelike hypersurfaces due to the Reeh-Schlieder theorem~\cite{ReehSchlieder, SummersWerner}. Despite the entanglement obtained from spacetime separated regions is generally very small, harvesting procedures are intimately related with the measurement problem of QFT \cite{Sorkin:1993gg,PhysRevD.104.025012}. Accordingly, they are sensitive to the structure of spacetime, and a proper description of detectors is crucial to understand the nature of the extracted entanglement, as we discuss below.

\subsection{Non-relativistic entanglement harvesting} \label{sec5.1}

The possibility for two causally disconnected detectors, interacting locally and briefly with the vacuum, to become entangled was initially pointed out in flat Minkowski spacetime \cite{Reznik:2002fz,Valentini:1991eah,PhysRevA.71.042104}.  The theoretical framework for such studies typically involves idealized models of detectors, most notably the Unruh-DeWitt (UDW) model~\cite{DeWitt1979}. 

\subsubsection{The Unruh-DeWitt detector model} \label{sec5.1.1}

In its simplest form, a Unruh-DeWitt detector is a \emph{two-level quantum system} (a qubit) that interacts locally with a real quantum scalar field\footnote{See Ref. \cite{PhysRevD.105.065016} for a generalization to complex scalar and Dirac fields and Ref. \cite{PhysRevD.108.085025} for a study of entanglement harvesting from the gravitational vacuum.} along a prescribed trajectory in spacetime. The interaction is governed by a time-dependent Hamiltonian of the form:
\begin{equation} \label{udw_interact}
    H_{\text{UDW}}(t) = \lambda \chi(t) \mu(t) \phi[x(t)],
\end{equation}
where:
\begin{itemize}
    \item[-] $t$ is the detector's proper time,
    \item[-] $\lambda$ is a small coupling constant,
    \item[-] $\chi(\tau)$ is a switching function controlling the duration and shape of the interaction,
    \item[-] $\mu(t)$ is the detector's monopole operator in the interaction picture,
    \item[-] $\phi[x(t)]$ is the scalar field operator evaluated along the detector's worldline $x(t)$.
\end{itemize}

The monopole operator $\mu(t)$ evolves freely and is typically expressed in terms of the ladder operators of the detector:
\begin{equation}
    \mu(t) = \sigma^+ e^{i \Omega t} + \sigma^- e^{-i \Omega t},
\end{equation}
where $\sigma^{\pm}$ are SU(2) ladder operators and $\Omega$ is the energy gap between the detector's ground and excited states.

The UDW model assumes detectors to interact \emph{locally} with quantum fields, making it ideal for studying phenomena sensitive to causal structure, locality, and vacuum entanglement.

\subsubsection{Basics of entanglement harvesting} \label{sec5.1.2}

In the context of \emph{entanglement harvesting}, two (or more) Unruh-DeWitt detectors are placed at different locations in spacetime, and they are allowed to interact locally with the same quantum field vacuum. Focusing on the standard framework involving two particle detectors (A and B), the total initial state is typically taken as:
\begin{equation}
    \rho_0 = \rho_{AB,0} \otimes \ket{0}\bra{0},
\end{equation}
where $\ket{0}$ is the quantum field vacuum, while $\rho_A$ and $\rho_B$ describe the initial states of the detectors and usually coincide with their respective ground states, namely
\be \label{gs_udw}
\rho_{AB,0}= \ket{g_A} \bra{g_A} \otimes \ket{g_B} \bra{g_B}.
\ee
From Eq. \eqref{udw_interact}, we can then write the total interaction Hamiltonian as \cite{PhysRevD.92.064042}
\be \label{totalint_udw}
H_{\rm int}= \sum_{\nu \in \{ A,B \}} \lambda_\nu \chi_\nu(t) \mu_\nu(t) \int d^3x F_\nu({\bf x}-{\bf x}_\nu) \phi({\bf x},\tau),
\ee
where the $F_\nu$ are the spatial smearing functions corresponding to each detector, while ${\bf x}_\nu$ denote the corresponding center-of-mass positions. From Eq. \eqref{totalint_udw}, in the limit of weak interaction we can obtain the time evolution of the total system via a perturbative Dyson expansion\footnote{This approach inevitably leads to violation of the microcausality condition when the detectors are not pointlike, as we will discuss below.} of the time evolution operator
\be \label{dysont_udw}
U= \mathbb{I}- i \int_{-\infty}^{\infty} dt H_{\rm int}(t)- \int_{-\infty}^{\infty} dt \int_{-\infty}^t dt^\prime H_{\rm int}(t) H_{\rm int}(t^\prime)+ \dots\, .
\ee
The final state of the system is then $\rho_f=U \rho_0U^\dagger$ and, to leading order in the coupling strength, we find 
\begin{equation} \label{densop_udwde}
    \rho_{AB} = \tr_{\phi}\left( \rho_f \right)= \begin{pmatrix}
        1-\mathcal{L}_{AA}-\mathcal{L}_{BB} & 0 & 0 & \mathcal{M} \\
        0 & \mathcal{L}_{AA} & \mathcal{L}_{AB} & 0 \\
        0 & \mathcal{L}_{BA} & \mathcal{L}_{BB} & 0 \\
        \mathcal{M}^* & 0 & 0 & 0 
    \end{pmatrix} + \mathcal{O}(\lambda_\nu^4),
\end{equation}
in the basis $\{ \ket{g_A} \otimes \ket{g_B}, \ket{e_A} \otimes \ket{g_B}, \ket{g_A} \otimes \ket{e_B}, \ket{e_A} \otimes \ket{e_B}  \}$, where $\ket{e_\nu}$ denotes the excited state of each detector. The explicit expressions for $\mathcal{L}_{\mu \nu}$ and $\mathcal{M}$ can be found after properly expanding the scalar field, assumed to be massless, in terms of plane-wave modes,
\be \label{scafiel_quant}
\phi({\bf x},t)= \int \frac{d^3k}{\sqrt{(2\pi)^3 2k}} \left[ \hat{a}_{\bf k}^\dagger e^{i ( kt- {\bf k}\cdot {\bf x})} + h.c.  \right],
\ee
where, as usual, $[\hat{a}_{\bf k},\hat{a}_{{\bf k}^\prime}]= \delta^{(3)}({\bf k}-{\bf k}^\prime)$, then obtaining
\begin{align}
   & \mathcal{L}_{\mu \nu}= \int d^3k L_\mu({\bf k}) L_\nu({\bf k})^*,\\
   & \mathcal{M}= \int d^3k M ({\bf k}),
\end{align}
with
\begin{align}
    & L_\mu({\bf k})= \lambda_\mu \frac{ e^{- i {\bf k} \cdot {\bf x}_\mu} \tilde{F}({\bf k})}{\sqrt{2k}} \int_{-\infty}^\infty dt_1 \chi_\mu(t_1) e^{i (k+ \Omega_\mu)t_1}, \label{l_gene} \\[6pt]
    & M({\bf k})= -\lambda_A \lambda_B e^{i {\bf k} \cdot({\bf x}_A-{\bf x}_B)} \frac{[ \tilde{F}({\bf k}) ]^2}{2k} \int_{-\infty}^\infty dt_1 \int_{-\infty}^{t_1} dt_2 e^{-ik(t_1-t_2)} \notag \\
    &\ \ \ \ \ \ \ \ \ \ \ \ \ \ \  \times \left[ \chi_A(t_1) \chi_B(t_2) e^{i(\Omega_At_1+\Omega_Bt_2)} + \chi_B(t_1) \chi_A(t_2) e^{i(\Omega_Bt_1+\Omega_At_2)}  \right], \label{m_gene}
\end{align}
and we have assumed $\tilde{F}_\nu({\bf k}) = \tilde{F}_\nu(-{\bf k}) = \tilde{F}({\bf k})$, where
\be \label{four_smea}
\tilde{F}_\nu({\bf k}) = \frac{1}{(2\pi)^{3/2}} \int d^3x F_\nu({\bf x}) e^{i {\bf k} \cdot {\bf x}}.
\ee
The above expressions are general, and can be then specified to various switching functions and spatial profiles, see e.g. Ref. \cite{PhysRevD.92.064042}.

In order to quantify the entanglement acquired by detectors after the interaction with the field $\phi$, the negativity $\mathcal{N}$ is usually employed \cite{PhysRevA.65.032314}, 
\be \label{negat_har}
\mathcal{N}(\rho_{AB})= \left \lvert \sum_{E_i < 0}  E_i \right \rvert,
\ee
where $E_i$ are the eigenvalues of $\rho_{AB}^{\Gamma_A}$, i.e., the partial transpose of $\rho_{AB}$ with respect to the subsystem A.

To second order in perturbation theory, there is only one eigenvalue of $\rho_{AB}^{\Gamma_A}$ that can be negative, namely
\be \label{neg_eig}
E_1= \frac{1}{2} [\mathcal{L}_{AA}+\mathcal{L}_{BB}-\sqrt{(\mathcal{L}_{AA}-\mathcal{L}_{BB})^2+4 \lvert \mathcal{M}\rvert^2}] + \mathcal{O}(\lambda_\nu^4).
\ee
Accordingly, we obtain
\be \label{secord_neg}
\mathcal{N}= \text{max}(0, \mathcal{N}^{(2)}),
\ee
where $\mathcal{N}^{(2)}=-E_1$. If the two detectors are identical and they are switched on for the same amount of time, we further obtain $\mathcal{L}_{AA}= \mathcal{L}_{BB} \equiv \mathcal{L}_{\mu \mu}$, thus leading to the simplified expression
\be \label{simpl_neg}
\mathcal{N}^{(2)}= \lvert \mathcal{M} \rvert-\mathcal{L}_{\mu \mu}.
\ee
Eq. \eqref{simpl_neg} then shows that, in order to have nonzero entanglement, the nonlocal term $\mathcal{M}$ has to be larger than the local contributions $\mathcal{L}_{\mu \mu}$. This typically depend on the details of the switching and the spatial profile of the detectors \cite{PhysRevD.92.064042,PhysRevD.108.085025,PhysRevD.95.105009,PhysRevD.103.065013}. Similarly, the presence of additional detectors has been shown to affect entanglement extraction \cite{Mendez-Avalos:2022obb}.

\begin{figure}
    \centering
    \includegraphics[scale=0.39]{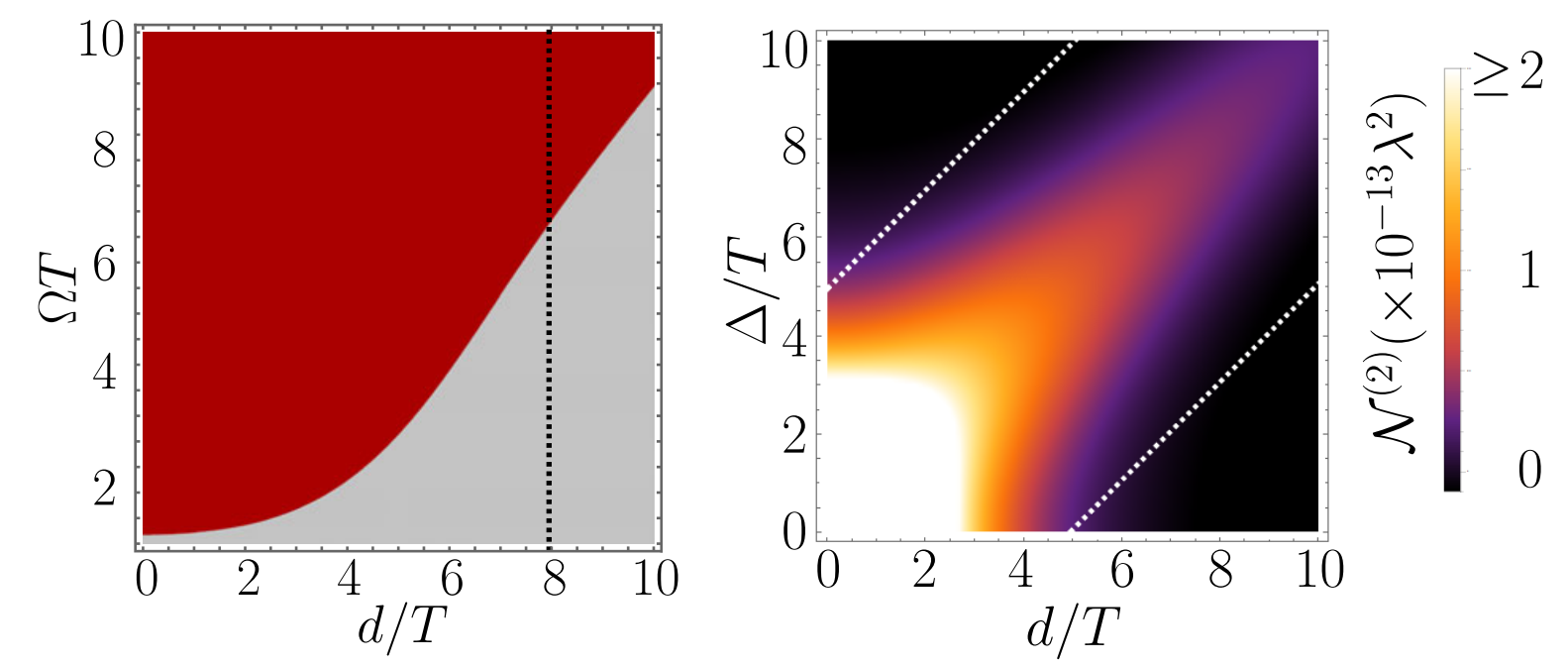}
    \caption{Three-dimensional entanglement harvesting assuming a Gaussian switching function and a Gaussian spatial profile for the two detectors. (Left) The red region shows the values of the spatial separation $d/T$ and the energy gap $\Omega T$ for which entanglement harvesting is possible, while within the grey region there is no entanglement harvesting. It is further assumed $\Delta/T=3$. (Right) Negativity $\mathcal{N}^{(2)}$ as function of $d/T$ and the time delay $\Delta/T$. It is further assumed $\Omega T=7$. In both plots the dashed lines represent the boundaries of the light cone and $\sigma/T=1$. Figure adapted from \cite{PhysRevD.95.105009}.}
    \label{fig_harv_GG}
\end{figure}

\paragraph{Example: Gaussian switching function and Gaussian smearing}

In order to quantify the amount of extracted entanglement in a concrete scenario, let us consider a non-negligible Gaussian smearing (quantified by $\sigma$) for the two detectors, implying 
\be \label{gau_smea}
F({\bf x})= \frac{1}{(\sqrt{\pi}\sigma)^3} e^{-\lvert {\bf x}\rvert^2/\sigma^2},
\ee
which readily gives $\tilde{F}({\bf k})=e^{-\lvert {\bf k} \rvert^2 \sigma^2/4}/(2\pi)^{3/2}$. 

We also introduce $d=\lvert {\bf x}_A-{\bf x}_B \rvert$ and $\Delta=t_B-t_A$, further assuming that the detectors are identical, namely $\Omega_A=\Omega_B\equiv \Omega$ and $\lambda_A=\lambda_B\equiv \lambda$. In order to simulate a smooth switching of the detectors, we select the Gaussian function \cite{PhysRevD.92.064042}
\be \label{gau_switch} 
\chi_\nu(t)= e^{-(t-t_\nu)/T^2},
\ee
where $T$ is a characteristic time scale. Using the ansatz of Eq. \eqref{gau_switch}, the time integrals in Eqs. \eqref{l_gene}-\eqref{m_gene} can be computed analytically.

We report in Fig. \ref{fig_harv_GG} some features of the harvesting procedure, highlighting the parameter region where entanglement can be efficiently harvested. We observe that, even if the detectors remain \emph{spacelike separated} during the interaction (i.e., there is no causal connection between them), their final state $\rho_{AB}$ can be \emph{entangled}. In this case, since the detectors cannot exchange information, entanglement solely arises from the harvesting of preexisting correlations in the quantum vacuum, thus corresponding to a \enquote{genuine} entanglement harvesting procedure.

In particular, we notice that it is always possible to obtain spacelike entanglement harvesting if the internal gap $\Omega$ of the detectors is sufficiently large, despite a larger $\Omega$ reduces the amount of the harvested entanglement. It can be further shown that detectors switched on in a smooth, more adiabatic manner are capable of harvesting entanglement in a more efficient way than suddenly switched detectors, both when the detectors are in causal contact and when they are spacelike separated \cite{PhysRevD.92.064042,PhysRevD.96.065008,PhysRevD.97.125002}. The harvesting protocol is also sensitive to the trajectories \cite{Salton2015} of the detectors, boundary
conditions of the field \cite{Brown:2014pda}, the nature of the detector-field couplings \cite{PhysRevD.96.085012}, as well as the geometry \cite{PhysRevD.79.044027} and topology \cite{PhysRevD.93.044001} of the background spacetime.


\subsection{Toward a fully relativistic approach} \label{sec5.2}

Despite their success in capturing essential features of field-detector interactions, UDW models are fundamentally non-relativistic: they assume pre-defined detector trajectories, fixed internal Hilbert spaces, and neglect dynamical backreaction. This may cause problems with covariance and causality \cite{PhysRevD.87.064038,PhysRevD.103.025007,PhysRevD.103.085002,PhysRevD.108.045015} and it has legitimately generated some doubts whether entanglement harvesting may provide trustworthy results \cite{Ruep:2021fjh}.

\subsubsection{Localized quantum fields as detectors} \label{sec5.2.1}

A solution to this issue may consist in replacing external detectors with localized quantum fields \cite{PhysRevD.109.045013}, which are appropriately modeled by constraining free quantum fields via external potentials. The localized field is then allowed to interact with a free field theory, thus acting as a fully relativistic localized probe.

Let us consider a scalar field $\phi_T$ in flat Minkowski spacetime, evolving under the action of a trapping potential $V$, according to the Lagrangian density
\be \label{loca_lag}
\mathcal{L}_T= \frac{1}{2} \partial_\mu \phi_T \partial_\mu \phi_T- \frac{m^2_T}{2} \phi_T^2 - V({\bf x}) \phi_T^2,
\ee
where $m_T$ is the field mass and the field coordinates are comoving with the source of the potential. The field is then quantized as usual according to:
\be \label{loca_exp}
\hat{\phi}_T= \sum_{\bf n} \hat{a}_{\bf n} e^{-i \omega_{\bf n}t} \Phi_{\bf n}({\bf x}) + \hat{a}_{\bf n}^\dagger e^{i \omega_{\bf n}t} \Phi_{\bf n}^*({\bf x}),
\ee
where the field modes satisfy
\be \label{mod_loca}
\left( -\nabla^2 + m_T^2 +2V({\bf x})  \right) \Phi_{\bf n} = \omega_{\bf n}^2 \Phi_{\bf n},
\ee
and they are appropriately normalized in $L^2(\mathbb{R}^3)$ via the Klein-Gordon scalar product, giving 
\be \label{loca_norml2}
\int d^3 x \ \lvert \Phi_{\bf n} \rvert^2 = \frac{1}{2 \omega_{\bf n}}.
\ee
A confining potential $V({\bf x})$ implies that each of the eigenfunctions $\Phi_{\bf n}({\bf x})$ is mostly localized
around the minima of $V({\bf x})$ and it decays to zero as $V({\bf x})$ increases. The confining nature of $V({\bf x})$ also allows to perform operations such as partial tracing over modes and independently describe the nature of each mode ${\bf n}$. 

In order to restrict the field to a finite region of space, one may employ a potential which is only finite in a compact convex connected set of $U \subset \mathbb{R}^3$, while it is infinite outside $U$ \cite{PhysRevD.109.045013}. The simplest example of such potentials is represented by a perfectly reflective cavity (infinite potential well)
\be \label{quad_pot_harv}
V({\bf x})= \begin{cases}
     0, \ \ \ \ \ {\bf x} \in U_d, \\
     \infty, \ \ \ \ \ {\bf x} \notin U_d.
\end{cases}
\ee
where $U_d=[0,d]^3$, effectively implementing Dirichlet boundary conditions. Looking for solutions of the form $\phi_T= e^{-iEt} \Phi_{\bf n}({\bf x})$, one finds 
\be \label{mode_quad_harv}
\Phi_{\bf n}({\bf x})= \frac{1}{\sqrt{2 \omega_{\bf n}}} f_{n_x}(x) f_{n_y}(y) f_{n_z}(z),
\ee
where ${\bf n}=(n_x,n_y,n_z)$ and 
\be \label{modes_quad_harv}
f_n(u)= \sqrt{\frac{2}{d}} \sin\left( \frac{\pi n u }{d} \right),
\ee
while 
\be \label{freq_quad_harv}
\omega_{\bf n}= \sqrt{m_T^2+ \frac{\pi^2}{d^2}(n_x^2+n_y^2+n_z^2)}.
\ee
From Eqs. \eqref{mode_quad_harv}-\eqref{freq_quad_harv}, we then obtain the proper quantization of the field via Eq. \eqref{loca_exp}.


\subsubsection{Fully relativistic entanglement harvesting } \label{Sec5.2.2}

Let us consider now two localized real scalar fields $\phi_A$ and $\phi_B$, evolving in Minkowski flat spacetime under the action of the cooresponding confining potentials $V_A({\bf x})$ and $V_B({\bf x})$, according to the above prescriptions. As a consequence of the confining procedure, both fields will have discrete modes, labeled by ${\bf n}_A$ and ${\bf n}_B$, with associated vacuum states $\ket{0_A}$ and $\ket{0_B}$.

Each of these fields is assumed to interact with a free Klein-Gordon field $\phi({\bf x})$, according to the interaction Hamiltonian density
\be \label{int_ham}
\mathcal{H}_I= \lambda \big(\zeta_A({\bf x}) \phi({\bf x}) \phi_A({\bf x})+\zeta_B({\bf x}) \phi({\bf x}) \phi_B({\bf x})\big),
\ee
where $\zeta_A({\bf x})$ and $\zeta_B({\bf x})$ are spacetime smearing functions, which are localized in time.

Let us assume now that, for both localized fields, we only have access to a limited number of modes, denoted by ${\bf N}_A$ and ${\bf N}_B$, respectively. Tracing over all the other modes of both fields, it can be shown that, at first perturbative order, the final state for the localized fields is exactly the same as one would obtain by considering harmonic oscillator particle detector models, with the corresponding Hamiltonian density
\be \label{eff_ham_harv}
\mathcal{H}_{\rm eff}= \lambda \big( \hat{Q}_{{\bf N}_A}^A({\bf x}) \hat{\phi}({\bf x}) +  \hat{Q}_{{\bf N}_B}^B({\bf x}) \hat{\phi}({\bf x}) \big),
\ee
where 
\begin{align}
    & \hat{Q}_{{\bf N}_A}^A({\bf x}) = \Lambda_A ({\bf x}) e^{-i\Omega_A t} \hat{a}^A_{{\bf N}_A} + \Lambda_A^* ({\bf x}) e^{i\Omega_A t} \hat{a}^{A \dagger}_{{{\bf N}_A}}, \\[4pt]
    & \hat{Q}_{{\bf N}_B}^B({\bf x}) = \Lambda_B ({\bf x}) e^{-i\Omega_B t} \hat{a}^B_{{\bf N}_B} + \Lambda_B^* ({\bf x}) e^{i\Omega_B t} \hat{a}^{ B \dagger}_{{{\bf N}_B} },
\end{align}
with $\Omega_i \equiv \omega_{{\bf N}_i}$ ($i=A,B$), and the effective smearing functions $\Lambda$ are given by 
\begin{align}
    & \Lambda_A({\bf x}) = \zeta_A({\bf x}) \Phi^A_{{\bf N}_A} ({\bf x}), \label{spacet_smea_A} \\
    & \Lambda_B({\bf x}) = \zeta_B({\bf x}) \Phi^B_{{\bf N}_B} ({\bf x}). \label{spacet_smea_B}
\end{align}
This outcome implies that, to leading order, previous studies adopting non-relativistic probes can be traced back to localized field scenarios. Furthermore, preserving the leading-order approximation, different modes of the probe fields do not interact with each other, so each pair of modes may acquire some amount of entanglement, and not only the two modes labeled by ${\bf N}_A$ and ${\bf N}_B$. This means that, in general, localized fields may harvest more entanglement with respect to harmonic oscillator detectors. Accordingly, UDW detectors only provide a \emph{lower bound} to the total entanglement that can be extracted via entanglement harvesting protocols.  \cite{PhysRevD.109.045018}. This confirms that entanglement harvesting is not merely a byproduct of non-relativistic modeling, but rather a genuine feature of quantum field correlations.

\paragraph{Example: localized fields in a cubic box} As an example of fully relativistic harvesting protocol, let us come back to the localized quantum fields $\phi_A$ and $\phi_B$ introduced in Sec. \ref{sec5.2.1}. Both fields interact with a free scalar field $\phi$, according to the Hamiltonian \eqref{eff_ham_harv}, where we set \cite{PhysRevD.109.045018}
\be \label{smea_cubic}
\zeta_A({\bf x}) = \zeta_B({\bf x}) = e^{-\frac{\pi t^2}{T^2}},
\ee
describing interactions that are adiabatically switched on, with $T$ controlling the effective timescale of the switching. The effective spacetime region where the localized fields interact with $\phi$ is then defined by the quantities in Eqs. \eqref{spacet_smea_A}-\eqref{spacet_smea_B}, and the localized field modes are displayed in Eq. \eqref{mode_quad_harv}.
\begin{figure}[ht]
    \centering
    \includegraphics[scale=0.45]{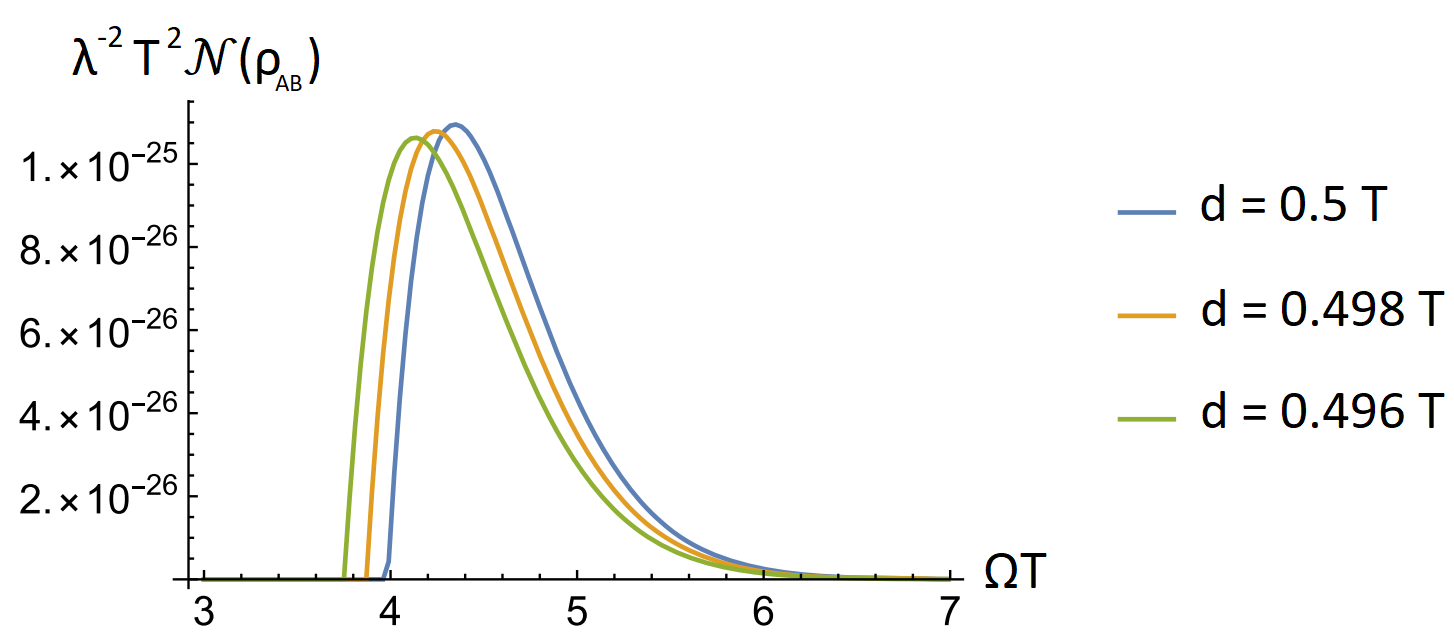}
    \caption{Rescaled entanglement negativity for the state $\rho_{AB}$, corresponding to two localized quantum detectors, confined in boxes of side $d$. The negativity is plotted as function of the energy $\Omega$ and the separation between the detectors' interaction regions is set to $L=4.5T$. Figure adapted from \cite{PhysRevD.109.045018}.}
    \label{fig_harv_fullrel}
\end{figure}

We assume the localized fields to start in their lowest energy state, according to $\rho_0 = \ket{0_A}\bra{0_A} \otimes \ket{0_B} \bra{0_B} \otimes \rho_\phi$, where $\rho_\phi$ is a zero-mean Gaussian state for the free field $\phi$. Furthermore, we suppose to only access the localized fields' excitations with lowest energy\footnote{The results discussed below can be generalized to a much more general class of initial states and accessible modes for the probe fields, see e.g. Ref. \cite{PhysRevD.109.065004}, where a path-integral approach is employed.}, namely ${\bf 1}_A={\bf 1}_B \equiv (1,1,1)$, with corresponding energy $\omega_{{\bf 1}_A}^A= \omega_{{\bf 1}_B}^B= \sqrt{m_T^2+ 3\pi^2/d^2}$.

In order to ensure negligible communication between the two detectors, we pick $d \simeq 0.5 T$, while the distance between the two cavities is assumed to be $L=4.5T$. The amount of harvested entanglement can then be computed according to the standard procedure described in Sec. \ref{sec5.1.2}. In particular, the initial density operator $\rho_0$ evolves as
\begin{align}
\rho_f&= U \rho_0 U^\dagger \notag \\
&= \rho_0 + U^{(1)} \rho_0 + \rho_0 U^{(1)\dagger} + U^{(1)} \rho_0 U^{(1)\dagger} + U^{(2)} \rho_0 + \rho_0 U^{(2) \dagger} + \mathcal{O}(\lambda^3),
\end{align}
where
\be \label{evop_fullyrel}
U= \mathcal{T} \exp \left( -i \int dV \mathcal{H}_{\rm eff}    \right)= \mathbb{I}+U^{(1)} + U^{(2)} + \mathcal{O}(\lambda^3),
\ee
with $dV$ the invariant spacetime volume element and the interaction Hamiltonian density $\mathcal{H}_{\rm eff}$ was introduced in Eq. \eqref{eff_ham_harv}.

Following the same steps of Sec. \ref{sec5.1.2} and setting $\omega_{{\bf 1}_A}= \omega_{{\bf 1}_B} \equiv \Omega$, it can be shown that the entanglement negativity is again described by Eqs. \eqref{secord_neg}-\eqref{simpl_neg}, where
\begin{align}
    & \mathcal{L}= \lambda^2 \int dV dV^\prime \Lambda_A({\bf x}) \Lambda_A({\bf x}^\prime) e^{-i \Omega (t-t^\prime)} W^+({\bf x},{\bf x}^\prime) \notag \\
    &\ \ \ =\lambda^2 \int dV dV^\prime \Lambda_B({\bf x}) \Lambda_B({\bf x}^\prime) e^{-i \Omega (t-t^\prime)} W^+({\bf x},{\bf x}^\prime), \\[4pt]
    & \mathcal{M}= - \lambda^2 \int dV dV^\prime \Lambda_A({\bf x}) \Lambda_B({\bf x}^\prime) e^{i \Omega (t+t^\prime)} G_F({\bf x},{\bf x}^\prime),
\end{align}
with $\Omega= \sqrt{m^2_T+ 3\pi/d^2}$, and we have introduced
\begin{align}
    & W^+({\bf x},{\bf x}^\prime) = \tr_\phi \left( \phi({\bf x}) \phi({\bf x}^\prime) \rho_\phi \right), \\
    & G_F({\bf x}, {\bf x}^\prime) = W^+({\bf x},{\bf x}^\prime) \theta(t-t^\prime) + W^+({\bf x}^\prime,{\bf x}) \theta(t^\prime -t),
\end{align}
denoting the Wightman function and the Feynman propagator of the field in the state $\rho_\phi$, respectively. The rescaled negativity corresponding to $\rho_{AB} \equiv \tr_\phi \left( \rho_f \right)$ is plotted in Fig. \ref{fig_harv_fullrel} as function of the energy $\Omega$ of the localized mode. 

In agreement with the non-relativistic behaviour depicted in Fig. \ref{fig_harv_GG}, we observe that there is a threshold in the energy gap $\Omega$ below which no entanglement can be harvested. At the same time, if the gap becomes too large, the extracted entanglement quickly decays to zero. The same behaviour has been shown in the case of localized field detectors under the influence of quadratic potentials or in the presence of a single detector field with two or more localization regions \cite{PhysRevD.109.045018}.

The presence of nonzero entanglement in spacelike scenarios then confirms that the harvesting protocol does not emerge from non-relativistic features of the detectors, representing instead a true prediction of QFT. Alternative approaches to the measurement problem in QFT were recently proposed in the framework of algebraic QFT \cite{Fewster:2018qbm,Fewster:2024pur}, and possible connections between detectors-based approaches and algebraic ones are currently object of study \cite{PhysRevD.109.045013,PhysRevD.109.065004,PhysRevD.105.065003,Papageorgiou:2023nvf,PhysRevD.111.045016}.

In light of the above-presented developments, we observe that the relevance of entanglement harvesting is not only limited to the future possibility of directly employ the extracted entanglement in quantum information scenarios. In fact, it also offers a fertile ground for exploring fundamental aspects of quantum field theory, such as the structure of vacuum entanglement, the operational meaning of locality, and the interplay between measurement, information, and geometry in relativistic settings.

\section{Position-space entanglement techniques} \label{secPOS}

Observations and measurements are typically performed in real space and entanglement between spatially separated regions may provide important insights into cosmology and the dynamics of quantum fields. Position-space entanglement in QFT was first considered in Ref. \cite{PhysRevD.34.373}, in the attempt to find a quantum origin to black hole entropy. In this paper, the entanglement entropy of a discrete scalar field was shown to obey an area law, provided the degrees of freedom residing inside an imaginary surface are traced out, thus suggesting a possible connection to the area law scaling of Bekenstein-Hawking entropy \cite{Bekenstein:1972tm,Bekenstein:1973ur,Bekenstein:1974ax,Hawking:1974rv,Hawking:1975vcx}.  We will focus on these cosmological developments, while the interested reader may consult the recent articles \cite{RevModPhys.90.035007,RevModPhys.90.045003} for a more detailed introduction to position-space entanglement in QFT. The possible connection between entanglement and black hole entropy has been recently reviewed in Refs. \cite{2008arXiv0806.0402D,Solodukhin:2011gn}, while in Ref. \cite{Nishioka:2009un} some holographic aspects are discussed.

\subsection{Entanglement in discrete field theories} \label{secPOS.1}

The evaluation of entanglement entropy for quantum fields in real space has to inevitably deal with the presence of divergences, which typically require regularization techniques to be tamed \cite{Casini:2009sr}.

This is not the case of discrete field theories, where an uncontroversial computation of the entropy is usually possible \cite{Riera:2006vj}. A common strategy consists then in studying quantum fields on a lattice, so that the lattice spacing allows for the ultraviolet regularization of the theory. In particular, discretized scalar fields have been proposed \cite{Das:2007mj} to mimic gravitational perturbations in black hole scenarios, focusing on static and spherically symmetric spacetimes.

Since a discretized scalar field on a lattice can be modeled as an open chain of coupled harmonic oscillators, we now briefly review the derivation of the ground state entanglement entropy in this latter case, assuming a Minkowski background as starting point and then moving to curved backgrounds.

\subsubsection{Entanglement entropy of discrete scalar fields} \label{secPOS1.1}

A system of $N$ coupled harmonic oscillators can be described by an Hamiltonian of the form
\be \label{hosc_ham}
H_O= \frac{1}{2} \sum_{i=1}^N p_i^2+ \frac{1}{2} \sum_{i,j=1}^N x_i K_{ij} x_j,
\ee
where the $p_i$ are the momenta of the oscillators, the $x_i$ their position, and the symmetric matrix $K$ describes the potential energy and interactions.

The normalized ground-state wave function of this system reads \cite{Srednicki:1993im}
\be \label{gswf_hosc}
\psi_0 (x_1, \dots, x_N)= \pi^{-N/4} (\text{det}\  \Omega)^{1/4} \exp{[-x \cdot \Omega \cdot x/2]},
\ee
with $\Omega := \sqrt{K}$. Our aim is now to trace over a fixed number of oscillators, in order to compute the entanglement entropy associated with the remaining part of the chain. For convenience, we write $\Omega$ in block form as
\be \label{oblk}
\Omega= \begin{pmatrix} A & B \\[4 pt]
B^T & C  \end{pmatrix},
\ee
where $A$ is a $n \times n$ matrix and $C$ is $(N-n) \times (N-n)$, assuming that the first $n$ oscillators are traced out.

Further defining $\beta:= \frac{1}{2} B^T A^{-1} B$ and $\gamma:= C- \beta$, we can write the ground state reduced density operator of the system in the position representation as
\be \label{redop_osc}
\rho_{\rm out} (x, x^\prime) \simeq \exp{[-(x \cdot \gamma \cdot x+ x^\prime \cdot \gamma \cdot x^\prime)/2+ x \cdot \beta \cdot x^\prime]},
\ee
where now both $x$ and $x^\prime$ are vectors with $N-n$ components. The spectrum of $\rho_{\rm out}$ is then found to be
\be \label{spec_redosc}
p_{n_{n+1},\dots,n_N}= \prod_{i=n+1}^N \left( 1-\xi_i \right) \xi_i^{n_i},\ \ \ \ \ \ n_i \in \mathbb{Z},
\ee
where $\xi_i= \frac{\lambda_i}{1+\sqrt{1-\lambda_i^2}}$ and $\lambda_i$ are the eigenvalues of the matrix $\gamma^{-1}\beta$. The corresponding von Neumann entropy is then given by
\begin{align} \label{enten_osc}
S(\rho_{\rm out})&\equiv -{\rm Tr}\left( \rho_{\rm out} \ln \rho_{\rm out} \right)  \notag \\
&=\sum_{j=n+1}^{N} \left( -\ln (1-\xi_j) - \frac{\xi_j}{1-\xi_j} \ln \xi_j\right),
\end{align}
showing that, in principle, the ground state of a system of coupled harmonic oscillators may be a highly entangled state.

The Hamiltonian of Eq. \eqref{hosc_ham} can be also employed to describe the degrees of freedom of a scalar field theory on a lattice of spherical shells. Starting from the (continuous) free scalar field Hamiltonian
\be \label{contham_scal}
H=\frac{1}{2} \int d^3x \left[ \pi^2({\bf x}) + \lvert \nabla \varphi({\bf x}) \rvert^2+ m^2 \varphi^2({\bf x})  \right],
\ee
where $\varphi$ is the field and $\mu$ its mass, we define
\begin{align}
    & \varphi_{lm}(r)= r \int d\Omega\  Z_{lm}(\theta,\phi) \varphi({\bf x}), \label{fie_discr} \\
    & \pi_{lm}(r) = r \int d\Omega\  Z_{lm} (\theta,\phi) \pi({\bf x})  \label{momfi_discr}.
\end{align}
In Eqs. \eqref{fie_discr}-\eqref{momfi_discr} we have introduced the radial coordinate $r \equiv \lvert {\bf x} \rvert$ and $Z_{lm}$ are real spherical harmonics, which form an orthonormal basis of harmonic functions on the sphere $\mathbf{S}^2$. Furthermore, it can be shown that the canonical commutation relations
\be \label{cancomm_discr}
\left[ \hat{\varphi}_{lm} (r), \hat{\pi}_{l^\prime m^\prime}(r^\prime) \right] = i \delta\left( r-r^\prime \right) \delta_{l l^\prime} \delta_{m m^\prime}
\ee
are satisfied. Inserting now Eqs. \eqref{fie_discr}-\eqref{momfi_discr} into the field Hamiltonian \eqref{contham_scal}, we obtain
\be \label{discr_ham}
H= \frac{1}{2} \sum_{lm} \int_0^\infty dr \left \{  \pi_{lm}^2(r)+r^2\left[ \frac{\partial}{\partial r} \left( \frac{\varphi_{lm}(r)}{r} \right)  \right]^2  + \left( \frac{l(l+1)}{r^2} +\mu^2 \right) \varphi_{lm}(r)  \right \},
\ee
where the only continuous variable left is the radial coordinate $r$. We can now regularize the theory by introducing a lattice of $N$ spherical shells with spacing $a$ and corresponding radii $r_i=ia$, with $1 \leq i \leq N$ ($i \in \mathbb{N}$).

The substitutions
\begin{align}
    & r \rightarrow j a, \label{rdiscr}\\[2.5pt]
    & \varphi_{lm}(ja) \rightarrow \varphi_{lm,j}, \label{fiediscr}\\[2.5pt]
    & \frac{\partial \varphi_{lm}(r)}{\partial r}\bigg \lvert_{r=ja} \rightarrow \frac{\varphi_{lm,j+1}-\varphi_{lm,j}}{a},\label{derdiscr}\\[2.5pt]
    & \pi_{lm}(ja) \rightarrow \frac{\pi_{lm,j}}{a}, \label{momdiscr} \\[2.5pt]
    & \int_0^{(N+1)a} dr \rightarrow a \sum_{j=1}^N,
\end{align}
then leads to the fully discretized Hamiltonian
\begin{align} \label{findiscr_ham}
H&= \frac{1}{2a} \sum_{lm} \sum_{j=1}^N \bigg[ \pi^2_{lm,j} + \left( j+ \frac{1}{2} \right)^2 \left( \frac{\varphi_{lm,j+1}}{j+1}-\frac{\varphi_{lm,j}}{j}  \right)^2  \notag \\[3pt]
&\ \ \ \ + \left( \frac{l(l+1)}{j^2}+\mu^2a^2  \right) \varphi^2_{lm,j}\bigg].
\end{align}
Eq. \eqref{findiscr_ham} can be expressed in the form $H = \sum_l H_l$, where
\be \label{ham_lpart}
H_l= \frac{1}{2a} \sum_{j=1}^N \left[ \pi^2_{l,j} + \left( j+\frac{1}{2}  \right)^2 \left( \frac{\varphi_{l,j+1}}{j+1} -\frac{\varphi_{l,j}}{j} \right)^2+ \left( \frac{l(l+1)}{j^2}  +\mu^2a^2 \right) \varphi^2_{l,j}   \right],
\ee
thus leading to the total entropy
\be \label{gs_entr}
S(n,N)= \sum_{l=0}^\infty (2l+1) S_l(n,N),
\ee
where $S_l(n,N)$ is the entanglement entropy corresponding to the ground state Hamiltonian $H_l$. We then notice that Eq. \eqref{ham_lpart} describes again a system of coupled harmonic oscillators, having the same form of Eq. \eqref{hosc_ham} with corresponding coupling matrix 
\begin{align} \label{coup_discrf}
    K_{ij}=& \left( \left( \frac{i+\frac{1}{2}}{i}  \right)^2+ \left(  \frac{i-\frac{1}{2}}{i} \right)^2+ \frac{l(l+1)}{i^2}+\mu^2a^2 \right) \delta_{ij} \notag \\[3pt]
           &- \frac{\left( i+ \frac{1}{2} \right)^2}{i(i+1)} \delta_{i+1,j}- \frac{\left( j+ \frac{1}{2} \right)^2}{j(j+1)} \delta_{i,j+1} 
\end{align}
We also notice that, for large $l$, the Hamiltonian $H_l$ becomes almost diagonal, implying that in the limit $l \to \infty$ the field degrees of freedom are decoupled, and thus the system is disentangled at its ground state. It can be shown that $S_l(N,n)$ decreases with $l$ fast enough so that the series \eqref{gs_entr} is convergent \cite{Srednicki:1993im, Katsinis:2017qzh}.

\subsubsection{Area law in black hole spacetimes} \label{secPOS1.2}

The above-presented results can be generalized to static, spherically symmetric spacetimes, described by the line element in Eq. \eqref{linel_stsph}. Exploiting spherical symmetry, we can expand again the scalar field in real spherical harmonics, assuming 
\be \label{spharm_exp}
\varphi(x)= \frac{1}{r}\sum_{lm} \varphi_{lm}(t,r) Z_{lm}(\theta, \phi).
\ee
A suitable strategy consists now in resorting to Lema\^itre coordinates 
\begin{align}
   &\zeta= t + \int dr \frac{\left[ 1-f(r) \right]^{-1/2}}{f(r)} \label{lemr} \\
   &\chi= t \pm \int dr \frac{\sqrt{1-f(r)}}{f(r)},  \label{lemt}
\end{align}
which are not singular at the horizon(s), i.e., when $f(r)=0$. Furthermore, we notice that $\zeta$ is spacelike everywhere, while this is not true for the radial coordinate $r$, which, for example, is timelike as $r<r_s \equiv 2GM$ for the Schwarzschild solution\footnote{Similarly, the Lema\^itre time coordinate $\chi$ is timelike everywhere, while $t$ is timelike only at $r> r_s$ for the Schwarzschild case.}, namely $f(r)=1-2GM/r$.
\begin{figure} 
 \centering
\includegraphics[scale=0.88]{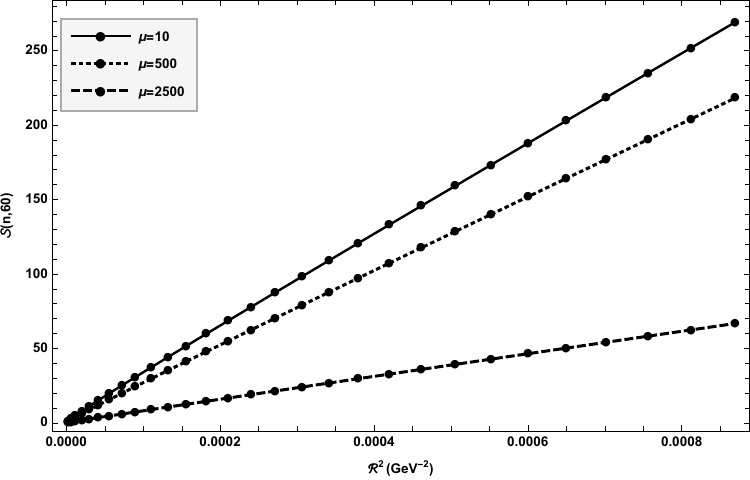}\hfill
    \caption{Ground state entanglement entropy $S(n,N)$ as function of $\mathcal{R}^2$ for a minimally coupled scalar field. The other parameters are: $a=10^{-3}$ GeV$^{-1}$, $l_{\rm max}=1000$, $N=60$ and $n < N/2$. We notice that $S \propto R^2$, implying that an area law is satisfied in the minimal coupling scenario. The proportionality coefficient between entropy and area decreases for larger field mass $\mu$ (GeV). Figure adapted from \cite{Belfiglio:2023sru}.}
    \label{fig_alawminim}
\end{figure}
Exploiting Eqs. \eqref{lemr}-\eqref{lemt}, we can write again the total Hamiltonian in the form 
\be \label{totham_ang}
H= \sum_{lm} H_{lm}
\ee
and, by fixing Lema\^itre time as $\chi=0$, we find 
\be \label{hamexp}
H_{lm}(0)= \frac{1}{2} \int_0^\infty dr \bigg[ \frac{\pi_{lm} r^{-2}}{1-f(r)} + r^2 \left( \partial_r \varphi_{lm}\right)^2 + \left( l(l+1)+ r^2 \mu^2 \right) \varphi_{lm}^2 \bigg].
\ee
The variables $\varphi_{lm}$ and $\pi_{lm}$ satisfy the Poisson brackets
\be \label{pois}
\{ \varphi_{lm}(r), \pi_{lm}(r^\prime) \}= \sqrt{1-f(r)} \delta(r-r^\prime)
\ee
and, by means of the canonical transformation
\be \label{cantrasf}
\pi_{lm} \rightarrow r \sqrt{1-f(r)} \pi_{lm},\ \ \ \ \varphi_{lm} \rightarrow \frac{\varphi_{lm}}{r},
\ee
we arrive at
\be \label{radham}
H_{lm}(0)= \frac{1}{2} \int_0^\infty dr \bigg \{ \pi_{lm}^2(r)+ r^2 \bigg[ \frac{\partial}{\partial r}\left( \frac{\varphi_{lm}}{r} \right) \bigg]^2 
+ \left( \frac{l(l+1)}{r^2}+\mu^2 \right) \varphi_{lm}^2 \bigg \},
\ee
which has the same form of the flat space Hamiltonian in Eq. \eqref{discr_ham}.

The final step for our discretization procedure consists in introducing a short-distance regulator, to provide an UV regularization of the radial coordinate $r$. In analogy with the flat space scenario, we discretize the field on a lattice of $N$ spherical shells, spaced by a fixed distance $a$. This procedure leads again to the Hamiltonian displayed in Eq. \eqref{findiscr_ham}, whose corresponding ground state entanglement entropy has been computed above.

In Fig. \ref{fig_alawminim}, the ground state von Neumann entropy corresponding to the scalar field $\varphi$ is plotted as function of $\mathcal{R}^2$, where $\mathcal{R} \equiv a(n+1/2)$ is the discretized radius and a proper cutoff $l_{\rm max}$ is imposed on spherical harmonics. We notice that the entanglement entropy satisfies an \emph{area law}, since $ S(n,N) \propto \mathcal{R}^2$. At the same time, larger scalar field masses result in smaller entanglement amount, in analogy to the momentum-space analysis of Sec. \ref{secMOMENT}, thus suggesting that it is typically harder to entangle scalar field particles with higher energy.

These results have been recently generalized to the case of nonminimally coupled scalar fields, showing that the presence of field-curvature coupling may lead to area law violations, both in regular black hole configurations \cite{Belfiglio:2023sru} or in the presence of quantum corrections at Planck lengthscales \cite{Belfiglio:2024qsa}. Deviations from the area law also arise in gravitational collapse scenarios \cite{Belfiglio:2025hzo}, where the dynamical formation of an horizon has been shown to result in a nontrivial scaling of the ground state entanglement entropy for discrete scalar fields.

\subsubsection{Building black hole thermodynamics from entanglement} \label{secPOS1.3}

The possible identification of black hole entropy in terms of entanglement, on account of the area law, has stimulated, in turn, the search for the subsystem analogs of other black hole thermodynamic quantities. In particular, a full quantum picture would necessarily require suitable counterparts for the black hole energy and temperature, in the attempt to formulate a quantum analog of the first law of black hole thermodynamics \cite{Page:2004xp,Bardeen:1973gs}.

To this end, various definitions of \emph{entanglement energy} and \emph{entanglement temperature} \cite{PhysRevD.55.7666,PhysRevD.58.064001,Wong:2013gua,Park:2015hcz} have been proposed in the literature, aiming to construct a thermodynamic picture arising from quantum measures (usually referred to as \emph{entanglement thermodynamics}), which may be then investigated in black hole scenarios.

Specifically, this correspondence may have relevant implications in the presence of horizons, which represent physical boundaries beyond which observers do not have access to information. Accordingly, while in flat spacetime the chosen boundary for area law calculations is merely artificial, entanglement characterization close to a spacetime horizon may capture relevant information about its thermodynamic properties, and horizon thermodynamics may in itself be of quantum origin \cite{PhysRevD.102.125025}.

In order to study the entanglement properties close to a black hole horizon, it is convenient to introduce the \emph{proper length} as \cite{PhysRevD.55.7666}
\be\label{eq:properlength}
	\rho=\int_{r_h}^r \frac{dr'}{\sqrt{f(r')}},
\ee
where the selected initial value $r_h$ represents the horizon radius, satisfying $f(r_h)=0$ and corresponding to $\rho=0$. The proper length coordinates can only describe the spacetime geometry in the region $f(r) > 0$ from the horizon and, so, in the limit $r_h\to 0$, the proper length reduces to the radial coordinate, namely $\rho\to r$.

Exploiting spherical symmetry, we can perform the usual partial wave expansion of the field, which in proper length coordinates is conveniently written as 
 \begin{align}
	\dot{\varphi}(\rho,\theta,\phi)&=\frac{f^{1/4}}{r}\sum_{lm}\dot{\varphi}_{lm}(\rho)Z_{lm}(\theta,\phi), \label{momp_exp}\\
	\varphi(\rho,\theta,\phi)&=\frac{f^{1/4}}{r}\sum_{lm}\varphi_{lm}(\rho)Z_{lm}(\theta,\phi). \label{posp_exp}
\end{align}
Further regularizing the theory by $\rho=ja$ , with $a$ the lattice spacing, for each lattice point $j$ along proper length, we obtain the corresponding lattice point along radial coordinate as $r_j=a^{-1}r(\rho)|_{\rho=ja}$, where $r(\rho)$ is derived from Eq. \eqref{eq:properlength}. Hence, we end up with the following Hamiltonian
\be \label{ham_massl_discr}
		H=\frac{1}{2a}\sum_{lmj}\left[\pi_{lm,j}^2+r_{j+\frac{1}{2}}^2f_{j+\frac{1}{2}}^{1/2}\bigg\{\frac{f_j^{1/4}\varphi_{lm,j}}{r_j}-\frac{f_{j+1}^{1/4}\varphi_{lm,j+1}}{r_{j+1}}\bigg\}^2+\frac{f_jl(l+1)}{r_j^2}\varphi_{lm,j}^2\right], 
\ee
where $f_j=f\left[r(\rho=ja)\right]$ and, for simplicity, we have assumed the field to be massless, namely $\mu=0$.

Again, we can trace back this Hamiltonian to that of Eq. \eqref{hosc_ham}, describing a system of coupled harmonic oscillators, thus writing
\be\label{eq:hlm}
	H_{lm}=\frac{1}{2}\left[\sum_i \pi_{lm,i}^2+\sum_{ij}K_{ij}\varphi_{lm,i}\varphi_{lm,j}\right],
\ee
and, then, decomposing Eq. \eqref{ham_massl_discr} by
\be\label{eq:lmHamil}
H=\frac{1}{a}\sum_{lm} H_{lm},    
\ee
with $K_{ij}$ the coupling matrix for each $lm$-lattice. Upon spatially partitioning the field into \emph{in} and \emph{out} degrees of freedom for each $lm$-mode, the total Hamiltonian of Eq. \eqref{eq:hlm} can be split into subsystems of $n$ and $N-n$ oscillators, thus giving
\be \label{totham_split}
    H=H_{in}+H_{out}+H_{int},
\ee
with
\begin{subequations} \label{hamsplit}
    \begin{align} 
    H_{in}&=\frac{1}{2}\left[\sum_{i=1}^n \pi_i^2+\sum_{ij=1}^nK_{_{ij}}\varphi_{i}\varphi_{j}\right],\\
    H_{out}&=\frac{1}{2}\left[\sum_{i=n+1}^N \pi_{i}^2+\sum_{ij=n+1}^N K_{_{ij}}\varphi_{i}\varphi_{j}\right],\\ H_{int}&=K_{_{n,n+1}}\varphi_{{n}}\varphi_{{n+1}},
    \end{align}
\end{subequations}
where we have dropped the $lm$-indices from each Hamiltonian in Eq. \eqref{hamsplit}, in order to simplify the notation. Let us now introduce the \emph{subsystem energy}
\be \label{subs_ene}
    E=\langle :H_{in}: \rangle_{in}=\tr\left[\rho_{in}:H_{in}:\right],
\ee
namely the expectation value of the normal-ordered Hamiltonian corresponding to the \emph{in} degrees of freedom, with 
\be\label{eq:hin}
    :H_{in}:\, =\frac{1}{2}\sum_{i=1}^n\left[\pi_i+i\sum_{j=1}^n\left(K_{in}^{1/2}\right)_{ij}\varphi_j\right]\times\left[\pi_i-i\sum_{j=1}^n\left(K_{in}^{1/2}\right)_{ij}\varphi_j\right].
\ee
Here, $K_{in}$ is the $n \times n$ sub-block of the coupling matrix $K$, restricted to the \emph{in} degrees of freedom. While $E$ serves as a measure for the disturbed vacuum energy of the \emph{in} subsystem, it is also one of the few candidates for \textit{entanglement energy} that were previously discussed in literature as suitable quantum analogs for energy, that could potentially reproduce the first law of black hole thermodynamics in tandem with the entanglement entropy \cite{PhysRevD.55.7666,PhysRevD.58.064001}. This quantity has been recently studied in singular \cite{PhysRevD.102.125025} and regular \cite{PhysRevD.111.024013} black hole scenarios, showing that in the continuum limit $a \rightarrow 0$ it may be traced back to the Komar energy \cite{PhysRev.113.934,PhysRevD.81.124006} associated with a given horizon. These results suggest that the entanglement features of quantum fields are capable of providing valuable insight into the thermodynamic properties of the underlying spacetime, thus not only limited to black hole entropy.

\paragraph{Example: Bardeen solution}

As a specific example, let us discuss the Bardeen solution, which represent the first regular black hole proposal, described by Eq. \eqref{linel_stsph} with the lapse function \eqref{bardlaps}.  Bardeen black holes allow for the presence of two horizons $r_\pm$ ($r_+>r_-$), corresponding to real, positive roots of $f(r)$:
\begin{align}
    r_+&=\frac{2GM}{3}\sqrt{\left\{1+2\cos{\frac{\theta}{3}}\right\}^2-\frac{9q^2}{4G^2M^2}}, \label{rpl_bard}\\
    r_-&=\frac{2GM}{3}\sqrt{\left\{1+2\cos{\left(\frac{\theta-2\pi}{3}\right)}\right\}^2-\frac{9q^2}{4G^2M^2}},\label{rmin_bard}\\
\end{align}
where we set
\be \label{bardangle}
    \theta=\cos^{-1}\left(1-\frac{27q^2}{8G^2M^2}\right).
\ee
The extremal limit $r_+\to r_-$ bounds the charge $q$ from above, namely
\be \label{extr_charge}
    0\leq q \leq \frac{4GM}{3\sqrt{3}}\sim0.77GM\,,
\ee
beyond which there can no longer be a horizon, i.e., $f(r)$ does not have positive roots. Sufficiently close to each horizon, we can approximate the lapse function by
\be\label{eq:lapseapprox}
f(r)\approx (r-r_h)f^\prime(r_h),
\ee
which, in the case of Bardeen solution, readily gives
\begin{align}
&r_j^{(\pm)}=\frac{r_\pm}{a}+\frac{j^2a}{4r_\pm}\left(1-\frac{3q^2}{r_\pm^2+q^2}\right), \label{rj_bard}\\
&f_j^{(\pm)}=\frac{j^2a^2}{4r_\pm^2}\left(1-\frac{3q^2}{r_\pm^2+q^2}\right)^2. \label{fj_bard}
\end{align}
Focusing on a single-oscillator subsystem very close to the horizon $r_+$ the following scaling relations for the entanglement entropy and energy are obtained:
\begin{align}
    &E=\frac{c_e}{a^2}E_{Komar}\,, \label{eq:enescal}\\
    &S=\frac{c_s}{a^2}S_{BH}\,, \label{eq:entroscal}
\end{align}
where
\begin{align}
        &E_{Komar}=\frac{r_+}{2}\left(1-\frac{3q^2}{r_+^2+q^2}\right), \label{ekom_bard}\\
        &S_{BH}=\pi r_+^2, \label{entr_bard}
\end{align}
are the Komar energy and the Bekenstein-Hawking entropy, respectively, while $c_e$, $c_s$ are constants. It is therefore easy to see that:
\be \label{ratio_bard_th}
    \frac{E}{2S}=\frac{\epsilon E_{Komar}}{2S_{BH}}=\epsilon T_H\,,\quad \epsilon=\frac{c_e}{2c_s},
\ee
where the Hawking temperature $T_H$ can be expressed as
\be \label{hortemp_bard}
    T_H=\frac{f'(r_+)}{4\pi}=\frac{1}{4\pi r_+}\left(1-\frac{3q^2}{r_+^2+q^2}\right).
\ee
Accordingly, we notice that $\epsilon$ captures deviations from the Komar relation, $E_{Komar}=2T_HS_{BH}$. Such deviations may, in principle, arise from sub-leading corrections or spurious edge effects at the horizon, or from considering the bipartition to be slightly away ($\rho\sim a$) from the horizon. As mentioned above, it can be shown that $\epsilon \rightarrow 1$ in the continuum limit $a \rightarrow 0$, thus confirming a possible interpretation of black hole thermodynamics in terms of entanglement \cite{PhysRevD.111.024013}.

\subsection{The entropy of Bekenstein-Hawking radiation}

As mentioned above, black holes behave as thermal objects. They have a temperature that leads to Hawking radiation. They also have an entropy given by the area of the horizon. These outcomes suggested that, from the point of view of the outside, black holes could be viewed as ordinary quantum systems, described by a large (but finite) number of degrees of freedom. 

Hawking famously challenged this interpretation, arguing that black hole formation and evaporation would increase the total von Neumann entropy of the universe, thereby violating unitarity of evolution \cite{Hawking:1975vcx} (see also \cite{Raju:2020smc}).

Assuming that a black hole is a fully thermodynamic system implies that its total entropy can be expressed as the sum of two contributions, one due to the horizon and the other from the quantum fields outside it, according to \cite{Bekenstein:1973ur}
\be \label{gen_entr}
S_{\rm gen}= \frac{A}{4G}+ S_{\rm out},
\ee
where $S_{\rm out}$ includes the claimed matter entropy outside the black hole. This generalized entropy, which we formally derive below, was shown to obey the second law of thermodynamics \cite{PhysRevD.85.104049}, namely $\Delta S_{\rm gen} \geq 0$, thus confirming its thermodynamic nature. 

This result is stronger than the area law discussed above, since it also includes the dynamics of the Hawking radiation and its corresponding entropy, which increases during black hole evaporation.


\subsubsection{Demonstrating Hawking's formula}

In order to properly introduce the Hawking radiation, let us focus for simplicity on the Schwarzschild metric, namely
\begin{equation}\label{schw}
ds^2 = \left( 1 - \frac{r_s}{r} \right) dt^2 - \frac{dr^2}{1 - \frac{r_s}{r}} - r^2 (d\theta^2+\sin^2 \theta d\phi^2) \ .
\end{equation}
Ignoring the angular components, we define new coordinates $r \to r_s(1 + \rho^2 / (4 r_s^2))$ and $t \to 2 r_s \tau$ to focus on the near-horizon limit $\rho \ll r_s$, yielding, $ds^2 \approx -\rho^2 d\tau^2 + d\rho^2$.

This is Rindler space, a portion of flat Minkowski spacetime. Using the transformation $x^0 = \rho \sinh \tau$, $x^1 = \rho \cosh \tau$, we indeed recover $ds^2 \approx - (dx^0)^2 + (dx^1)^2$.

Accordingly, the horizon at $r = r_s$ appears locally smooth and regular to free-falling observers, consistently with the equivalence principle. However, observers maintaining fixed $r$ are accelerating and require force to hover above the horizon.

Remarkably, an observer with constant acceleration in flat spacetime perceives a thermal bath, according to the Unruh effect \cite{PhysRevD.14.870}. A heuristic derivation of this fact relies on the Wick rotation \cite{Bisognano:1976za}, then giving $x_E^0 = \rho \sin \theta$ and $x^1 = \rho \cos \theta$. 

This maps the Euclidean geometry to $\mathbb{R}^2$, and motion at fixed $\rho$ becomes circular with period $2\pi$. This periodicity corresponds to thermal behavior, with proper temperature
\begin{equation}\label{PropT}
T_{\rm proper} = \frac{1}{2\pi \rho} = \frac{a}{2\pi} = \frac{\hbar a}{2\pi k_B c} \ ,
\end{equation}
where $a$ is the proper acceleration. Eq. \eqref{PropT} then gives the temperature measured by an ideal thermometer carried by an accelerated observer.

Near the horizon, the proper temperature diverges as $\rho \to 0$, but redshift ensures that the temperature measured far away is finite. In equilibrium, the Tolman relation holds:
\begin{equation}
T_{\rm proper}(r) \sqrt{g_{\tau\tau}(r)} = \text{constant} \ ,
\end{equation}
implying that locations at higher gravitational potential feel colder to a local observer. Applying this to the full Schwarzschild geometry and rescaling time back to $t$ yields the Hawking temperature:
\begin{equation}\label{thawking}
T = T_{\rm proper}(r \gg r_s) = \frac{1}{4\pi r_s} \ .
\end{equation}

The connection between Euclidean time and temperature can be made rigorous using the path integral formalism. In quantum field theory, the thermal partition function,
\begin{equation}
Z = Tr \left[ e^{-\beta H} \right],
\end{equation}
corresponds to a Euclidean path integral on a manifold with periodic imaginary time. This manifests thermal periodicity in correlation functions.

For black holes, we Wick rotate $t = i t_E$ in Eq. \eqref{schw}, obtaining the Euclidean Schwarzschild geometry\footnote{See e.g. Refs. \cite{PhysRevD.15.2752,PhysRevD.45.2762} for further details on Euclidean black hole geometries.}
\be
ds_E^2 = \left(1 - \frac{r_s}{r} \right) dt_E^2 + \frac{dr^2}{1 - \frac{r_s}{r}} + r^2 d\Omega_2^2,
\ee
with $t_E = t_E + \beta$. To avoid a conical singularity at $r = r_s$, the period must be $\beta = 4\pi r_s$. The partition function is given by $Z(\beta) \sim e^{-I_{\rm classical}} Z_{\rm quantum}$, 
where $I_{\rm classical}$ is the on-shell gravitational action, and $Z_{\rm quantum}$ accounts for quantum field contributions. The geometry is smooth at $r = r_s$, ensuring that quantum fields experience no singularity, consistently with the classical prediction that an infalling observer encounters nothing unusual at the horizon.

If we then apply the thermodynamic identity, $S = \left(1 - \beta \partial_\beta \right) \log Z(\beta)$, this
yields the generalized entropy, that we introduced heuristically in Eq. \eqref{gen_entr}.

\subsubsection{Fine and coarse-grained entropy}

The Hawking process can be roughly interpreted as pair creation of entangled particles near the horizon, with one
particle escaping to infinity and the other falling toward the
singularity. The corresponding entanglement entropy associated with this process can be computed following the same steps discussed in Sec. \ref{secMOMENT}, tracing out the inaccessible contribution inside the black hole and then evaluating the von Neumann entropy associated with the reduced density operator for the outside degrees of freedom. It becomes then essential to distinguish between two notions of entropy. 

\begin{itemize}
    \item[-] We denote as \emph{fine-grained entropy} the von Neumann entropy of Hawking radiation. As discussed in Sec. \ref{secENTA}, this quantity vanishes for pure states and remains invariant under unitary evolution $\rho \to U \rho U^{-1}$. If one computes the von Neumann entropy associated with Hawking radiation, one then observe that this quantity increases with the number of emitted quanta.
\item[-] Let us suppose now that we measure only a subset of observables $\{A_i\}$ associated with a system described by the density operator $\rho$. We then consider all density matrices $\tilde{\rho}$ satisfying
\begin{equation}
\tr[\tilde{\rho} A_i] = \tr[\rho A_i] .
\end{equation}

The \emph{coarse-grained entropy} is the maximum von Neumann entropy among all such $\tilde{\rho}$:
\begin{equation}
S_{\text{coarse}} = \max_{\tilde{\rho}} \left( - \tr[\tilde{\rho} \log \tilde{\rho}] \right) .
\end{equation}

A canonical example is the thermodynamic entropy, where the coarse observables $A_i$ are macroscopic quantities such as energy and volume. This entropy version satisfies the second law of thermodynamics, increasing under unitary evolution. Accordingly, the total entropy in Eq. \eqref{gen_entr}, which increases rapidly when the black hole first forms and the horizon area starts growing, can only represent a coarse-grained entropy contribution.

\end{itemize}

We observe that the coarse-grained entropy cannot be smaller than the fine-grained one, namely $S \leq S_{\rm coarse}$. This is consequence of the fact that we can always consider $\rho$ as a candidate $\tilde{\rho}$.


\subsection{Deriving the Page curve} \label{sec6.3}

Let us consider a black hole formed by the collapse of a pure state, for example, a large-amplitude gravitational wave~\cite{Christodoulou:2008nj}. The resulting black hole emits radiation that is approximately thermal. How can this thermal character be reconciled with the purity of the initial state?

\subsubsection{The Hawking information paradox} \label{sec6.3.1}

The essential mechanism that explains such a thermalization is that the vacuum in quantum field theory is an entangled state. When a horizon forms, the vacuum modes get effectively partitioned into two components: one that falls into the black hole and one that escapes to infinity. These are sometimes referred to as the ``interior'' and ``exterior'' Hawking quanta, respectively. While the full vacuum state is pure, tracing out either half results in a mixed state. This is a general consequence of the entanglement structure of the vacuum and follows from unitarity and Lorentz invariance~\cite{Bisognano:1976za}.

This is often heuristically described by saying that particle pairs are continuously created near the horizon: one escapes, and the other falls into the black hole. The outgoing Hawking quantum is therefore entangled with its infalling partner, and when considered in isolation, it appears as a mixed state, thermal at the Hawking temperature (see Eq.~\ref{thawking}). Accordingly, the presence of this entanglement does not violate unitarity by itself: any sufficiently complex quantum system initially in a pure state can evolve to appear thermal at intermediate times. In the early stages of black hole evaporation, the von Neumann entropy of the radiation is expected to rise, since more and more quanta are produced and the outgoing radiation is entangled with the remaining black hole degrees of freedom \cite{Mathur:2009hf}.
\begin{figure}
    \centering
    \includegraphics[scale=0.45]{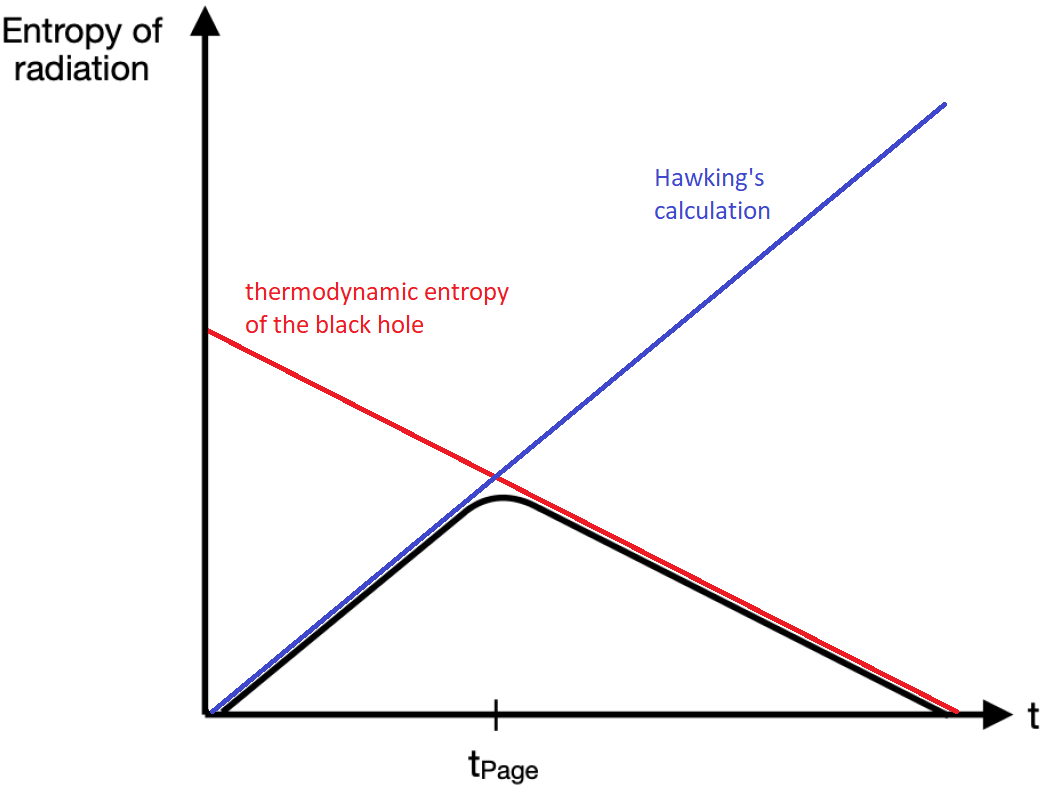}
    \caption{The entropy of the outgoing Hawking radiation cannot exceed the thermodynamic entropy of the black hole, in order to preserve the unitarity of evolution. Accordingly, the entropy of radiation should start decreasing during the latest stages of black hole evaporation, thus following the Page curve (black). Figure adapted from \cite{RevModPhys.93.035002}. }
    \label{fig_pagecv}
\end{figure}

However, a tension arises later in the evaporation process. As the black hole shrinks, its thermodynamic entropy, given by the Bekenstein-Hawking formula, decreases. Eventually, the entropy of the radiation, $S_{\rm rad}$, surpasses the black hole’s thermodynamic entropy, $S_{\rm BH}$. If this happens, the fine-grained entropy of the radiation can no longer be purified by the remaining black hole, since the number of available degrees of freedom in the black hole is insufficient.

To phrase this more precisely: if the combined black hole and radiation system remains in a pure state, then 
\begin{itemize}
    \item[-] their respective fine-grained entropies must match, namely $S_{\rm rad} = S_{\rm BH}^{\rm fine}$,
    \item[-] the fine-grained entropy of the black hole cannot exceed its coarse-grained thermodynamic entropy, $
S_{\rm BH}^{\rm fine} \leq S_{\rm BH}^{\rm coarse} = A/4G$.
\end{itemize}

Hence, if $S_{\rm rad} > S_{\rm BH}^{\rm coarse}$, the assumption that the global state is pure leads to a contradiction. We also observe that:

\begin{itemize}
\item[-] As the black hole evaporates, its mass decreases. This backreaction is included in the semiclassical analysis and does not resolve the paradox.
\item[-] When the black hole shrinks to Planckian size near the end of its evaporation, the semiclassical approximation breaks down. However, the paradox arises earlier, when the black hole is still semiclassical and large.
\item[-] The argument is robust, relying solely on fundamental properties of the fine-grained entropy. It has been shown that small corrections to the Hawking process—such as minor modifications of the near-horizon dynamics—are insufficient to resolve the paradox~\cite{Mathur:2009hf,Almheiri:2012rt,Almheiri:2013hfa}. The resolution, if any, must be non-perturbative in the gravitational coupling $G$.
\end{itemize}

Accordingly, in order to preserve unitarity, the entropy of radiation should start decreasing at the time corresponding to $S_{\rm rad}= S_{\rm BH}^{\rm coarse}$. This is called Page time, and the expected dynamics of the radiation entropy should follow the so-called Page curve, which we show in Fig. \ref{fig_pagecv}.

One might argue that semiclassical gravity can only offer approximate results and it is not reliable for a precise computation of the von Neumann entropy. While this was the prevailing view for many years, we will review below that there is instead a semiclassical prescription capable of capturing the fine-grained entropy.

\subsubsection{Generalized entropy of an evaporating black hole} \label{sec6.3.2}

As discussed above, the Bekenstein-Hawking entropy should be interpreted as the coarse-grained entropy of the black hole. Remarkably, there exists also a gravitational prescription for computing the fine-grained (von Neumann) entropy \cite{Ryu:2006bv,Hubeny:2007xt,Faulkner:2013ana,Engelhardt:2014gca}. 

This prescription again involves a generalized entropy, defined as the sum of an area term and the entropy of quantum fields in the exterior region. However, in this case the dividing surface is chosen in such a way to minimize the entropy, according to
\begin{equation} \label{Appr}
S \sim \min \left[ \frac{ \mathrm{Area} }{4 G} + S_{\mathrm{outside}} \right].
\end{equation}

More precisely, we look for an \enquote{extremal} surface that minimizes the entropy in the spatial direction but maximizes it in time. This leads to a more refined definition of the fine-grained entropy via the so-called maximin construction \cite{Wall:2012uf,Akers:2019lzs}: one first selects a Cauchy slice and computes the minimal surface on that slice, then maximizes over all such slices.

This prescription then gives
\begin{equation} \label{RT}
S = \min_X \left\{ \mathrm{ext}_X \left[ \frac{ \mathrm{Area}(X) }{4 G} + S_{\mathrm{semi}}(\Sigma_X) \right] \right\} \,,
\end{equation}
where $X$ is a surface of codimension 2 (it has two dimensions fewer than those of the spacetime it lives in), $\Sigma_X$ is the region bounded by $X$ and $S_{\rm semi}(\Sigma_X)$ is the (semiclassical) von Neumann entropy of quantum fields living on $\Sigma_X$.  Accordingly, the surface that extremizes $S$ is often called \enquote{quantum extremal surface} because of the presence of contributions from quantum fields, despite being a classical geometric surface within the spacetime.

Looking at Eq. \eqref{RT}, the quantity inside the brackets is then the  generalized entropy, namely
\begin{equation} \label{sgendef}
S_{\mathrm{gen}}(X) = \frac{ \mathrm{Area}(X) }{4 G} + S_{\mathrm{semi}}(\Sigma_X) \,.
\end{equation}

This entropy was initially studied in the context of the AdS/CFT correspondence \cite{Maldacena:1997re,Ryu:2006bv,Nishioka:2009un} (see also \cite{Hubeny:2014bla} for a recent review), which represents the most successful realization of the holographic principle \cite{RevModPhys.74.825}.
When applied to black hole scenarios, the extremization procedure allows the surface $X$ to move across the horizon into the black hole interior. Consequently, the fine-grained entropy depends on the geometry behind the horizon: black holes with identical exteriors but different interiors can have different von Neumann entropies. 

We now apply the prescription \eqref{RT} to the full time evolution of an evaporating black hole and discuss how the resultant entropy varies throughout the black hole lifetime.

\begin{figure}
    \centering
    \includegraphics[scale=0.5]{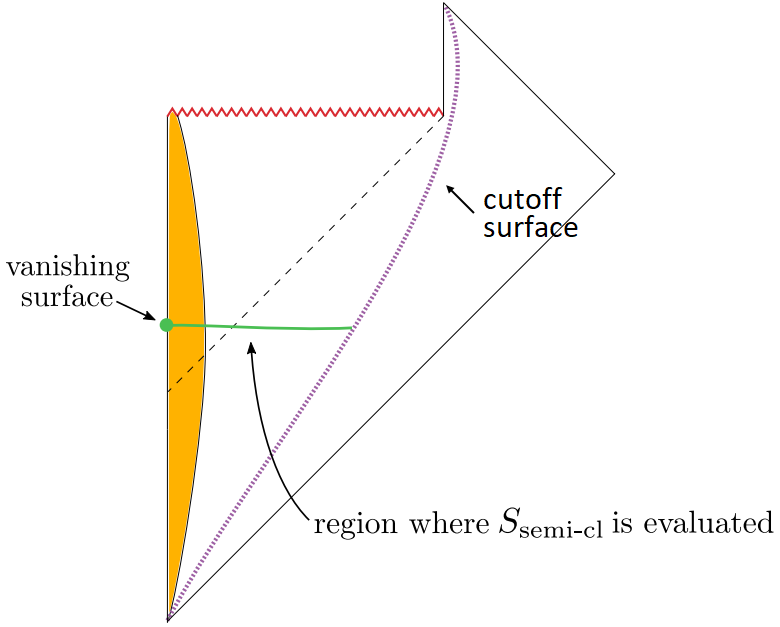}
    \caption{The minimal surface for the black hole at early times shrinks down to zero. The generalized entropy then reduces to the bulk entropy of the region enclosed by the cutoff surface. Figure adapted from \cite{RevModPhys.93.035002}.}
    \label{fig_noisl}
\end{figure}

Consider the early-time regime, immediately after black hole formation but before any Hawking quanta have escaped. In this case, no extremal surface is found by deforming $X$ into the interior, and one is forced to shrink the surface to a point. Accordingly, a vanishing extremal surface implies that the first term in Eq. \eqref{RT} is zero and the fine-grained entropy reduces to the entropy of matter enclosed by the cutoff surface, as shown by Fig. \ref{fig_noisl}. Assuming that the collapsing matter shell is in a pure state, this entropy vanishes\footnote{We neglect the UV-divergent entanglement contribution near the cutoff, assuming it to be time-independent and subtracting it implicitly.}. We underline that, in the absence of Hawking radiation, this fine-grained entropy remains constant in time. This contrasts with the coarse-grained (Bekenstein-Hawking) entropy, which starts at zero and grows to $4\pi r_s^2$.

When the black hole starts evaporating, a non-vanishing extremal surface appears, whose position depends on the amount of Hawking radiation already escaped from the black hole, and thus on time. It turns out that the precise location of this surface is close to the event horizon, and it can be determined by computing the so-called scrambling time \cite{Hayden:2007cs,Sekino:2008he,PhysRevD.88.124041}.
\begin{figure}
    \centering
    \includegraphics[scale=0.40]{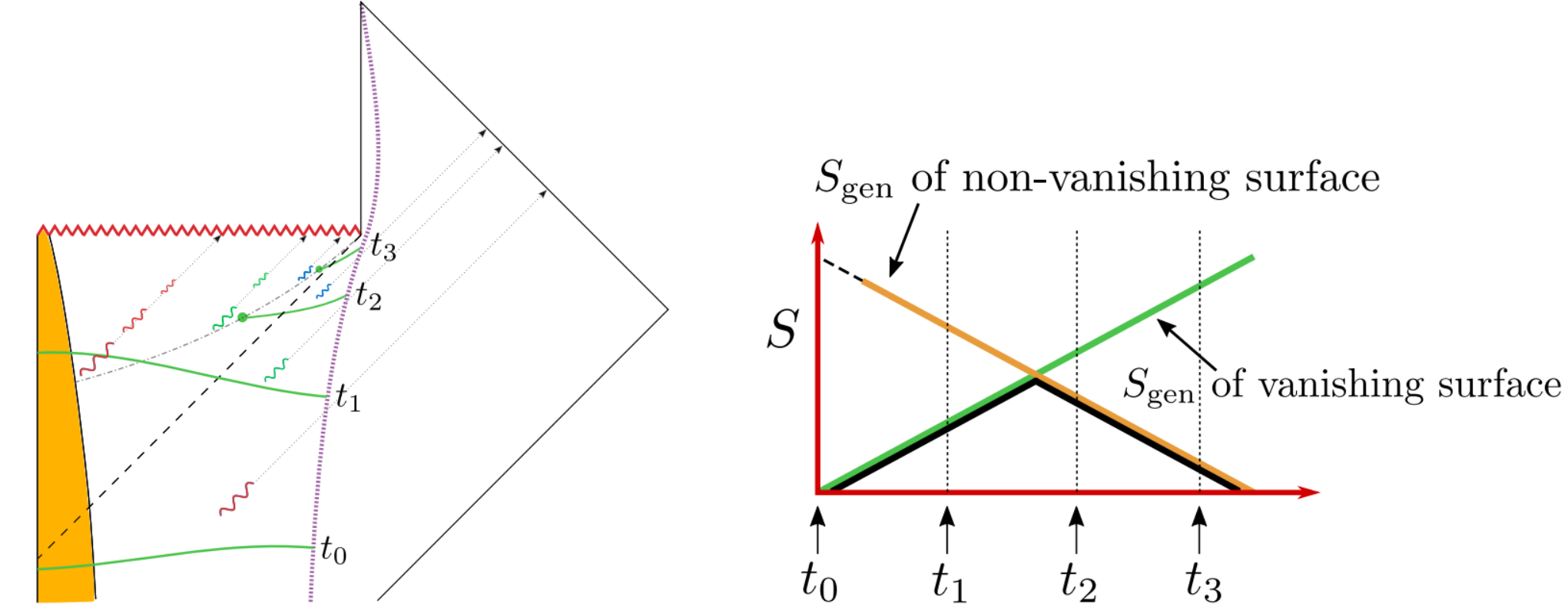}
    \caption{Page curve (black) for the fine-grained entropy of a black hole. The minimization procedure implies that the entropy is initially obtained from the growing contribution (green) due to the interior Hawking quanta, while the decreasing contribution (yellow) associated with the horizon area describes the entropy during the latest stages of black hole evaporation. Figure adapted from \cite{RevModPhys.93.035002}.}
    \label{fig_2extr}
\end{figure}
Due to the appearance of a non-vanishing extremal surface, the generalized entropy in Eq. \eqref{sgendef} has now a contribution from the von Neumann entropy $S_{\rm semi}$ of quantum fields, in addition with the area term. However, the non-vanishing surface only captures few Hawking quanta, so 
\begin{equation}
S_{\mathrm{gen}} \approx \frac{ \mathrm{Area}_{\mathrm{Horizon}}(t) }{4 G} \,,
\end{equation}
whose dynamics closely resembles that of thermodynamic entropy, thus giving a decreasing contribution during evaporation. Accordingly, the application of Eq. \eqref{RT} requires minimization over all available extremal surfaces. In Fig. \ref{fig_2extr}, we sketch the contributions arising from the two such surfaces: at early times, only the vanishing surface exists, giving a contribution that starts at zero and grows monotonically until the black hole evaporates. Some short time after the black hole forms, the non-vanishing surface appears, starting with a large value given by the current area of the black hole and
decreasing as the black hole shrinks.

We then observe that the Page curve for the fine-grained entropy of the black hole is captured by the transition between a growing contribution due to the pileup of interior Hawking quanta (vanishing extremal surface) and a decreasing contribution associated with the black hole shrinking (non-vanishing extremal surface).

\subsubsection{Black hole islands} \label{sec6.3.3}

We discussed how the fine-grained entropy of black holes, as computed via the gravitational prescription \eqref{RT}, correctly reproduces the Page curve. However, this result does not directly resolve the information paradox, which concerns the entropy growth of the Hawking radiation itself. 

Indeed, semiclassical evolution predicts an ever-increasing entropy $S_{\mathrm{semi}}(\Sigma_{\mathrm{Rad}})$ in the region outside the gravitational cutoff which contains the emitted radiation \cite{PhysRevD.14.2460}. 

This radiation then propagates in a region where gravitational effects may be small. However, since gravity was essential in preparing the quantum state of the system, the appropriate tool to compute its entropy remains the gravitational fine-grained entropy formula.  Eq. \eqref{RT} can be indeed applied even when there is no black hole present: \emph{the key idea is that the entropy should be extremized over all possible surfaces $X$, which may vary in position and connectivity}.

While we initially considered connected surfaces $\Sigma_X$, it is natural to also analyze disconnected configurations. Such configurations enlarge the boundary area, and thus can only be favored if they simultaneously reduce the semi-classical von Neumann entropy. This possibility arises in the presence of entangled matter distributed across distant regions.

This scenario precisely matches that of Hawking radiation, which is entangled with quantum fields residing inside the black hole.  Hence, we work in the following way:

\begin{itemize}
    \item[-] we include portions of the black hole interior, referring to them with the name \emph{islands}. Due to the presence of such islands, the semi-classical entropy can be reduced, albeit at the cost of adding a geometric area term. 
    \item[-] during the latest stages of black hole evaporation, the reduction in entropy outweighs the added area contribution, and the minimal generalized entropy is achieved by including these interior regions.
\end{itemize} 

The complete gravitational fine-grained entropy of the Hawking radiation is thus computed via the \emph{island formula} 
\begin{equation}
S_{\mathrm{Rad}} = \min_X \left\{ \mathrm{ext}_X \left[ \frac{\mathrm{Area}(X)}{4 G} + S_{\mathrm{semi}}(\Sigma_{\mathrm{Rad}} \cup \Sigma_{\mathrm{Island}}) \right] \right\},
\label{island}
\end{equation}
that represents a generalization of Eq. \eqref{RT}. Indeed, $\mathrm{Area}(X)$ is the area of the boundary of the island, and the min/ext operation is performed over all possible extremal surfaces $X$ \cite{Almheiri:2019hni,Penington:2019kki,Almheiri:2019qdq}. 
\begin{figure}
    \centering
    \includegraphics[scale=0.42]{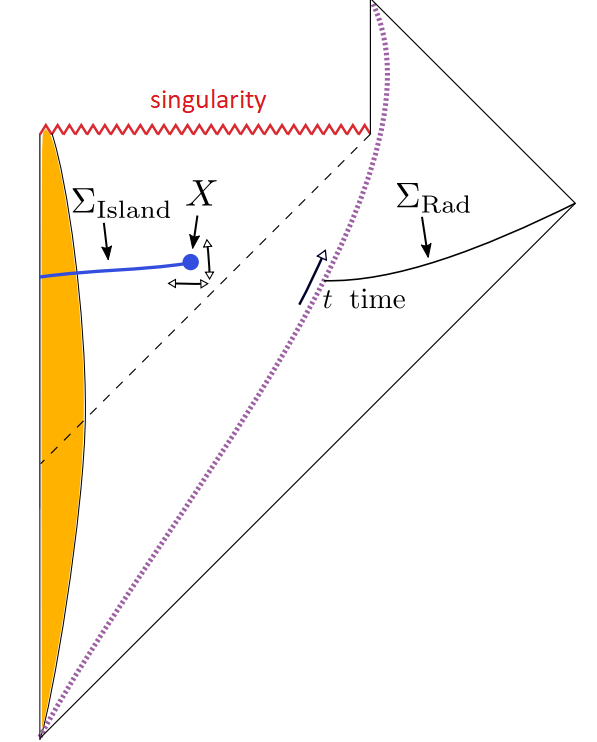}
    \hskip0.1in
    \includegraphics[scale=0.36]{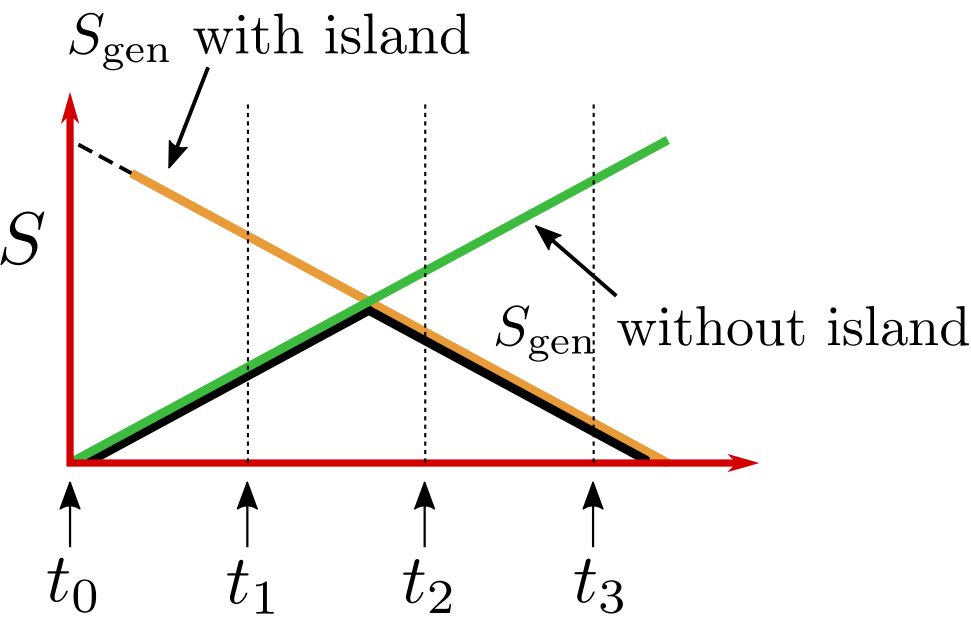}
    \caption{Page curve (black) for the fine-grained entropy of Hawking radiation. The no-island contribution (green) gives a growing entropy due to the outgoing Hawking quanta. However, a non-vanishing island that extremizes the entropy appears some time after black hole formation. This contribution (yellow) starts at a large value, given by the area of the horizon at early times, and decreases to zero during evaporation. Figure adapted from \cite{RevModPhys.93.035002}.}
    \label{isla_fig}
\end{figure}
On the left-hand side, $S_{\mathrm{Rad}}$ represents the full quantum entropy of the Hawking radiation. On the right-hand side, $S_{\mathrm{semi}}$ is the von Neumann entropy of the combined radiation and island systems, computed in the semi-classical approximation. It is crucial to distinguish these two: the left-hand side is the entropy of the exact state, while the right-hand side involves a semi-classical surrogate. Accordingly, the formula does not provide the exact radiation state explicitly; rather, it yields its entropy indirectly.

The practical implementation of the island formula proceeds as follows. One aims to compute the entropy of all Hawking radiation lying outside the cutoff surface, extending to spatial infinity. This region is denoted by $\Sigma_{\mathrm{Rad}}$. We then extremize the right-hand side of Eq. \eqref{island} with respect to the position of the island(s), inside the black hole. Fig. \ref{isla_fig} shows the case of a single island, centered around the origin. Finally, we minimize over all possible extremal positions and choices of islands.

This gives the Page curve: the rising piece comes from the no-island contribution $S_{\mathrm{semi}}(\Sigma_{\mathrm{Rad}})$, while the novel decreasing piece arises from the island contribution, which is governed by the shrinking area, since the von Neumann entropy term remains small at all times due to purification. 

Accordingly, we have again reproduced the Page curve, in agreement with the generalized entropy discussed in Sec. \ref{sec6.3.2}: this is exactly what we expected from a unitary evolution starting from a pure state, i.e., the entropy of the black hole and the entropy of the radiation region should be equal, thus confirming that information is not lost during evaporation processes \cite{RevModPhys.93.035002}.

In recent years, the information paradox has been studied in various black hole spacetimes, including Schwarzschild-de Sitter \cite{Goswami:2022ylc}, Reissner-Nordstr\"om \cite{Wang:2021woy} and Reissner-Nordstr\"om-AdS \cite{PhysRevD.110.046008} solutions, rotating Kerr black holes \cite{PhysRevD.110.066012}, regular configurations \cite{PhysRevD.108.104059} and primordial black holes \cite{PhysRevD.111.083015}, highlighting how the Page curve is affected by modifications of the black hole geometry. Islands have been also considered in de Sitter \cite{Balasubramanian:2020xqf} and quasi-de Sitter \cite{Seo:2022ezk} spacetimes and they may also have implications on inflation and primordial perturbations \cite{PhysRevD.107.123509}.


\section{Summary and outlook} \label{secCONCL}

Relativistic quantum information has emerged as a novel field between quantum field theory, general relativity, and quantum information science. Through the study of entanglement in curved spacetimes and non-inertial frames, relativistic quantum information offers a unique perspective on the structure of quantum correlations under the influence of gravity and acceleration. 

In this review, we explored a selection of key developments in the field, showing how momentum and position-space techniques may allow to quantify the entanglement entropy of quantum fields in dynamical spacetimes and in black hole scenarios characterized by the presence of event horizons and spacetime singularities. Another central theme has been the extraction and quantification of entanglement from the vacuum state of quantum fields, through the entanglement harvesting protocol.

We initially introduced the notion of entanglement for pure and mixed quantum states, presenting the von Neumann entropy and the negativity as basic entanglement measures, widely employed for information theoretic purposes.

Afterwards, quantum fields in curved spacetimes have been reviewed. In particular, field quantization has been reported, with great attention devoted to the GPP mechanism, which allows to produce particles from vacuum due to the sole gravitational interaction between a quantum field and the background. The processes of momentum and energy exchanges are thus revised, respectively dealing with inhomogeneties and anisotropic configurations. Consequences on entanglement and observations are explored in detail, providing outcomes for bosons and fermions. In this respect, the Einstein-Cartan theory is also used as a naive example to go further the standard general relativistic setup. Analogously, squeezing entropy and quantum-to-classical transition, together with dynamics of non-Gaussianities, are investigated in the framework of inflationary perturbations, stimulating the search for quantum signatures in the CMB.

In order to probe the entanglement features of quantum fields, the harvesting protocol has received increasing attention in recent years. Localized detectors may indeed shed light on the fundamental nature of the quantum vacuum, also providing entanglement resources directly usable for quantum information purposes. Particularly, the non-relativistic and relativistic approaches are discussed in detail, showing how detectors can be used and placed to argue characteristics of entanglement.  Furthermore, we have seen how spacetime geometry and causal structure modulate the efficiency of entanglement harvesting, thus making the extraction of quantum correlations highly sensitive to the background spacetime.

The geometrical nature of entanglement also emerges when computing the position-space entanglement of quantum fields in black hole spacetimes, particularly when the entangling surface is represented by the black hole horizon. In strongly gravitating scenarios, the behavior of entanglement indeed exhibits profound connections with thermodynamic properties and holographic principles. The emergence of area laws for entanglement entropy - whether derived from ground states of discretized quantum fields or through holographic arguments - points to a deep and still largely unexplored relationship between geometry and quantum information.

Furthermore, we have presented a gravitational formula to obtain the von Neumann entropy of black holes and Hawking radiation, which allows to reproduce the Page curve and thus solve the information paradox associated with black hole evaporation, even if a computation of the precise radiation quantum state is still missing.

Despite significant progress, many open questions remain. These include the proper formulation of entanglement in dynamically evolving spacetimes, the full relativistic generalization of information-theoretic protocols, and the interplay between entanglement structure and semiclassical or quantum gravitational effects. Furthermore, understanding how entanglement contributes to the emergence of spacetime geometry - as hinted by various tensor network and quantum gravity approaches (see e.g. \cite{Faulkner:2013ica,PhysRevLett.114.221601,Czech:2015kbp,Faulkner:2017tkh,Czech:2017ryf}) - represents a frontier that could redefine the foundations of both quantum theory and general relativity.

The above-presented efforts have also stimulated the search for some analogue models \cite{Barcelo:2005fc}, aiming to mimic the dynamics of quantum fields in curved spacetime via laboratory test beds and, in particular, to observe analogue gravitational particle creation and the corresponding entanglement generation via experimental setups. Since the GPP mechanism is independent of the microscopic description of the medium, various quantum fluids have been proposed to reproduce such a phenomenon in laboratory settings. In particular, a qualitative comparison with cosmological particle production was recently reported for a two-dimensional atomic Bose-Einstein condensate \cite{Hung:2012nc}, and phonon pair creation was observed in an ion trap, accompanied by the generation of spatial entanglement \cite{PhysRevLett.123.180502}. Similar effects were observed in superconducting circuits \cite{Wilson:2011rsw}, optical fibers \cite{Vezzoli2019} and quantum fluids of light \cite{Steinhauer:2021fhb}. Moreover, laboratory analogues of Hawking radiation and the corresponding entanglement production have been recently proposed via Bose-Einstein condensates \cite{Steinhauer:2015saa,Bec_hawk24} and optical systems \cite{PhysRevLett.128.091301,PhysRevD.106.105021}.

In conclusion, relativistic quantum information continues to provide both conceptual insights and concrete tools for probing the quantum structure of spacetime in both weak and strong gravity regimes, opening new avenues in theoretical physics toward the study of entanglement in gravity. Accordingly, by further bridging the gap between information theory and fundamental physics, it holds promise not only for deepening our understanding of the universe, but also for enabling novel technologies grounded in relativistic quantum phenomena, as e.g. harvesting and/or decoherence, of great interest for the cosmological community.

\bibliographystyle{elsarticle-num-names}

\begin{thebibliography}{231}
\expandafter\ifx\csname natexlab\endcsname\relax\def\natexlab#1{#1}\fi
\providecommand{\url}[1]{\texttt{#1}}
\providecommand{\href}[2]{#2}
\providecommand{\path}[1]{#1}
\providecommand{\DOIprefix}{doi:}
\providecommand{\ArXivprefix}{arXiv:}
\providecommand{\URLprefix}{URL: }
\providecommand{\Pubmedprefix}{pmid:}
\providecommand{\doi}[1]{\href{http://dx.doi.org/#1}{\path{#1}}}
\providecommand{\Pubmed}[1]{\href{pmid:#1}{\path{#1}}}
\providecommand{\bibinfo}[2]{#2}
\ifx\xfnm\relax \def\xfnm[#1]{\unskip,\space#1}\fi
\bibitem[{Schrödinger(1935)}]{Schrödinger_1935}
\bibinfo{author}{E.~Schrödinger},
\newblock \bibinfo{title}{Discussion of probability relations between separated systems},
\newblock \bibinfo{journal}{Mathematical Proceedings of the Cambridge Philosophical Society} \bibinfo{volume}{31} (\bibinfo{year}{1935}) \bibinfo{pages}{555–563}. \DOIprefix\doi{10.1017/S0305004100013554}.
\bibitem[{Horodecki et~al.(2009)Horodecki, Horodecki, Horodecki, and Horodecki}]{RevModPhys.81.865}
\bibinfo{author}{R.~Horodecki}, \bibinfo{author}{P.~Horodecki}, \bibinfo{author}{M.~Horodecki}, \bibinfo{author}{K.~Horodecki},
\newblock \bibinfo{title}{Quantum entanglement},
\newblock \bibinfo{journal}{Rev. Mod. Phys.} \bibinfo{volume}{81} (\bibinfo{year}{2009}) \bibinfo{pages}{865--942}. \URLprefix \url{https://link.aps.org/doi/10.1103/RevModPhys.81.865}. \DOIprefix\doi{10.1103/RevModPhys.81.865}.
\bibitem[{Erhard et~al.(2020)Erhard, Krenn, and Zeilinger}]{Erhard2020}
\bibinfo{author}{M.~Erhard}, \bibinfo{author}{M.~Krenn}, \bibinfo{author}{A.~Zeilinger},
\newblock \bibinfo{title}{Advances in high-dimensional quantum entanglement},
\newblock \bibinfo{journal}{Nature Reviews Physics} \bibinfo{volume}{2} (\bibinfo{year}{2020}) \bibinfo{pages}{365--381}. \URLprefix \url{https://doi.org/10.1038/s42254-020-0193-5}. \DOIprefix\doi{10.1038/s42254-020-0193-5}.
\bibitem[{Nielsen and Chuang(2012)}]{Nielsen:2012yss}
\bibinfo{author}{M.~A. Nielsen}, \bibinfo{author}{I.~L. Chuang}, \bibinfo{title}{{Quantum Computation and Quantum Information}}, \bibinfo{publisher}{Cambridge University Press}, \bibinfo{year}{2012}. \DOIprefix\doi{10.1017/cbo9780511976667}.
\bibitem[{Sood and Pooja(2024)}]{10167529}
\bibinfo{author}{S.~K. Sood}, \bibinfo{author}{Pooja},
\newblock \bibinfo{title}{Quantum computing review: A decade of research},
\newblock \bibinfo{journal}{IEEE Transactions on Engineering Management} \bibinfo{volume}{71} (\bibinfo{year}{2024}) \bibinfo{pages}{6662--6676}. \DOIprefix\doi{10.1109/TEM.2023.3284689}.
\bibitem[{Buhrman et~al.(2010)Buhrman, Cleve, Massar, and de~Wolf}]{RevModPhys.82.665}
\bibinfo{author}{H.~Buhrman}, \bibinfo{author}{R.~Cleve}, \bibinfo{author}{S.~Massar}, \bibinfo{author}{R.~de~Wolf},
\newblock \bibinfo{title}{Nonlocality and communication complexity},
\newblock \bibinfo{journal}{Rev. Mod. Phys.} \bibinfo{volume}{82} (\bibinfo{year}{2010}) \bibinfo{pages}{665--698}. \URLprefix \url{https://link.aps.org/doi/10.1103/RevModPhys.82.665}. \DOIprefix\doi{10.1103/RevModPhys.82.665}.
\bibitem[{Giovannetti et~al.(2011)Giovannetti, Lloyd, and Maccone}]{Giovannetti2011}
\bibinfo{author}{V.~Giovannetti}, \bibinfo{author}{S.~Lloyd}, \bibinfo{author}{L.~Maccone},
\newblock \bibinfo{title}{Advances in quantum metrology},
\newblock \bibinfo{journal}{Nature Photonics} \bibinfo{volume}{5} (\bibinfo{year}{2011}) \bibinfo{pages}{222--229}. \URLprefix \url{https://doi.org/10.1038/nphoton.2011.35}. \DOIprefix\doi{10.1038/nphoton.2011.35}.
\bibitem[{Degen et~al.(2017)Degen, Reinhard, and Cappellaro}]{RevModPhys.89.035002}
\bibinfo{author}{C.~L. Degen}, \bibinfo{author}{F.~Reinhard}, \bibinfo{author}{P.~Cappellaro},
\newblock \bibinfo{title}{Quantum sensing},
\newblock \bibinfo{journal}{Rev. Mod. Phys.} \bibinfo{volume}{89} (\bibinfo{year}{2017}) \bibinfo{pages}{035002}. \URLprefix \url{https://link.aps.org/doi/10.1103/RevModPhys.89.035002}. \DOIprefix\doi{10.1103/RevModPhys.89.035002}.
\bibitem[{Defienne et~al.(2024)Defienne, Bowen, Chekhova, Lemos, Oron, Ramelow, Treps, and Faccio}]{Defienne2024}
\bibinfo{author}{H.~Defienne}, \bibinfo{author}{W.~P. Bowen}, \bibinfo{author}{M.~Chekhova}, \bibinfo{author}{G.~B. Lemos}, \bibinfo{author}{D.~Oron}, \bibinfo{author}{S.~Ramelow}, \bibinfo{author}{N.~Treps}, \bibinfo{author}{D.~Faccio},
\newblock \bibinfo{title}{Advances in quantum imaging},
\newblock \bibinfo{journal}{Nature Photonics} \bibinfo{volume}{18} (\bibinfo{year}{2024}) \bibinfo{pages}{1024--1036}. \URLprefix \url{https://doi.org/10.1038/s41566-024-01516-w}. \DOIprefix\doi{10.1038/s41566-024-01516-w}.
\bibitem[{Reznik(2003)}]{Reznik:2002fz}
\bibinfo{author}{B.~Reznik},
\newblock \bibinfo{title}{{Entanglement from the vacuum}},
\newblock \bibinfo{journal}{Found. Phys.} \bibinfo{volume}{33} (\bibinfo{year}{2003}) \bibinfo{pages}{167--176}. \DOIprefix\doi{10.1023/A:1022875910744}. \href{http://arxiv.org/abs/quant-ph/0212044}{{\tt arXiv:quant-ph/0212044}}.
\bibitem[{Pozas-Kerstjens and Mart\'{\i}n-Mart\'{\i}nez(2015)}]{PhysRevD.92.064042}
\bibinfo{author}{A.~Pozas-Kerstjens}, \bibinfo{author}{E.~Mart\'{\i}n-Mart\'{\i}nez},
\newblock \bibinfo{title}{Harvesting correlations from the quantum vacuum},
\newblock \bibinfo{journal}{Phys. Rev. D} \bibinfo{volume}{92} (\bibinfo{year}{2015}) \bibinfo{pages}{064042}. \URLprefix \url{https://link.aps.org/doi/10.1103/PhysRevD.92.064042}. \DOIprefix\doi{10.1103/PhysRevD.92.064042}.
\bibitem[{Unruh(1976)}]{PhysRevD.14.870}
\bibinfo{author}{W.~G. Unruh},
\newblock \bibinfo{title}{Notes on black-hole evaporation},
\newblock \bibinfo{journal}{Phys. Rev. D} \bibinfo{volume}{14} (\bibinfo{year}{1976}) \bibinfo{pages}{870--892}. \URLprefix \url{https://link.aps.org/doi/10.1103/PhysRevD.14.870}. \DOIprefix\doi{10.1103/PhysRevD.14.870}.
\bibitem[{Unruh and Wald(1984)}]{PhysRevD.29.1047}
\bibinfo{author}{W.~G. Unruh}, \bibinfo{author}{R.~M. Wald},
\newblock \bibinfo{title}{What happens when an accelerating observer detects a rindler particle},
\newblock \bibinfo{journal}{Phys. Rev. D} \bibinfo{volume}{29} (\bibinfo{year}{1984}) \bibinfo{pages}{1047--1056}. \URLprefix \url{https://link.aps.org/doi/10.1103/PhysRevD.29.1047}. \DOIprefix\doi{10.1103/PhysRevD.29.1047}.
\bibitem[{Perche et~al.(2024)Perche, Polo-G\'omez, Torres, and Mart\'{\i}n-Mart\'{\i}nez}]{PhysRevD.109.045018}
\bibinfo{author}{T.~R. Perche}, \bibinfo{author}{J.~Polo-G\'omez}, \bibinfo{author}{B.~d. S.~L. Torres}, \bibinfo{author}{E.~Mart\'{\i}n-Mart\'{\i}nez},
\newblock \bibinfo{title}{Fully relativistic entanglement harvesting},
\newblock \bibinfo{journal}{Phys. Rev. D} \bibinfo{volume}{109} (\bibinfo{year}{2024}) \bibinfo{pages}{045018}. \URLprefix \url{https://link.aps.org/doi/10.1103/PhysRevD.109.045018}. \DOIprefix\doi{10.1103/PhysRevD.109.045018}.
\bibitem[{Parker(1969)}]{PhysRev.183.1057}
\bibinfo{author}{L.~Parker},
\newblock \bibinfo{title}{Quantized fields and particle creation in expanding universes. i},
\newblock \bibinfo{journal}{Phys. Rev.} \bibinfo{volume}{183} (\bibinfo{year}{1969}) \bibinfo{pages}{1057--1068}. \URLprefix \url{https://link.aps.org/doi/10.1103/PhysRev.183.1057}. \DOIprefix\doi{10.1103/PhysRev.183.1057}.
\bibitem[{Parker(1971)}]{PhysRevD.3.346}
\bibinfo{author}{L.~Parker},
\newblock \bibinfo{title}{Quantized fields and particle creation in expanding universes. ii},
\newblock \bibinfo{journal}{Phys. Rev. D} \bibinfo{volume}{3} (\bibinfo{year}{1971}) \bibinfo{pages}{346--356}. \URLprefix \url{https://link.aps.org/doi/10.1103/PhysRevD.3.346}. \DOIprefix\doi{10.1103/PhysRevD.3.346}.
\bibitem[{Duncan(1978)}]{PhysRevD.17.964}
\bibinfo{author}{A.~Duncan},
\newblock \bibinfo{title}{Explicit dimensional renormalization of quantum field theory in curved space-time},
\newblock \bibinfo{journal}{Phys. Rev. D} \bibinfo{volume}{17} (\bibinfo{year}{1978}) \bibinfo{pages}{964--971}. \URLprefix \url{https://link.aps.org/doi/10.1103/PhysRevD.17.964}. \DOIprefix\doi{10.1103/PhysRevD.17.964}.
\bibitem[{Ball et~al.(2006)Ball, Fuentes-Schuller, and Schuller}]{Ball:2005xa}
\bibinfo{author}{J.~L. Ball}, \bibinfo{author}{I.~Fuentes-Schuller}, \bibinfo{author}{F.~P. Schuller},
\newblock \bibinfo{title}{{Entanglement in an expanding spacetime}},
\newblock \bibinfo{journal}{Phys. Lett. A} \bibinfo{volume}{359} (\bibinfo{year}{2006}) \bibinfo{pages}{550--554}. \DOIprefix\doi{10.1016/j.physleta.2006.07.028}. \href{http://arxiv.org/abs/quant-ph/0506113}{{\tt arXiv:quant-ph/0506113}}.
\bibitem[{Fuentes et~al.(2010)Fuentes, Mann, Martin-Martinez, and Moradi}]{Fuentes:2010dt}
\bibinfo{author}{I.~Fuentes}, \bibinfo{author}{R.~B. Mann}, \bibinfo{author}{E.~Martin-Martinez}, \bibinfo{author}{S.~Moradi},
\newblock \bibinfo{title}{{Entanglement of Dirac fields in an expanding spacetime}},
\newblock \bibinfo{journal}{Phys. Rev. D} \bibinfo{volume}{82} (\bibinfo{year}{2010}) \bibinfo{pages}{045030}. \DOIprefix\doi{10.1103/PhysRevD.82.045030}. \href{http://arxiv.org/abs/1007.1569}{{\tt arXiv:1007.1569}}.
\bibitem[{Starobinsky(1980)}]{Starobinsky:1980te}
\bibinfo{author}{A.~A. Starobinsky},
\newblock \bibinfo{title}{{A New Type of Isotropic Cosmological Models Without Singularity}},
\newblock \bibinfo{journal}{Phys. Lett. B} \bibinfo{volume}{91} (\bibinfo{year}{1980}) \bibinfo{pages}{99--102}. \DOIprefix\doi{10.1016/0370-2693(80)90670-X}.
\bibitem[{Guth(1981)}]{Guth:1980zm}
\bibinfo{author}{A.~H. Guth},
\newblock \bibinfo{title}{{The Inflationary Universe: A Possible Solution to the Horizon and Flatness Problems}},
\newblock \bibinfo{journal}{Phys. Rev. D} \bibinfo{volume}{23} (\bibinfo{year}{1981}) \bibinfo{pages}{347--356}. \DOIprefix\doi{10.1103/PhysRevD.23.347}.
\bibitem[{Bassett et~al.(2006)Bassett, Tsujikawa, and Wands}]{RevModPhys.78.537}
\bibinfo{author}{B.~A. Bassett}, \bibinfo{author}{S.~Tsujikawa}, \bibinfo{author}{D.~Wands},
\newblock \bibinfo{title}{Inflation dynamics and reheating},
\newblock \bibinfo{journal}{Rev. Mod. Phys.} \bibinfo{volume}{78} (\bibinfo{year}{2006}) \bibinfo{pages}{537--589}. \URLprefix \url{https://link.aps.org/doi/10.1103/RevModPhys.78.537}. \DOIprefix\doi{10.1103/RevModPhys.78.537}.
\bibitem[{Akrami et~al.(2020)}]{Planck:2018jri}
\bibinfo{author}{Y.~Akrami}, et~al. (\bibinfo{collaboration}{Planck}),
\newblock \bibinfo{title}{{Planck 2018 results. X. Constraints on inflation}},
\newblock \bibinfo{journal}{Astron. Astrophys.} \bibinfo{volume}{641} (\bibinfo{year}{2020}) \bibinfo{pages}{A10}. \DOIprefix\doi{10.1051/0004-6361/201833887}. \href{http://arxiv.org/abs/1807.06211}{{\tt arXiv:1807.06211}}.
\bibitem[{Schander and Thiemann(2021)}]{Schander:2021pgt}
\bibinfo{author}{S.~Schander}, \bibinfo{author}{T.~Thiemann},
\newblock \bibinfo{title}{{Backreaction in Cosmology}},
\newblock \bibinfo{journal}{Front. Astron. Space Sci.} \bibinfo{volume}{0} (\bibinfo{year}{2021}) \bibinfo{pages}{113}. \DOIprefix\doi{10.3389/fspas.2021.692198}. \href{http://arxiv.org/abs/2106.06043}{{\tt arXiv:2106.06043}}.
\bibitem[{Lombardo and Lopez~Nacir(2005)}]{Lombardo:2005iz}
\bibinfo{author}{F.~C. Lombardo}, \bibinfo{author}{D.~Lopez~Nacir},
\newblock \bibinfo{title}{{Decoherence during inflation: The Generation of classical inhomogeneities}},
\newblock \bibinfo{journal}{Phys. Rev. D} \bibinfo{volume}{72} (\bibinfo{year}{2005}) \bibinfo{pages}{063506}. \DOIprefix\doi{10.1103/PhysRevD.72.063506}. \href{http://arxiv.org/abs/gr-qc/0506051}{{\tt arXiv:gr-qc/0506051}}.
\bibitem[{Franco and Calzetta(2011)}]{Franco:2011fg}
\bibinfo{author}{M.~Franco}, \bibinfo{author}{E.~Calzetta},
\newblock \bibinfo{title}{{Decoherence in the cosmic background radiation}},
\newblock \bibinfo{journal}{Class. Quant. Grav.} \bibinfo{volume}{28} (\bibinfo{year}{2011}) \bibinfo{pages}{145024}. \DOIprefix\doi{10.1088/0264-9381/28/14/145024}. \href{http://arxiv.org/abs/1103.0188}{{\tt arXiv:1103.0188}}.
\bibitem[{Liu et~al.(2016)Liu, Sou, and Wang}]{Liu:2016aaf}
\bibinfo{author}{J.~Liu}, \bibinfo{author}{C.-M. Sou}, \bibinfo{author}{Y.~Wang},
\newblock \bibinfo{title}{{Cosmic Decoherence: Massive Fields}},
\newblock \bibinfo{journal}{JHEP} \bibinfo{volume}{10} (\bibinfo{year}{2016}) \bibinfo{pages}{072}. \DOIprefix\doi{10.1007/JHEP10(2016)072}. \href{http://arxiv.org/abs/1608.07909}{{\tt arXiv:1608.07909}}.
\bibitem[{Martin and Vennin(2018)}]{Martin:2018lin}
\bibinfo{author}{J.~Martin}, \bibinfo{author}{V.~Vennin},
\newblock \bibinfo{title}{{Non Gaussianities from Quantum Decoherence during Inflation}},
\newblock \bibinfo{journal}{JCAP} \bibinfo{volume}{06} (\bibinfo{year}{2018}) \bibinfo{pages}{037}. \DOIprefix\doi{10.1088/1475-7516/2018/06/037}. \href{http://arxiv.org/abs/1805.05609}{{\tt arXiv:1805.05609}}.
\bibitem[{Kiefer et~al.(1998)Kiefer, Polarski, and Starobinsky}]{Kiefer:1998qe}
\bibinfo{author}{C.~Kiefer}, \bibinfo{author}{D.~Polarski}, \bibinfo{author}{A.~A. Starobinsky},
\newblock \bibinfo{title}{{Quantum to classical transition for fluctuations in the early universe}},
\newblock \bibinfo{journal}{Int. J. Mod. Phys. D} \bibinfo{volume}{7} (\bibinfo{year}{1998}) \bibinfo{pages}{455--462}. \DOIprefix\doi{10.1142/S0218271898000292}. \href{http://arxiv.org/abs/gr-qc/9802003}{{\tt arXiv:gr-qc/9802003}}.
\bibitem[{Burgess et~al.(2008)Burgess, Holman, and Hoover}]{Burgess:2006jn}
\bibinfo{author}{C.~P. Burgess}, \bibinfo{author}{R.~Holman}, \bibinfo{author}{D.~Hoover},
\newblock \bibinfo{title}{{Decoherence of inflationary primordial fluctuations}},
\newblock \bibinfo{journal}{Phys. Rev. D} \bibinfo{volume}{77} (\bibinfo{year}{2008}) \bibinfo{pages}{063534}. \DOIprefix\doi{10.1103/PhysRevD.77.063534}. \href{http://arxiv.org/abs/astro-ph/0601646}{{\tt arXiv:astro-ph/0601646}}.
\bibitem[{Ashtekar et~al.(2020)Ashtekar, Corichi, and Kesavan}]{PhysRevD.102.023512}
\bibinfo{author}{A.~Ashtekar}, \bibinfo{author}{A.~Corichi}, \bibinfo{author}{A.~Kesavan},
\newblock \bibinfo{title}{Emergence of classical behavior in the early universe},
\newblock \bibinfo{journal}{Phys. Rev. D} \bibinfo{volume}{102} (\bibinfo{year}{2020}) \bibinfo{pages}{023512}. \URLprefix \url{https://link.aps.org/doi/10.1103/PhysRevD.102.023512}. \DOIprefix\doi{10.1103/PhysRevD.102.023512}.
\bibitem[{Brahma et~al.(2020)Brahma, Alaryani, and Brandenberger}]{PhysRevD.102.043529}
\bibinfo{author}{S.~Brahma}, \bibinfo{author}{O.~Alaryani}, \bibinfo{author}{R.~Brandenberger},
\newblock \bibinfo{title}{Entanglement entropy of cosmological perturbations},
\newblock \bibinfo{journal}{Phys. Rev. D} \bibinfo{volume}{102} (\bibinfo{year}{2020}) \bibinfo{pages}{043529}. \URLprefix \url{https://link.aps.org/doi/10.1103/PhysRevD.102.043529}. \DOIprefix\doi{10.1103/PhysRevD.102.043529}.
\bibitem[{Martin-Martinez and Menicucci(2012)}]{Martin-Martinez:2012chf}
\bibinfo{author}{E.~Martin-Martinez}, \bibinfo{author}{N.~C. Menicucci},
\newblock \bibinfo{title}{{Cosmological quantum entanglement}},
\newblock \bibinfo{journal}{Class. Quant. Grav.} \bibinfo{volume}{29} (\bibinfo{year}{2012}) \bibinfo{pages}{224003}. \DOIprefix\doi{10.1088/0264-9381/29/22/224003}. \href{http://arxiv.org/abs/1204.4918}{{\tt arXiv:1204.4918}}.
\bibitem[{Balasubramanian et~al.(2012)Balasubramanian, McDermott, and Van~Raamsdonk}]{PhysRevD.86.045014}
\bibinfo{author}{V.~Balasubramanian}, \bibinfo{author}{M.~B. McDermott}, \bibinfo{author}{M.~Van~Raamsdonk},
\newblock \bibinfo{title}{Momentum-space entanglement and renormalization in quantum field theory},
\newblock \bibinfo{journal}{Phys. Rev. D} \bibinfo{volume}{86} (\bibinfo{year}{2012}) \bibinfo{pages}{045014}. \URLprefix \url{https://link.aps.org/doi/10.1103/PhysRevD.86.045014}. \DOIprefix\doi{10.1103/PhysRevD.86.045014}.
\bibitem[{Brahma et~al.(2023)Brahma, Calder\'on-Figueroa, Hassan, and Mi}]{Brahma:2023hki}
\bibinfo{author}{S.~Brahma}, \bibinfo{author}{J.~Calder\'on-Figueroa}, \bibinfo{author}{M.~Hassan}, \bibinfo{author}{X.~Mi},
\newblock \bibinfo{title}{{Momentum-space entanglement entropy in de Sitter spacetime}},
\newblock \bibinfo{journal}{Phys. Rev. D} \bibinfo{volume}{108} (\bibinfo{year}{2023}) \bibinfo{pages}{043522}. \DOIprefix\doi{10.1103/PhysRevD.108.043522}. \href{http://arxiv.org/abs/2302.13894}{{\tt arXiv:2302.13894}}.
\bibitem[{Frieman(1989)}]{Frieman:1985fr}
\bibinfo{author}{J.~A. Frieman},
\newblock \bibinfo{title}{{Particle Creation in Inhomogeneous Space-times}},
\newblock \bibinfo{journal}{Phys. Rev. D} \bibinfo{volume}{39} (\bibinfo{year}{1989}) \bibinfo{pages}{389}. \DOIprefix\doi{10.1103/PhysRevD.39.389}.
\bibitem[{Cespedes and Verdaguer(1990)}]{Cespedes:1989kh}
\bibinfo{author}{J.~Cespedes}, \bibinfo{author}{E.~Verdaguer},
\newblock \bibinfo{title}{{Particle Production in Inhomogeneous Cosmologies}},
\newblock \bibinfo{journal}{Phys. Rev. D} \bibinfo{volume}{41} (\bibinfo{year}{1990}) \bibinfo{pages}{1022}. \DOIprefix\doi{10.1103/PhysRevD.41.1022}.
\bibitem[{Belfiglio et~al.(2022)Belfiglio, Luongo, and Mancini}]{Belfiglio:2022cnd}
\bibinfo{author}{A.~Belfiglio}, \bibinfo{author}{O.~Luongo}, \bibinfo{author}{S.~Mancini},
\newblock \bibinfo{title}{{Geometric corrections to cosmological entanglement}},
\newblock \bibinfo{journal}{Phys. Rev. D} \bibinfo{volume}{105} (\bibinfo{year}{2022}) \bibinfo{pages}{123523}. \DOIprefix\doi{10.1103/PhysRevD.105.123523}. \href{http://arxiv.org/abs/2201.12299}{{\tt arXiv:2201.12299}}.
\bibitem[{Belfiglio et~al.(2023)Belfiglio, Luongo, and Mancini}]{Belfiglio:2022yvs}
\bibinfo{author}{A.~Belfiglio}, \bibinfo{author}{O.~Luongo}, \bibinfo{author}{S.~Mancini},
\newblock \bibinfo{title}{{Inflationary entanglement}},
\newblock \bibinfo{journal}{Phys. Rev. D} \bibinfo{volume}{107} (\bibinfo{year}{2023}) \bibinfo{pages}{103512}. \DOIprefix\doi{10.1103/PhysRevD.107.103512}. \href{http://arxiv.org/abs/2212.06448}{{\tt arXiv:2212.06448}}.
\bibitem[{Belfiglio et~al.(2024)Belfiglio, Luongo, and Mancini}]{Belfiglio:2023moe}
\bibinfo{author}{A.~Belfiglio}, \bibinfo{author}{O.~Luongo}, \bibinfo{author}{S.~Mancini},
\newblock \bibinfo{title}{{Superhorizon entanglement from inflationary particle production}},
\newblock \bibinfo{journal}{Phys. Rev. D} \bibinfo{volume}{109} (\bibinfo{year}{2024}) \bibinfo{pages}{123520}. \DOIprefix\doi{10.1103/PhysRevD.109.123520}. \href{http://arxiv.org/abs/2312.11419}{{\tt arXiv:2312.11419}}.
\bibitem[{Weinberg(1989)}]{RevModPhys.61.1}
\bibinfo{author}{S.~Weinberg},
\newblock \bibinfo{title}{The cosmological constant problem},
\newblock \bibinfo{journal}{Rev. Mod. Phys.} \bibinfo{volume}{61} (\bibinfo{year}{1989}) \bibinfo{pages}{1--23}. \URLprefix \url{https://link.aps.org/doi/10.1103/RevModPhys.61.1}. \DOIprefix\doi{10.1103/RevModPhys.61.1}.
\bibitem[{Martin(2012)}]{Martin:2012bt}
\bibinfo{author}{J.~Martin},
\newblock \bibinfo{title}{{Everything You Always Wanted To Know About The Cosmological Constant Problem (But Were Afraid To Ask)}},
\newblock \bibinfo{journal}{Comptes Rendus Physique} \bibinfo{volume}{13} (\bibinfo{year}{2012}) \bibinfo{pages}{566--665}. \DOIprefix\doi{10.1016/j.crhy.2012.04.008}. \href{http://arxiv.org/abs/1205.3365}{{\tt arXiv:1205.3365}}.
\bibitem[{Kolb and Long(2024)}]{RevModPhys.96.045005}
\bibinfo{author}{E.~W. Kolb}, \bibinfo{author}{A.~J. Long},
\newblock \bibinfo{title}{Cosmological gravitational particle production and its implications for cosmological relics},
\newblock \bibinfo{journal}{Rev. Mod. Phys.} \bibinfo{volume}{96} (\bibinfo{year}{2024}) \bibinfo{pages}{045005}. \URLprefix \url{https://link.aps.org/doi/10.1103/RevModPhys.96.045005}. \DOIprefix\doi{10.1103/RevModPhys.96.045005}.
\bibitem[{Bekenstein(1972)}]{Bekenstein:1972tm}
\bibinfo{author}{J.~D. Bekenstein},
\newblock \bibinfo{title}{{Black holes and the second law}},
\newblock \bibinfo{journal}{Lett. Nuovo Cim.} \bibinfo{volume}{4} (\bibinfo{year}{1972}) \bibinfo{pages}{737--740}. \DOIprefix\doi{10.1007/BF02757029}.
\bibitem[{Bekenstein(1973)}]{Bekenstein:1973ur}
\bibinfo{author}{J.~D. Bekenstein},
\newblock \bibinfo{title}{{Black holes and entropy}},
\newblock \bibinfo{journal}{Phys. Rev. D} \bibinfo{volume}{7} (\bibinfo{year}{1973}) \bibinfo{pages}{2333--2346}. \DOIprefix\doi{10.1103/PhysRevD.7.2333}.
\bibitem[{Bekenstein(1974)}]{Bekenstein:1974ax}
\bibinfo{author}{J.~D. Bekenstein},
\newblock \bibinfo{title}{{Generalized second law of thermodynamics in black hole physics}},
\newblock \bibinfo{journal}{Phys. Rev. D} \bibinfo{volume}{9} (\bibinfo{year}{1974}) \bibinfo{pages}{3292--3300}. \DOIprefix\doi{10.1103/PhysRevD.9.3292}.
\bibitem[{Hawking(1974)}]{Hawking:1974rv}
\bibinfo{author}{S.~W. Hawking},
\newblock \bibinfo{title}{{Black hole explosions}},
\newblock \bibinfo{journal}{Nature} \bibinfo{volume}{248} (\bibinfo{year}{1974}) \bibinfo{pages}{30--31}. \DOIprefix\doi{10.1038/248030a0}.
\bibitem[{Hawking(1975)}]{Hawking:1975vcx}
\bibinfo{author}{S.~W. Hawking},
\newblock \bibinfo{title}{{Particle Creation by Black Holes}},
\newblock \bibinfo{journal}{Commun. Math. Phys.} \bibinfo{volume}{43} (\bibinfo{year}{1975}) \bibinfo{pages}{199--220}. \DOIprefix\doi{10.1007/BF02345020}, \bibinfo{note}{[Erratum: Commun.Math.Phys. 46, 206 (1976)]}.
\bibitem[{Wald(2001)}]{Wald:1999vt}
\bibinfo{author}{R.~M. Wald},
\newblock \bibinfo{title}{{The thermodynamics of black holes}},
\newblock \bibinfo{journal}{Living Rev. Rel.} \bibinfo{volume}{4} (\bibinfo{year}{2001}) \bibinfo{pages}{6}. \DOIprefix\doi{10.12942/lrr-2001-6}. \href{http://arxiv.org/abs/gr-qc/9912119}{{\tt arXiv:gr-qc/9912119}}.
\bibitem[{Page(2005)}]{Page:2004xp}
\bibinfo{author}{D.~N. Page},
\newblock \bibinfo{title}{{Hawking radiation and black hole thermodynamics}},
\newblock \bibinfo{journal}{New J. Phys.} \bibinfo{volume}{7} (\bibinfo{year}{2005}) \bibinfo{pages}{203}. \DOIprefix\doi{10.1088/1367-2630/7/1/203}. \href{http://arxiv.org/abs/hep-th/0409024}{{\tt arXiv:hep-th/0409024}}.
\bibitem[{{Das} et~al.(2008){Das}, {Shankaranarayanan}, and {Sur}}]{2008arXiv0806.0402D}
\bibinfo{author}{S.~{Das}}, \bibinfo{author}{S.~{Shankaranarayanan}}, \bibinfo{author}{S.~{Sur}},
\newblock \bibinfo{title}{{Black hole entropy from entanglement: A review}},
\newblock \bibinfo{journal}{arXiv e-prints}  (\bibinfo{year}{2008}) \bibinfo{pages}{arXiv:0806.0402}. \DOIprefix\doi{10.48550/arXiv.0806.0402}. \href{http://arxiv.org/abs/0806.0402}{{\tt arXiv:0806.0402}}.
\bibitem[{Solodukhin(2011)}]{Solodukhin:2011gn}
\bibinfo{author}{S.~N. Solodukhin},
\newblock \bibinfo{title}{{Entanglement entropy of black holes}},
\newblock \bibinfo{journal}{Living Rev. Rel.} \bibinfo{volume}{14} (\bibinfo{year}{2011}) \bibinfo{pages}{8}. \DOIprefix\doi{10.12942/lrr-2011-8}. \href{http://arxiv.org/abs/1104.3712}{{\tt arXiv:1104.3712}}.
\bibitem[{Chandran and Shankaranarayanan(2020)}]{PhysRevD.102.125025}
\bibinfo{author}{S.~M. Chandran}, \bibinfo{author}{S.~Shankaranarayanan},
\newblock \bibinfo{title}{One-to-one correspondence between entanglement mechanics and black hole thermodynamics},
\newblock \bibinfo{journal}{Phys. Rev. D} \bibinfo{volume}{102} (\bibinfo{year}{2020}) \bibinfo{pages}{125025}. \URLprefix \url{https://link.aps.org/doi/10.1103/PhysRevD.102.125025}. \DOIprefix\doi{10.1103/PhysRevD.102.125025}.
\bibitem[{Jacobson and Satz(2013)}]{PhysRevD.87.084047}
\bibinfo{author}{T.~Jacobson}, \bibinfo{author}{A.~Satz},
\newblock \bibinfo{title}{Black hole entanglement entropy and the renormalization group},
\newblock \bibinfo{journal}{Phys. Rev. D} \bibinfo{volume}{87} (\bibinfo{year}{2013}) \bibinfo{pages}{084047}. \URLprefix \url{https://link.aps.org/doi/10.1103/PhysRevD.87.084047}. \DOIprefix\doi{10.1103/PhysRevD.87.084047}.
\bibitem[{Cooperman and Luty(2014)}]{Cooperman:2013iqr}
\bibinfo{author}{J.~H. Cooperman}, \bibinfo{author}{M.~A. Luty},
\newblock \bibinfo{title}{{Renormalization of Entanglement Entropy and the Gravitational Effective Action}},
\newblock \bibinfo{journal}{JHEP} \bibinfo{volume}{12} (\bibinfo{year}{2014}) \bibinfo{pages}{045}. \DOIprefix\doi{10.1007/JHEP12(2014)045}. \href{http://arxiv.org/abs/1302.1878}{{\tt arXiv:1302.1878}}.
\bibitem[{Balasubramanian et~al.(2024)Balasubramanian, Lawrence, Mag\'an, and Sasieta}]{PhysRevX.14.011024}
\bibinfo{author}{V.~Balasubramanian}, \bibinfo{author}{A.~Lawrence}, \bibinfo{author}{J.~M. Mag\'an}, \bibinfo{author}{M.~Sasieta},
\newblock \bibinfo{title}{Microscopic origin of the entropy of black holes in general relativity},
\newblock \bibinfo{journal}{Phys. Rev. X} \bibinfo{volume}{14} (\bibinfo{year}{2024}) \bibinfo{pages}{011024}. \URLprefix \url{https://link.aps.org/doi/10.1103/PhysRevX.14.011024}. \DOIprefix\doi{10.1103/PhysRevX.14.011024}.
\bibitem[{Belfiglio et~al.(2024)Belfiglio, Luongo, and Mancini}]{Belfiglio:2023sru}
\bibinfo{author}{A.~Belfiglio}, \bibinfo{author}{O.~Luongo}, \bibinfo{author}{S.~Mancini},
\newblock \bibinfo{title}{{Entanglement area law violation from field-curvature coupling}},
\newblock \bibinfo{journal}{Phys. Lett. B} \bibinfo{volume}{848} (\bibinfo{year}{2024}) \bibinfo{pages}{138398}. \DOIprefix\doi{10.1016/j.physletb.2023.138398}. \href{http://arxiv.org/abs/2306.08357}{{\tt arXiv:2306.08357}}.
\bibitem[{Almheiri et~al.(2021)Almheiri, Hartman, Maldacena, Shaghoulian, and Tajdini}]{RevModPhys.93.035002}
\bibinfo{author}{A.~Almheiri}, \bibinfo{author}{T.~Hartman}, \bibinfo{author}{J.~Maldacena}, \bibinfo{author}{E.~Shaghoulian}, \bibinfo{author}{A.~Tajdini},
\newblock \bibinfo{title}{The entropy of hawking radiation},
\newblock \bibinfo{journal}{Rev. Mod. Phys.} \bibinfo{volume}{93} (\bibinfo{year}{2021}) \bibinfo{pages}{035002}. \URLprefix \url{https://link.aps.org/doi/10.1103/RevModPhys.93.035002}. \DOIprefix\doi{10.1103/RevModPhys.93.035002}.
\bibitem[{Hawking(1976)}]{PhysRevD.14.2460}
\bibinfo{author}{S.~W. Hawking},
\newblock \bibinfo{title}{Breakdown of predictability in gravitational collapse},
\newblock \bibinfo{journal}{Phys. Rev. D} \bibinfo{volume}{14} (\bibinfo{year}{1976}) \bibinfo{pages}{2460--2473}. \URLprefix \url{https://link.aps.org/doi/10.1103/PhysRevD.14.2460}. \DOIprefix\doi{10.1103/PhysRevD.14.2460}.
\bibitem[{Page(1993)}]{Page:1993wv}
\bibinfo{author}{D.~N. Page},
\newblock \bibinfo{title}{{Information in black hole radiation}},
\newblock \bibinfo{journal}{Phys. Rev. Lett.} \bibinfo{volume}{71} (\bibinfo{year}{1993}) \bibinfo{pages}{3743--3746}. \DOIprefix\doi{10.1103/PhysRevLett.71.3743}. \href{http://arxiv.org/abs/hep-th/9306083}{{\tt arXiv:hep-th/9306083}}.
\bibitem[{Page(2013)}]{Page:2013dx}
\bibinfo{author}{D.~N. Page},
\newblock \bibinfo{title}{{Time Dependence of Hawking Radiation Entropy}},
\newblock \bibinfo{journal}{JCAP} \bibinfo{volume}{09} (\bibinfo{year}{2013}) \bibinfo{pages}{028}. \DOIprefix\doi{10.1088/1475-7516/2013/09/028}. \href{http://arxiv.org/abs/1301.4995}{{\tt arXiv:1301.4995}}.
\bibitem[{Penington(2020)}]{Penington:2019npb}
\bibinfo{author}{G.~Penington},
\newblock \bibinfo{title}{{Entanglement Wedge Reconstruction and the Information Paradox}},
\newblock \bibinfo{journal}{JHEP} \bibinfo{volume}{09} (\bibinfo{year}{2020}) \bibinfo{pages}{002}. \DOIprefix\doi{10.1007/JHEP09(2020)002}. \href{http://arxiv.org/abs/1905.08255}{{\tt arXiv:1905.08255}}.
\bibitem[{Almheiri et~al.(2019{\natexlab{a}})Almheiri, Engelhardt, Marolf, and Maxfield}]{Almheiri:2019psf}
\bibinfo{author}{A.~Almheiri}, \bibinfo{author}{N.~Engelhardt}, \bibinfo{author}{D.~Marolf}, \bibinfo{author}{H.~Maxfield},
\newblock \bibinfo{title}{{The entropy of bulk quantum fields and the entanglement wedge of an evaporating black hole}},
\newblock \bibinfo{journal}{JHEP} \bibinfo{volume}{12} (\bibinfo{year}{2019}{\natexlab{a}}) \bibinfo{pages}{063}. \DOIprefix\doi{10.1007/JHEP12(2019)063}. \href{http://arxiv.org/abs/1905.08762}{{\tt arXiv:1905.08762}}.
\bibitem[{Almheiri et~al.(2019{\natexlab{b}})Almheiri, Mahajan, and Maldacena}]{Almheiri:2019yqk}
\bibinfo{author}{A.~Almheiri}, \bibinfo{author}{R.~Mahajan}, \bibinfo{author}{J.~Maldacena},
\newblock \bibinfo{title}{{Islands outside the horizon}}  (\bibinfo{year}{2019}{\natexlab{b}}). \href{http://arxiv.org/abs/1910.11077}{{\tt arXiv:1910.11077}}.
\bibitem[{Almheiri et~al.(2020)Almheiri, Mahajan, Maldacena, and Zhao}]{Almheiri:2019hni}
\bibinfo{author}{A.~Almheiri}, \bibinfo{author}{R.~Mahajan}, \bibinfo{author}{J.~Maldacena}, \bibinfo{author}{Y.~Zhao},
\newblock \bibinfo{title}{{The Page curve of Hawking radiation from semiclassical geometry}},
\newblock \bibinfo{journal}{JHEP} \bibinfo{volume}{03} (\bibinfo{year}{2020}) \bibinfo{pages}{149}. \DOIprefix\doi{10.1007/JHEP03(2020)149}. \href{http://arxiv.org/abs/1908.10996}{{\tt arXiv:1908.10996}}.
\bibitem[{Maldacena(1998)}]{Maldacena:1997re}
\bibinfo{author}{J.~M. Maldacena},
\newblock \bibinfo{title}{{The Large N limit of superconformal field theories and supergravity}},
\newblock \bibinfo{journal}{Adv. Theor. Math. Phys.} \bibinfo{volume}{2} (\bibinfo{year}{1998}) \bibinfo{pages}{231--252}. \DOIprefix\doi{10.4310/ATMP.1998.v2.n2.a1}. \href{http://arxiv.org/abs/hep-th/9711200}{{\tt arXiv:hep-th/9711200}}.
\bibitem[{Ryu and Takayanagi(2006)}]{Ryu:2006bv}
\bibinfo{author}{S.~Ryu}, \bibinfo{author}{T.~Takayanagi},
\newblock \bibinfo{title}{{Holographic derivation of entanglement entropy from AdS/CFT}},
\newblock \bibinfo{journal}{Phys. Rev. Lett.} \bibinfo{volume}{96} (\bibinfo{year}{2006}) \bibinfo{pages}{181602}. \DOIprefix\doi{10.1103/PhysRevLett.96.181602}. \href{http://arxiv.org/abs/hep-th/0603001}{{\tt arXiv:hep-th/0603001}}.
\bibitem[{Plenio and Virmani(2007)}]{Plenio:2007zz}
\bibinfo{author}{M.~B. Plenio}, \bibinfo{author}{S.~S. Virmani},
\newblock \bibinfo{title}{{An Introduction to Entanglement Theory}},
\newblock \bibinfo{journal}{Quant. Inf. Comput.} \bibinfo{volume}{7} (\bibinfo{year}{2007}) \bibinfo{pages}{001--051}. \href{http://arxiv.org/abs/quant-ph/0504163}{{\tt arXiv:quant-ph/0504163}}.
\bibitem[{Serafini(2023)}]{Serafini:2023rrn}
\bibinfo{author}{A.~Serafini}, \bibinfo{title}{{Quantum Continuous Variables: A Primer of Theoretical Methods (2nd ed.)}}, \bibinfo{publisher}{CRC Press}, \bibinfo{year}{2023}. \DOIprefix\doi{10.1201/9781003250975}.
\bibitem[{Holevo and Werner(2001)}]{PhysRevA.63.032312}
\bibinfo{author}{A.~S. Holevo}, \bibinfo{author}{R.~F. Werner},
\newblock \bibinfo{title}{Evaluating capacities of bosonic gaussian channels},
\newblock \bibinfo{journal}{Phys. Rev. A} \bibinfo{volume}{63} (\bibinfo{year}{2001}) \bibinfo{pages}{032312}. \URLprefix \url{https://link.aps.org/doi/10.1103/PhysRevA.63.032312}. \DOIprefix\doi{10.1103/PhysRevA.63.032312}.
\bibitem[{Birrell and Davies(1984)}]{Birrell:1982ix}
\bibinfo{author}{N.~D. Birrell}, \bibinfo{author}{P.~C.~W. Davies}, \bibinfo{title}{{Quantum Fields in Curved Space}}, Cambridge Monographs on Mathematical Physics, \bibinfo{publisher}{Cambridge Univ. Press}, \bibinfo{address}{Cambridge, UK}, \bibinfo{year}{1984}. \DOIprefix\doi{10.1017/CBO9780511622632}.
\bibitem[{Parker and Toms(2009)}]{Parker:2009uva}
\bibinfo{author}{L.~E. Parker}, \bibinfo{author}{D.~Toms}, \bibinfo{title}{{Quantum Field Theory in Curved Spacetime}: {Quantized Field and Gravity}}, Cambridge Monographs on Mathematical Physics, \bibinfo{publisher}{Cambridge University Press}, \bibinfo{year}{2009}. \DOIprefix\doi{10.1017/CBO9780511813924}.
\bibitem[{Hollands and Wald(2015)}]{Hollands:2014eia}
\bibinfo{author}{S.~Hollands}, \bibinfo{author}{R.~M. Wald},
\newblock \bibinfo{title}{{Quantum fields in curved spacetime}},
\newblock \bibinfo{journal}{Phys. Rept.} \bibinfo{volume}{574} (\bibinfo{year}{2015}) \bibinfo{pages}{1--35}. \DOIprefix\doi{10.1016/j.physrep.2015.02.001}. \href{http://arxiv.org/abs/1401.2026}{{\tt arXiv:1401.2026}}.
\bibitem[{Wald(2010)}]{wald2010general}
\bibinfo{author}{R.~M. Wald}, \bibinfo{title}{General Relativity}, \bibinfo{publisher}{University of Chicago Press}, \bibinfo{year}{2010}. \URLprefix \url{https://books.google.it/books?id=9S-hzg6-moYC}.
\bibitem[{Ford(2021)}]{Ford:2021syk}
\bibinfo{author}{L.~H. Ford},
\newblock \bibinfo{title}{{Cosmological particle production: a review}},
\newblock \bibinfo{journal}{Rept. Prog. Phys.} \bibinfo{volume}{84} (\bibinfo{year}{2021}). \DOIprefix\doi{10.1088/1361-6633/ac1b23}. \href{http://arxiv.org/abs/2112.02444}{{\tt arXiv:2112.02444}}.
\bibitem[{Wald(1995)}]{Wald:1995yp}
\bibinfo{author}{R.~M. Wald}, \bibinfo{title}{{Quantum Field Theory in Curved Space-Time and Black Hole Thermodynamics}}, Chicago Lectures in Physics, \bibinfo{publisher}{University of Chicago Press}, \bibinfo{address}{Chicago, IL}, \bibinfo{year}{1995}.
\bibitem[{Hehl et~al.(1976)Hehl, von~der Heyde, Kerlick, and Nester}]{RevModPhys.48.393}
\bibinfo{author}{F.~W. Hehl}, \bibinfo{author}{P.~von~der Heyde}, \bibinfo{author}{G.~D. Kerlick}, \bibinfo{author}{J.~M. Nester},
\newblock \bibinfo{title}{General relativity with spin and torsion: Foundations and prospects},
\newblock \bibinfo{journal}{Rev. Mod. Phys.} \bibinfo{volume}{48} (\bibinfo{year}{1976}) \bibinfo{pages}{393--416}. \URLprefix \url{https://link.aps.org/doi/10.1103/RevModPhys.48.393}. \DOIprefix\doi{10.1103/RevModPhys.48.393}.
\bibitem[{Deser and van Nieuwenhuizen(1974)}]{PhysRevD.10.411}
\bibinfo{author}{S.~Deser}, \bibinfo{author}{P.~van Nieuwenhuizen},
\newblock \bibinfo{title}{Nonrenormalizability of the quantized dirac-einstein system},
\newblock \bibinfo{journal}{Phys. Rev. D} \bibinfo{volume}{10} (\bibinfo{year}{1974}) \bibinfo{pages}{411--420}. \URLprefix \url{https://link.aps.org/doi/10.1103/PhysRevD.10.411}. \DOIprefix\doi{10.1103/PhysRevD.10.411}.
\bibitem[{Deser and Isham(1976)}]{PhysRevD.14.2505}
\bibinfo{author}{S.~Deser}, \bibinfo{author}{C.~J. Isham},
\newblock \bibinfo{title}{Canonical vierbein form of general relativity},
\newblock \bibinfo{journal}{Phys. Rev. D} \bibinfo{volume}{14} (\bibinfo{year}{1976}) \bibinfo{pages}{2505--2510}. \URLprefix \url{https://link.aps.org/doi/10.1103/PhysRevD.14.2505}. \DOIprefix\doi{10.1103/PhysRevD.14.2505}.
\bibitem[{Shapiro(2002)}]{Shapiro:2001rz}
\bibinfo{author}{I.~L. Shapiro},
\newblock \bibinfo{title}{{Physical aspects of the space-time torsion}},
\newblock \bibinfo{journal}{Phys. Rept.} \bibinfo{volume}{357} (\bibinfo{year}{2002}) \bibinfo{pages}{113}. \DOIprefix\doi{10.1016/S0370-1573(01)00030-8}. \href{http://arxiv.org/abs/hep-th/0103093}{{\tt arXiv:hep-th/0103093}}.
\bibitem[{Benisty et~al.(2019)Benisty, Guendelman, Saridakis, Stoecker, Struckmeier, and Vasak}]{Benisty:2019jqz}
\bibinfo{author}{D.~Benisty}, \bibinfo{author}{E.~I. Guendelman}, \bibinfo{author}{E.~N. Saridakis}, \bibinfo{author}{H.~Stoecker}, \bibinfo{author}{J.~Struckmeier}, \bibinfo{author}{D.~Vasak},
\newblock \bibinfo{title}{{Inflation from fermions with curvature-dependent mass}},
\newblock \bibinfo{journal}{Phys. Rev. D} \bibinfo{volume}{100} (\bibinfo{year}{2019}) \bibinfo{pages}{043523}. \DOIprefix\doi{10.1103/PhysRevD.100.043523}. \href{http://arxiv.org/abs/1905.03731}{{\tt arXiv:1905.03731}}.
\bibitem[{Pierini et~al.(2016)Pierini, Moradi, and Mancini}]{Pierini:2015jma}
\bibinfo{author}{R.~Pierini}, \bibinfo{author}{S.~Moradi}, \bibinfo{author}{S.~Mancini},
\newblock \bibinfo{title}{{The role of spin in entanglement generated by expanding spacetime}},
\newblock \bibinfo{journal}{Int. J. Theor. Phys.} \bibinfo{volume}{55} (\bibinfo{year}{2016}) \bibinfo{pages}{3059--3078}. \DOIprefix\doi{10.1007/s10773-016-2937-7}. \href{http://arxiv.org/abs/1507.06811}{{\tt arXiv:1507.06811}}.
\bibitem[{Bassett et~al.(2002)Bassett, Peloso, Sorbo, and Tsujikawa}]{Bassett:2001jg}
\bibinfo{author}{B.~A. Bassett}, \bibinfo{author}{M.~Peloso}, \bibinfo{author}{L.~Sorbo}, \bibinfo{author}{S.~Tsujikawa},
\newblock \bibinfo{title}{{Fermion production from preheating amplified metric perturbations}},
\newblock \bibinfo{journal}{Nucl. Phys. B} \bibinfo{volume}{622} (\bibinfo{year}{2002}) \bibinfo{pages}{393--415}. \DOIprefix\doi{10.1016/S0550-3213(01)00608-3}. \href{http://arxiv.org/abs/hep-ph/0109176}{{\tt arXiv:hep-ph/0109176}}.
\bibitem[{Schwarzschild(1916)}]{Schwarzschild:1916uq}
\bibinfo{author}{K.~Schwarzschild},
\newblock \bibinfo{title}{{On the gravitational field of a mass point according to Einstein's theory}},
\newblock \bibinfo{journal}{Sitzungsber. Preuss. Akad. Wiss. Berlin (Math. Phys. )} \bibinfo{volume}{1916} (\bibinfo{year}{1916}) \bibinfo{pages}{189--196}. \href{http://arxiv.org/abs/physics/9905030}{{\tt arXiv:physics/9905030}}.
\bibitem[{Birkhoff(1927)}]{GDB1927}
\bibinfo{author}{G.~D. Birkhoff}, \bibinfo{title}{Relativity and Modern Physics}, \bibinfo{publisher}{Harvard University Press}, \bibinfo{address}{Cambridge, MA and London, England}, \bibinfo{year}{1927}. \URLprefix \url{https://doi.org/10.4159/harvard.9780674734487}. \DOIprefix\doi{doi:10.4159/harvard.9780674734487}.
\bibitem[{Lan et~al.(2023)Lan, Yang, Guo, and Miao}]{Lan:2023cvz}
\bibinfo{author}{C.~Lan}, \bibinfo{author}{H.~Yang}, \bibinfo{author}{Y.~Guo}, \bibinfo{author}{Y.-G. Miao},
\newblock \bibinfo{title}{{Regular Black Holes: A Short Topic Review}},
\newblock \bibinfo{journal}{Int. J. Theor. Phys.} \bibinfo{volume}{62} (\bibinfo{year}{2023}) \bibinfo{pages}{202}. \DOIprefix\doi{10.1007/s10773-023-05454-1}. \href{http://arxiv.org/abs/2303.11696}{{\tt arXiv:2303.11696}}.
\bibitem[{{Bardeen}(1968)}]{1968qtr..conf...87B}
\bibinfo{author}{J.~{Bardeen}},
\newblock \bibinfo{title}{{Non-singular general relativistic gravitational collapse}},
\newblock in: \bibinfo{booktitle}{Proceedings of the 5th International Conference on Gravitation and the Theory of Relativity}, \bibinfo{year}{1968}, p.~\bibinfo{pages}{87}.
\bibitem[{Ayon-Beato and Garcia(2000)}]{Ayon-Beato:2000mjt}
\bibinfo{author}{E.~Ayon-Beato}, \bibinfo{author}{A.~Garcia},
\newblock \bibinfo{title}{{The Bardeen model as a nonlinear magnetic monopole}},
\newblock \bibinfo{journal}{Phys. Lett. B} \bibinfo{volume}{493} (\bibinfo{year}{2000}) \bibinfo{pages}{149--152}. \DOIprefix\doi{10.1016/S0370-2693(00)01125-4}. \href{http://arxiv.org/abs/gr-qc/0009077}{{\tt arXiv:gr-qc/0009077}}.
\bibitem[{Arnowitt et~al.(1959)Arnowitt, Deser, and Misner}]{PhysRev.116.1322}
\bibinfo{author}{R.~Arnowitt}, \bibinfo{author}{S.~Deser}, \bibinfo{author}{C.~W. Misner},
\newblock \bibinfo{title}{Dynamical structure and definition of energy in general relativity},
\newblock \bibinfo{journal}{Phys. Rev.} \bibinfo{volume}{116} (\bibinfo{year}{1959}) \bibinfo{pages}{1322--1330}. \URLprefix \url{https://link.aps.org/doi/10.1103/PhysRev.116.1322}. \DOIprefix\doi{10.1103/PhysRev.116.1322}.
\bibitem[{Manko and Ruiz(2016)}]{Manko:2016ixk}
\bibinfo{author}{V.~S. Manko}, \bibinfo{author}{E.~Ruiz},
\newblock \bibinfo{title}{{On electromagnetic energy in Bardeen and ABG spacetimes}},
\newblock \bibinfo{journal}{Phys. Lett. B} \bibinfo{volume}{760} (\bibinfo{year}{2016}) \bibinfo{pages}{759--762}. \DOIprefix\doi{10.1016/j.physletb.2016.07.066}. \href{http://arxiv.org/abs/1605.00792}{{\tt arXiv:1605.00792}}.
\bibitem[{Hayward(2006)}]{PhysRevLett.96.031103}
\bibinfo{author}{S.~A. Hayward},
\newblock \bibinfo{title}{Formation and evaporation of nonsingular black holes},
\newblock \bibinfo{journal}{Phys. Rev. Lett.} \bibinfo{volume}{96} (\bibinfo{year}{2006}) \bibinfo{pages}{031103}. \URLprefix \url{https://link.aps.org/doi/10.1103/PhysRevLett.96.031103}. \DOIprefix\doi{10.1103/PhysRevLett.96.031103}.
\bibitem[{Neves and Saa(2014)}]{Neves:2014aba}
\bibinfo{author}{J.~C.~S. Neves}, \bibinfo{author}{A.~Saa},
\newblock \bibinfo{title}{{Regular rotating black holes and the weak energy condition}},
\newblock \bibinfo{journal}{Phys. Lett. B} \bibinfo{volume}{734} (\bibinfo{year}{2014}) \bibinfo{pages}{44--48}. \DOIprefix\doi{10.1016/j.physletb.2014.05.026}. \href{http://arxiv.org/abs/1402.2694}{{\tt arXiv:1402.2694}}.
\bibitem[{Penrose(1963)}]{PhysRevLett.10.66}
\bibinfo{author}{R.~Penrose},
\newblock \bibinfo{title}{Asymptotic properties of fields and space-times},
\newblock \bibinfo{journal}{Phys. Rev. Lett.} \bibinfo{volume}{10} (\bibinfo{year}{1963}) \bibinfo{pages}{66--68}. \URLprefix \url{https://link.aps.org/doi/10.1103/PhysRevLett.10.66}. \DOIprefix\doi{10.1103/PhysRevLett.10.66}.
\bibitem[{Kumar and Shankaranarayanan(2017)}]{PhysRevD.95.065023}
\bibinfo{author}{S.~S. Kumar}, \bibinfo{author}{S.~Shankaranarayanan},
\newblock \bibinfo{title}{Role of spatial higher order derivatives in momentum space entanglement},
\newblock \bibinfo{journal}{Phys. Rev. D} \bibinfo{volume}{95} (\bibinfo{year}{2017}) \bibinfo{pages}{065023}. \URLprefix \url{https://link.aps.org/doi/10.1103/PhysRevD.95.065023}. \DOIprefix\doi{10.1103/PhysRevD.95.065023}.
\bibitem[{Rai and Boyanovsky(2020)}]{PhysRevD.102.063532}
\bibinfo{author}{M.~Rai}, \bibinfo{author}{D.~Boyanovsky},
\newblock \bibinfo{title}{Origin of entropy of gravitationally produced dark matter: The entanglement entropy},
\newblock \bibinfo{journal}{Phys. Rev. D} \bibinfo{volume}{102} (\bibinfo{year}{2020}) \bibinfo{pages}{063532}. \URLprefix \url{https://link.aps.org/doi/10.1103/PhysRevD.102.063532}. \DOIprefix\doi{10.1103/PhysRevD.102.063532}.
\bibitem[{Pierini et~al.(2017)Pierini, Moradi, and Mancini}]{Pierini:2016rkk}
\bibinfo{author}{R.~Pierini}, \bibinfo{author}{S.~Moradi}, \bibinfo{author}{S.~Mancini},
\newblock \bibinfo{title}{{Spacetime anisotropy affects cosmological entanglement}},
\newblock \bibinfo{journal}{Nucl. Phys. B} \bibinfo{volume}{924} (\bibinfo{year}{2017}) \bibinfo{pages}{684--698}. \DOIprefix\doi{10.1016/j.nuclphysb.2017.09.025}. \href{http://arxiv.org/abs/1606.03005}{{\tt arXiv:1606.03005}}.
\bibitem[{Pierini et~al.(2019)Pierini, Moradi, and Mancini}]{Pierini:2018wki}
\bibinfo{author}{R.~Pierini}, \bibinfo{author}{S.~Moradi}, \bibinfo{author}{S.~Mancini},
\newblock \bibinfo{title}{{Entanglement in anisotropic expanding spacetime}},
\newblock \bibinfo{journal}{Eur. Phys. J. D} \bibinfo{volume}{73} (\bibinfo{year}{2019}) \bibinfo{pages}{33}. \DOIprefix\doi{10.1140/epjd/e2019-90463-y}. \href{http://arxiv.org/abs/1812.02835}{{\tt arXiv:1812.02835}}.
\bibitem[{Ma and Bertschinger(1995)}]{Ma:1995ey}
\bibinfo{author}{C.-P. Ma}, \bibinfo{author}{E.~Bertschinger},
\newblock \bibinfo{title}{{Cosmological perturbation theory in the synchronous and conformal Newtonian gauges}},
\newblock \bibinfo{journal}{Astrophys. J.} \bibinfo{volume}{455} (\bibinfo{year}{1995}) \bibinfo{pages}{7--25}. \DOIprefix\doi{10.1086/176550}. \href{http://arxiv.org/abs/astro-ph/9506072}{{\tt arXiv:astro-ph/9506072}}.
\bibitem[{Belfiglio et~al.(2021)Belfiglio, Luongo, and Mancini}]{PhysRevD.104.043523}
\bibinfo{author}{A.~Belfiglio}, \bibinfo{author}{O.~Luongo}, \bibinfo{author}{S.~Mancini},
\newblock \bibinfo{title}{Entanglement production in einstein-cartan theory},
\newblock \bibinfo{journal}{Phys. Rev. D} \bibinfo{volume}{104} (\bibinfo{year}{2021}) \bibinfo{pages}{043523}. \URLprefix \url{https://link.aps.org/doi/10.1103/PhysRevD.104.043523}. \DOIprefix\doi{10.1103/PhysRevD.104.043523}.
\bibitem[{Kibble(1961)}]{Kibble:1961ba}
\bibinfo{author}{T.~W.~B. Kibble},
\newblock \bibinfo{title}{{Lorentz invariance and the gravitational field}},
\newblock \bibinfo{journal}{J. Math. Phys.} \bibinfo{volume}{2} (\bibinfo{year}{1961}) \bibinfo{pages}{212--221}. \DOIprefix\doi{10.1063/1.1703702}.
\bibitem[{SCIAMA(1964)}]{RevModPhys.36.463}
\bibinfo{author}{D.~W. SCIAMA},
\newblock \bibinfo{title}{The physical structure of general relativity},
\newblock \bibinfo{journal}{Rev. Mod. Phys.} \bibinfo{volume}{36} (\bibinfo{year}{1964}) \bibinfo{pages}{463--469}. \URLprefix \url{https://link.aps.org/doi/10.1103/RevModPhys.36.463}. \DOIprefix\doi{10.1103/RevModPhys.36.463}.
\bibitem[{Buchbinder and Shapiro(1985)}]{Buchbinder:1985ux}
\bibinfo{author}{I.~L. Buchbinder}, \bibinfo{author}{I.~L. Shapiro},
\newblock \bibinfo{title}{{ON THE RENORMALIZATION OF MODELS OF QUANTUM FIELD THEORY IN AN EXTERNAL GRAVITATIONAL FIELD WITH TORSION}},
\newblock \bibinfo{journal}{Phys. Lett. B} \bibinfo{volume}{151} (\bibinfo{year}{1985}) \bibinfo{pages}{263--266}. \DOIprefix\doi{10.1016/0370-2693(85)90848-2}.
\bibitem[{Buchbinder and Shapiro(1990)}]{Buchbinder:1990ku}
\bibinfo{author}{I.~L. Buchbinder}, \bibinfo{author}{I.~L. Shapiro},
\newblock \bibinfo{title}{{On the renormalization group equations in curved space-time with torsion}},
\newblock \bibinfo{journal}{Class. Quant. Grav.} \bibinfo{volume}{7} (\bibinfo{year}{1990}) \bibinfo{pages}{1197--1206}. \DOIprefix\doi{10.1088/0264-9381/7/7/015}.
\bibitem[{Belyaev et~al.(2017)Belyaev, Thomas, and Shapiro}]{PhysRevD.95.095033}
\bibinfo{author}{A.~S. Belyaev}, \bibinfo{author}{M.~C. Thomas}, \bibinfo{author}{I.~L. Shapiro},
\newblock \bibinfo{title}{Torsion as a dark matter candidate from the higgs portal},
\newblock \bibinfo{journal}{Phys. Rev. D} \bibinfo{volume}{95} (\bibinfo{year}{2017}) \bibinfo{pages}{095033}. \URLprefix \url{https://link.aps.org/doi/10.1103/PhysRevD.95.095033}. \DOIprefix\doi{10.1103/PhysRevD.95.095033}.
\bibitem[{Poplawski(2010)}]{Poplawski:2010kb}
\bibinfo{author}{N.~J. Poplawski},
\newblock \bibinfo{title}{{Cosmology with torsion: An alternative to cosmic inflation}},
\newblock \bibinfo{journal}{Phys. Lett. B} \bibinfo{volume}{694} (\bibinfo{year}{2010}) \bibinfo{pages}{181--185}. \DOIprefix\doi{10.1016/j.physletb.2010.09.056}. \href{http://arxiv.org/abs/1007.0587}{{\tt arXiv:1007.0587}}, \bibinfo{note}{[Erratum: Phys.Lett.B 701, 672--672 (2011)]}.
\bibitem[{Cabral et~al.(2020)Cabral, Lobo, and Rubiera-Garcia}]{Cabral:2020mzw}
\bibinfo{author}{F.~Cabral}, \bibinfo{author}{F.~S.~N. Lobo}, \bibinfo{author}{D.~Rubiera-Garcia},
\newblock \bibinfo{title}{{The cosmological principle in theories with torsion: The case of Einstein-Cartan-Dirac-Maxwell gravity}},
\newblock \bibinfo{journal}{JCAP} \bibinfo{volume}{10} (\bibinfo{year}{2020}) \bibinfo{pages}{057}. \DOIprefix\doi{10.1088/1475-7516/2020/10/057}. \href{http://arxiv.org/abs/2004.13693}{{\tt arXiv:2004.13693}}.
\bibitem[{Mukhanov et~al.(1992)Mukhanov, Feldman, and Brandenberger}]{Mukhanov:1990me}
\bibinfo{author}{V.~F. Mukhanov}, \bibinfo{author}{H.~A. Feldman}, \bibinfo{author}{R.~H. Brandenberger},
\newblock \bibinfo{title}{{Theory of cosmological perturbations. Part 1. Classical perturbations. Part 2. Quantum theory of perturbations. Part 3. Extensions}},
\newblock \bibinfo{journal}{Phys. Rept.} \bibinfo{volume}{215} (\bibinfo{year}{1992}) \bibinfo{pages}{203--333}. \DOIprefix\doi{10.1016/0370-1573(92)90044-Z}.
\bibitem[{Riotto(2003)}]{Riotto:2002yw}
\bibinfo{author}{A.~Riotto},
\newblock \bibinfo{title}{{Inflation and the theory of cosmological perturbations}},
\newblock \bibinfo{journal}{ICTP Lect. Notes Ser.} \bibinfo{volume}{14} (\bibinfo{year}{2003}) \bibinfo{pages}{317--413}. \href{http://arxiv.org/abs/hep-ph/0210162}{{\tt arXiv:hep-ph/0210162}}.
\bibitem[{Brandenberger(2004)}]{Brandenberger:2003vk}
\bibinfo{author}{R.~H. Brandenberger},
\newblock \bibinfo{title}{{Lectures on the theory of cosmological perturbations}},
\newblock \bibinfo{journal}{Lect. Notes Phys.} \bibinfo{volume}{646} (\bibinfo{year}{2004}) \bibinfo{pages}{127--167}. \href{http://arxiv.org/abs/hep-th/0306071}{{\tt arXiv:hep-th/0306071}}.
\bibitem[{Bardeen(1980)}]{Bardeen:1980kt}
\bibinfo{author}{J.~M. Bardeen},
\newblock \bibinfo{title}{{Gauge Invariant Cosmological Perturbations}},
\newblock \bibinfo{journal}{Phys. Rev. D} \bibinfo{volume}{22} (\bibinfo{year}{1980}) \bibinfo{pages}{1882--1905}. \DOIprefix\doi{10.1103/PhysRevD.22.1882}.
\bibitem[{Bunch and Davies(1978)}]{Bunch:1978yq}
\bibinfo{author}{T.~S. Bunch}, \bibinfo{author}{P.~C.~W. Davies},
\newblock \bibinfo{title}{{Quantum Field Theory in de Sitter Space: Renormalization by Point Splitting}},
\newblock \bibinfo{journal}{Proc. Roy. Soc. Lond. A} \bibinfo{volume}{360} (\bibinfo{year}{1978}) \bibinfo{pages}{117--134}. \DOIprefix\doi{10.1098/rspa.1978.0060}.
\bibitem[{Danielsson and Olsson(2004)}]{Danielsson:2003wb}
\bibinfo{author}{U.~H. Danielsson}, \bibinfo{author}{M.~E. Olsson},
\newblock \bibinfo{title}{{On thermalization in de Sitter space}},
\newblock \bibinfo{journal}{JHEP} \bibinfo{volume}{03} (\bibinfo{year}{2004}) \bibinfo{pages}{036}. \DOIprefix\doi{10.1088/1126-6708/2004/03/036}. \href{http://arxiv.org/abs/hep-th/0309163}{{\tt arXiv:hep-th/0309163}}.
\bibitem[{Greene et~al.(2006)Greene, Parikh, and van~der Schaar}]{Greene:2005wk}
\bibinfo{author}{B.~Greene}, \bibinfo{author}{M.~Parikh}, \bibinfo{author}{J.~P. van~der Schaar},
\newblock \bibinfo{title}{{Universal correction to the inflationary vacuum}},
\newblock \bibinfo{journal}{JHEP} \bibinfo{volume}{04} (\bibinfo{year}{2006}) \bibinfo{pages}{057}. \DOIprefix\doi{10.1088/1126-6708/2006/04/057}. \href{http://arxiv.org/abs/hep-th/0512243}{{\tt arXiv:hep-th/0512243}}.
\bibitem[{Martineau(2007)}]{Martineau:2006ki}
\bibinfo{author}{P.~Martineau},
\newblock \bibinfo{title}{{On the decoherence of primordial fluctuations during inflation}},
\newblock \bibinfo{journal}{Class. Quant. Grav.} \bibinfo{volume}{24} (\bibinfo{year}{2007}) \bibinfo{pages}{5817--5834}. \DOIprefix\doi{10.1088/0264-9381/24/23/006}. \href{http://arxiv.org/abs/astro-ph/0601134}{{\tt arXiv:astro-ph/0601134}}.
\bibitem[{Burgess et~al.(2015)Burgess, Holman, Tasinato, and Williams}]{Burgess:2014eoa}
\bibinfo{author}{C.~P. Burgess}, \bibinfo{author}{R.~Holman}, \bibinfo{author}{G.~Tasinato}, \bibinfo{author}{M.~Williams},
\newblock \bibinfo{title}{{EFT Beyond the Horizon: Stochastic Inflation and How Primordial Quantum Fluctuations Go Classical}},
\newblock \bibinfo{journal}{JHEP} \bibinfo{volume}{03} (\bibinfo{year}{2015}) \bibinfo{pages}{090}. \DOIprefix\doi{10.1007/JHEP03(2015)090}. \href{http://arxiv.org/abs/1408.5002}{{\tt arXiv:1408.5002}}.
\bibitem[{Nelson(2016)}]{Nelson:2016kjm}
\bibinfo{author}{E.~Nelson},
\newblock \bibinfo{title}{{Quantum Decoherence During Inflation from Gravitational Nonlinearities}},
\newblock \bibinfo{journal}{JCAP} \bibinfo{volume}{03} (\bibinfo{year}{2016}) \bibinfo{pages}{022}. \DOIprefix\doi{10.1088/1475-7516/2016/03/022}. \href{http://arxiv.org/abs/1601.03734}{{\tt arXiv:1601.03734}}.
\bibitem[{Martin et~al.(2022)Martin, Micheli, and Vennin}]{Martin:2021znx}
\bibinfo{author}{J.~Martin}, \bibinfo{author}{A.~Micheli}, \bibinfo{author}{V.~Vennin},
\newblock \bibinfo{title}{{Discord and decoherence}},
\newblock \bibinfo{journal}{JCAP} \bibinfo{volume}{04} (\bibinfo{year}{2022}) \bibinfo{pages}{051}. \DOIprefix\doi{10.1088/1475-7516/2022/04/051}. \href{http://arxiv.org/abs/2112.05037}{{\tt arXiv:2112.05037}}.
\bibitem[{Shandera et~al.(2018)Shandera, Agarwal, and Kamal}]{PhysRevD.98.083535}
\bibinfo{author}{S.~Shandera}, \bibinfo{author}{N.~Agarwal}, \bibinfo{author}{A.~Kamal},
\newblock \bibinfo{title}{Open quantum cosmological system},
\newblock \bibinfo{journal}{Phys. Rev. D} \bibinfo{volume}{98} (\bibinfo{year}{2018}) \bibinfo{pages}{083535}. \URLprefix \url{https://link.aps.org/doi/10.1103/PhysRevD.98.083535}. \DOIprefix\doi{10.1103/PhysRevD.98.083535}.
\bibitem[{Brandenberger et~al.(1992)Brandenberger, Mukhanov, and Prokopec}]{PhysRevLett.69.3606}
\bibinfo{author}{R.~Brandenberger}, \bibinfo{author}{V.~Mukhanov}, \bibinfo{author}{T.~Prokopec},
\newblock \bibinfo{title}{Entropy of a classical stochastic field and cosmological perturbations},
\newblock \bibinfo{journal}{Phys. Rev. Lett.} \bibinfo{volume}{69} (\bibinfo{year}{1992}) \bibinfo{pages}{3606--3609}. \URLprefix \url{https://link.aps.org/doi/10.1103/PhysRevLett.69.3606}. \DOIprefix\doi{10.1103/PhysRevLett.69.3606}.
\bibitem[{Brandenberger et~al.(1993)Brandenberger, Mukhanov, and Prokopec}]{PhysRevD.48.2443}
\bibinfo{author}{R.~Brandenberger}, \bibinfo{author}{V.~Mukhanov}, \bibinfo{author}{T.~Prokopec},
\newblock \bibinfo{title}{Entropy of the gravitational field},
\newblock \bibinfo{journal}{Phys. Rev. D} \bibinfo{volume}{48} (\bibinfo{year}{1993}) \bibinfo{pages}{2443--2455}. \URLprefix \url{https://link.aps.org/doi/10.1103/PhysRevD.48.2443}. \DOIprefix\doi{10.1103/PhysRevD.48.2443}.
\bibitem[{Gasperini and Giovannini(1993)}]{Gasperini:1993mq}
\bibinfo{author}{M.~Gasperini}, \bibinfo{author}{M.~Giovannini},
\newblock \bibinfo{title}{{Quantum squeezing and cosmological entropy production}},
\newblock \bibinfo{journal}{Class. Quant. Grav.} \bibinfo{volume}{10} (\bibinfo{year}{1993}) \bibinfo{pages}{L133--L136}. \DOIprefix\doi{10.1088/0264-9381/10/9/004}. \href{http://arxiv.org/abs/gr-qc/9307024}{{\tt arXiv:gr-qc/9307024}}.
\bibitem[{Berjon et~al.(2021)Berjon, Okon, and Sudarsky}]{PhysRevD.103.043521}
\bibinfo{author}{J.~Berjon}, \bibinfo{author}{E.~Okon}, \bibinfo{author}{D.~Sudarsky},
\newblock \bibinfo{title}{Critical review of prevailing explanations for the emergence of classicality in cosmology},
\newblock \bibinfo{journal}{Phys. Rev. D} \bibinfo{volume}{103} (\bibinfo{year}{2021}) \bibinfo{pages}{043521}. \URLprefix \url{https://link.aps.org/doi/10.1103/PhysRevD.103.043521}. \DOIprefix\doi{10.1103/PhysRevD.103.043521}.
\bibitem[{Hsiang and Hu(2022)}]{Hsiang:2021kgh}
\bibinfo{author}{J.-T. Hsiang}, \bibinfo{author}{B.-L. Hu},
\newblock \bibinfo{title}{{No Intrinsic Decoherence of Inflationary Cosmological Perturbations}},
\newblock \bibinfo{journal}{Universe} \bibinfo{volume}{8} (\bibinfo{year}{2022}) \bibinfo{pages}{27}. \DOIprefix\doi{10.3390/universe8010027}. \href{http://arxiv.org/abs/2112.04092}{{\tt arXiv:2112.04092}}.
\bibitem[{Agullo et~al.(2022)Agullo, Bonga, and Metidieri}]{Agullo:2022ttg}
\bibinfo{author}{I.~Agullo}, \bibinfo{author}{B.~Bonga}, \bibinfo{author}{P.~R. Metidieri},
\newblock \bibinfo{title}{{Does inflation squeeze cosmological perturbations?}},
\newblock \bibinfo{journal}{JCAP} \bibinfo{volume}{09} (\bibinfo{year}{2022}) \bibinfo{pages}{032}. \DOIprefix\doi{10.1088/1475-7516/2022/09/032}. \href{http://arxiv.org/abs/2203.07066}{{\tt arXiv:2203.07066}}.
\bibitem[{Maldacena and Pimentel(2013)}]{Maldacena:2012xp}
\bibinfo{author}{J.~Maldacena}, \bibinfo{author}{G.~L. Pimentel},
\newblock \bibinfo{title}{{Entanglement entropy in de Sitter space}},
\newblock \bibinfo{journal}{JHEP} \bibinfo{volume}{02} (\bibinfo{year}{2013}) \bibinfo{pages}{038}. \DOIprefix\doi{10.1007/JHEP02(2013)038}. \href{http://arxiv.org/abs/1210.7244}{{\tt arXiv:1210.7244}}.
\bibitem[{Martin and Vennin(2021)}]{Martin:2021qkg}
\bibinfo{author}{J.~e. Martin}, \bibinfo{author}{V.~Vennin},
\newblock \bibinfo{title}{{Real-space entanglement in the Cosmic Microwave Background}},
\newblock \bibinfo{journal}{JCAP} \bibinfo{volume}{10} (\bibinfo{year}{2021}) \bibinfo{pages}{036}. \DOIprefix\doi{10.1088/1475-7516/2021/10/036}. \href{http://arxiv.org/abs/2106.15100}{{\tt arXiv:2106.15100}}.
\bibitem[{Espinosa-Portal\'es and Vennin(2022)}]{Espinosa-Portales:2022yok}
\bibinfo{author}{L.~Espinosa-Portal\'es}, \bibinfo{author}{V.~Vennin},
\newblock \bibinfo{title}{{Real-space Bell inequalities in de~Sitter}},
\newblock \bibinfo{journal}{JCAP} \bibinfo{volume}{07} (\bibinfo{year}{2022}) \bibinfo{pages}{037}. \DOIprefix\doi{10.1088/1475-7516/2022/07/037}. \href{http://arxiv.org/abs/2203.03505}{{\tt arXiv:2203.03505}}.
\bibitem[{Chandran et~al.(2024)Chandran, Rajeev, and Shankaranarayanan}]{PhysRevD.109.023503}
\bibinfo{author}{S.~M. Chandran}, \bibinfo{author}{K.~Rajeev}, \bibinfo{author}{S.~Shankaranarayanan},
\newblock \bibinfo{title}{Real-space quantum-to-classical transition of time dependent background fluctuations},
\newblock \bibinfo{journal}{Phys. Rev. D} \bibinfo{volume}{109} (\bibinfo{year}{2024}) \bibinfo{pages}{023503}. \URLprefix \url{https://link.aps.org/doi/10.1103/PhysRevD.109.023503}. \DOIprefix\doi{10.1103/PhysRevD.109.023503}.
\bibitem[{Maldacena(2003)}]{Maldacena:2002vr}
\bibinfo{author}{J.~M. Maldacena},
\newblock \bibinfo{title}{{Non-Gaussian features of primordial fluctuations in single field inflationary models}},
\newblock \bibinfo{journal}{JHEP} \bibinfo{volume}{05} (\bibinfo{year}{2003}) \bibinfo{pages}{013}. \DOIprefix\doi{10.1088/1126-6708/2003/05/013}. \href{http://arxiv.org/abs/astro-ph/0210603}{{\tt arXiv:astro-ph/0210603}}.
\bibitem[{Chen et~al.(2007)Chen, Huang, Kachru, and Shiu}]{Chen:2006nt}
\bibinfo{author}{X.~Chen}, \bibinfo{author}{M.-x. Huang}, \bibinfo{author}{S.~Kachru}, \bibinfo{author}{G.~Shiu},
\newblock \bibinfo{title}{{Observational signatures and non-Gaussianities of general single field inflation}},
\newblock \bibinfo{journal}{JCAP} \bibinfo{volume}{01} (\bibinfo{year}{2007}) \bibinfo{pages}{002}. \DOIprefix\doi{10.1088/1475-7516/2007/01/002}. \href{http://arxiv.org/abs/hep-th/0605045}{{\tt arXiv:hep-th/0605045}}.
\bibitem[{Brahma et~al.(2022)Brahma, Berera, and Calder\'on-Figueroa}]{Brahma:2022yxu}
\bibinfo{author}{S.~Brahma}, \bibinfo{author}{A.~Berera}, \bibinfo{author}{J.~Calder\'on-Figueroa},
\newblock \bibinfo{title}{{Quantum corrections to the primordial tensor spectrum: open EFTs \& Markovian decoupling of UV modes}},
\newblock \bibinfo{journal}{JHEP} \bibinfo{volume}{08} (\bibinfo{year}{2022}) \bibinfo{pages}{225}. \DOIprefix\doi{10.1007/JHEP08(2022)225}. \href{http://arxiv.org/abs/2206.05797}{{\tt arXiv:2206.05797}}.
\bibitem[{Salcedo et~al.(2024)Salcedo, Colas, and Pajer}]{Salcedo:2024smn}
\bibinfo{author}{S.~A. Salcedo}, \bibinfo{author}{T.~Colas}, \bibinfo{author}{E.~Pajer},
\newblock \bibinfo{title}{{The open effective field theory of inflation}},
\newblock \bibinfo{journal}{JHEP} \bibinfo{volume}{10} (\bibinfo{year}{2024}) \bibinfo{pages}{248}. \DOIprefix\doi{10.1007/JHEP10(2024)248}. \href{http://arxiv.org/abs/2404.15416}{{\tt arXiv:2404.15416}}.
\bibitem[{Burgess et~al.(2023)Burgess, Holman, Kaplanek, Martin, and Vennin}]{Burgess:2022nwu}
\bibinfo{author}{C.~P. Burgess}, \bibinfo{author}{R.~Holman}, \bibinfo{author}{G.~Kaplanek}, \bibinfo{author}{J.~Martin}, \bibinfo{author}{V.~Vennin},
\newblock \bibinfo{title}{{Minimal decoherence from inflation}},
\newblock \bibinfo{journal}{JCAP} \bibinfo{volume}{07} (\bibinfo{year}{2023}) \bibinfo{pages}{022}. \DOIprefix\doi{10.1088/1475-7516/2023/07/022}. \href{http://arxiv.org/abs/2211.11046}{{\tt arXiv:2211.11046}}.
\bibitem[{Banerjee et~al.(2023)Banerjee, Choudhury, Chowdhury, Knaute, Panda, and Shirish}]{Banerjee:2021lqu}
\bibinfo{author}{S.~Banerjee}, \bibinfo{author}{S.~Choudhury}, \bibinfo{author}{S.~Chowdhury}, \bibinfo{author}{J.~Knaute}, \bibinfo{author}{S.~Panda}, \bibinfo{author}{K.~Shirish},
\newblock \bibinfo{title}{{Thermalization in quenched open quantum cosmology}},
\newblock \bibinfo{journal}{Nucl. Phys. B} \bibinfo{volume}{996} (\bibinfo{year}{2023}) \bibinfo{pages}{116368}. \DOIprefix\doi{10.1016/j.nuclphysb.2023.116368}. \href{http://arxiv.org/abs/2104.10692}{{\tt arXiv:2104.10692}}.
\bibitem[{Hsiang and Hu(2021)}]{Hsiang:2021qqo}
\bibinfo{author}{J.-T. Hsiang}, \bibinfo{author}{B.-L. Hu},
\newblock \bibinfo{title}{{Intrinsic Entropy of Squeezed Quantum Fields and Nonequilibrium Quantum Dynamics of Cosmological Perturbations}},
\newblock \bibinfo{journal}{Entropy} \bibinfo{volume}{23} (\bibinfo{year}{2021}) \bibinfo{pages}{1544}. \DOIprefix\doi{10.3390/e23111544}. \href{http://arxiv.org/abs/2110.02757}{{\tt arXiv:2110.02757}}.
\bibitem[{Salton et~al.(2015)Salton, Mann, and Menicucci}]{Salton2015}
\bibinfo{author}{G.~Salton}, \bibinfo{author}{R.~B. Mann}, \bibinfo{author}{N.~C. Menicucci},
\newblock \bibinfo{title}{Acceleration-assisted entanglement harvesting and rangefinding},
\newblock \bibinfo{journal}{New Journal of Physics} \bibinfo{volume}{17} (\bibinfo{year}{2015}) \bibinfo{pages}{035001}.
\bibitem[{Reeh and Schlieder(1961)}]{ReehSchlieder}
\bibinfo{author}{H.~Reeh}, \bibinfo{author}{S.~Schlieder},
\newblock \bibinfo{title}{Bemerkungen zur unit\"{a}r\"{a}quivalenz von lorentzinvarianten feldern},
\newblock \bibinfo{journal}{Nuovo Cimento} \bibinfo{volume}{22} (\bibinfo{year}{1961}) \bibinfo{pages}{1051--1068}.
\bibitem[{Summers and Werner(1987)}]{SummersWerner}
\bibinfo{author}{S.~J. Summers}, \bibinfo{author}{R.~Werner},
\newblock \bibinfo{title}{Bell’s inequalities and quantum field theory. i. general setting},
\newblock \bibinfo{journal}{Journal of Mathematical Physics} \bibinfo{volume}{28} (\bibinfo{year}{1987}) \bibinfo{pages}{2440--2447}.
\bibitem[{Sorkin(1993)}]{Sorkin:1993gg}
\bibinfo{author}{R.~D. Sorkin},
\newblock \bibinfo{title}{{Impossible measurements on quantum fields}},
\newblock in: \bibinfo{booktitle}{{Directions in General Relativity: An International Symposium in Honor of the 60th Birthdays of Dieter Brill and Charles Misner}}, \bibinfo{year}{1993}, pp. \bibinfo{pages}{293--305}. \href{http://arxiv.org/abs/gr-qc/9302018}{{\tt arXiv:gr-qc/9302018}}.
\bibitem[{Borsten et~al.(2021)Borsten, Jubb, and Kells}]{PhysRevD.104.025012}
\bibinfo{author}{L.~Borsten}, \bibinfo{author}{I.~Jubb}, \bibinfo{author}{G.~Kells},
\newblock \bibinfo{title}{Impossible measurements revisited},
\newblock \bibinfo{journal}{Phys. Rev. D} \bibinfo{volume}{104} (\bibinfo{year}{2021}) \bibinfo{pages}{025012}. \URLprefix \url{https://link.aps.org/doi/10.1103/PhysRevD.104.025012}. \DOIprefix\doi{10.1103/PhysRevD.104.025012}.
\bibitem[{Valentini(1991)}]{Valentini:1991eah}
\bibinfo{author}{A.~Valentini},
\newblock \bibinfo{title}{{Non-local correlations in quantum electrodynamics}},
\newblock \bibinfo{journal}{Phys. Lett. A} \bibinfo{volume}{153} (\bibinfo{year}{1991}) \bibinfo{pages}{321--325}. \DOIprefix\doi{10.1016/0375-9601(91)90952-5}.
\bibitem[{Reznik et~al.(2005)Reznik, Retzker, and Silman}]{PhysRevA.71.042104}
\bibinfo{author}{B.~Reznik}, \bibinfo{author}{A.~Retzker}, \bibinfo{author}{J.~Silman},
\newblock \bibinfo{title}{Violating bell's inequalities in vacuum},
\newblock \bibinfo{journal}{Phys. Rev. A} \bibinfo{volume}{71} (\bibinfo{year}{2005}) \bibinfo{pages}{042104}. \URLprefix \url{https://link.aps.org/doi/10.1103/PhysRevA.71.042104}. \DOIprefix\doi{10.1103/PhysRevA.71.042104}.
\bibitem[{DeWitt(1979)}]{DeWitt1979}
\bibinfo{author}{B.~S. DeWitt},
\newblock \bibinfo{title}{Quantum gravity: the new synthesis},
\newblock in: \bibinfo{editor}{S.~Hawking}, \bibinfo{editor}{W.~Israel} (Eds.), \bibinfo{booktitle}{General Relativity: An Einstein Centenary Survey}, \bibinfo{publisher}{Cambridge University Press}, \bibinfo{year}{1979}, pp. \bibinfo{pages}{680--745}.
\bibitem[{Perche et~al.(2022)Perche, Lima, and Mart\'{\i}n-Mart\'{\i}nez}]{PhysRevD.105.065016}
\bibinfo{author}{T.~R. Perche}, \bibinfo{author}{C.~Lima}, \bibinfo{author}{E.~Mart\'{\i}n-Mart\'{\i}nez},
\newblock \bibinfo{title}{Harvesting entanglement from complex scalar and fermionic fields with linearly coupled particle detectors},
\newblock \bibinfo{journal}{Phys. Rev. D} \bibinfo{volume}{105} (\bibinfo{year}{2022}) \bibinfo{pages}{065016}. \URLprefix \url{https://link.aps.org/doi/10.1103/PhysRevD.105.065016}. \DOIprefix\doi{10.1103/PhysRevD.105.065016}.
\bibitem[{Perche et~al.(2023)Perche, Ragula, and Mart\'{\i}n-Mart\'{\i}nez}]{PhysRevD.108.085025}
\bibinfo{author}{T.~R. Perche}, \bibinfo{author}{B.~Ragula}, \bibinfo{author}{E.~Mart\'{\i}n-Mart\'{\i}nez},
\newblock \bibinfo{title}{Harvesting entanglement from the gravitational vacuum},
\newblock \bibinfo{journal}{Phys. Rev. D} \bibinfo{volume}{108} (\bibinfo{year}{2023}) \bibinfo{pages}{085025}. \URLprefix \url{https://link.aps.org/doi/10.1103/PhysRevD.108.085025}. \DOIprefix\doi{10.1103/PhysRevD.108.085025}.
\bibitem[{Vidal and Werner(2002)}]{PhysRevA.65.032314}
\bibinfo{author}{G.~Vidal}, \bibinfo{author}{R.~F. Werner},
\newblock \bibinfo{title}{Computable measure of entanglement},
\newblock \bibinfo{journal}{Phys. Rev. A} \bibinfo{volume}{65} (\bibinfo{year}{2002}) \bibinfo{pages}{032314}. \URLprefix \url{https://link.aps.org/doi/10.1103/PhysRevA.65.032314}. \DOIprefix\doi{10.1103/PhysRevA.65.032314}.
\bibitem[{Pozas-Kerstjens et~al.(2017)Pozas-Kerstjens, Louko, and Mart\'{\i}n-Mart\'{\i}nez}]{PhysRevD.95.105009}
\bibinfo{author}{A.~Pozas-Kerstjens}, \bibinfo{author}{J.~Louko}, \bibinfo{author}{E.~Mart\'{\i}n-Mart\'{\i}nez},
\newblock \bibinfo{title}{Degenerate detectors are unable to harvest spacelike entanglement},
\newblock \bibinfo{journal}{Phys. Rev. D} \bibinfo{volume}{95} (\bibinfo{year}{2017}) \bibinfo{pages}{105009}. \URLprefix \url{https://link.aps.org/doi/10.1103/PhysRevD.95.105009}. \DOIprefix\doi{10.1103/PhysRevD.95.105009}.
\bibitem[{Foo et~al.(2021)Foo, Mann, and Zych}]{PhysRevD.103.065013}
\bibinfo{author}{J.~Foo}, \bibinfo{author}{R.~B. Mann}, \bibinfo{author}{M.~Zych},
\newblock \bibinfo{title}{Entanglement amplification between superposed detectors in flat and curved spacetimes},
\newblock \bibinfo{journal}{Phys. Rev. D} \bibinfo{volume}{103} (\bibinfo{year}{2021}) \bibinfo{pages}{065013}. \URLprefix \url{https://link.aps.org/doi/10.1103/PhysRevD.103.065013}. \DOIprefix\doi{10.1103/PhysRevD.103.065013}.
\bibitem[{Mendez-Avalos et~al.(2022)Mendez-Avalos, Henderson, Gallock-Yoshimura, and Mann}]{Mendez-Avalos:2022obb}
\bibinfo{author}{D.~Mendez-Avalos}, \bibinfo{author}{L.~J. Henderson}, \bibinfo{author}{K.~Gallock-Yoshimura}, \bibinfo{author}{R.~B. Mann},
\newblock \bibinfo{title}{{Entanglement harvesting of three Unruh-DeWitt detectors}},
\newblock \bibinfo{journal}{Gen. Rel. Grav.} \bibinfo{volume}{54} (\bibinfo{year}{2022}) \bibinfo{pages}{87}. \DOIprefix\doi{10.1007/s10714-022-02956-x}. \href{http://arxiv.org/abs/2206.11902}{{\tt arXiv:2206.11902}}.
\bibitem[{Simidzija and Mart\'{\i}n-Mart\'{\i}nez(2017)}]{PhysRevD.96.065008}
\bibinfo{author}{P.~Simidzija}, \bibinfo{author}{E.~Mart\'{\i}n-Mart\'{\i}nez},
\newblock \bibinfo{title}{Nonperturbative analysis of entanglement harvesting from coherent field states},
\newblock \bibinfo{journal}{Phys. Rev. D} \bibinfo{volume}{96} (\bibinfo{year}{2017}) \bibinfo{pages}{065008}. \URLprefix \url{https://link.aps.org/doi/10.1103/PhysRevD.96.065008}. \DOIprefix\doi{10.1103/PhysRevD.96.065008}.
\bibitem[{Simidzija et~al.(2018)Simidzija, Jonsson, and Mart\'{\i}n-Mart\'{\i}nez}]{PhysRevD.97.125002}
\bibinfo{author}{P.~Simidzija}, \bibinfo{author}{R.~H. Jonsson}, \bibinfo{author}{E.~Mart\'{\i}n-Mart\'{\i}nez},
\newblock \bibinfo{title}{General no-go theorem for entanglement extraction},
\newblock \bibinfo{journal}{Phys. Rev. D} \bibinfo{volume}{97} (\bibinfo{year}{2018}) \bibinfo{pages}{125002}. \URLprefix \url{https://link.aps.org/doi/10.1103/PhysRevD.97.125002}. \DOIprefix\doi{10.1103/PhysRevD.97.125002}.
\bibitem[{Brown et~al.(2014)Brown, Donnelly, Kempf, Mann, Martin-Martinez, and Menicucci}]{Brown:2014pda}
\bibinfo{author}{E.~G. Brown}, \bibinfo{author}{W.~Donnelly}, \bibinfo{author}{A.~Kempf}, \bibinfo{author}{R.~B. Mann}, \bibinfo{author}{E.~Martin-Martinez}, \bibinfo{author}{N.~C. Menicucci},
\newblock \bibinfo{title}{{Quantum seismology}},
\newblock \bibinfo{journal}{New J. Phys.} \bibinfo{volume}{16} (\bibinfo{year}{2014}) \bibinfo{pages}{105020}. \DOIprefix\doi{10.1088/1367-2630/16/10/105020}. \href{http://arxiv.org/abs/1407.0071}{{\tt arXiv:1407.0071}}.
\bibitem[{Sachs et~al.(2017)Sachs, Mann, and Mart\'{\i}n-Mart\'{\i}nez}]{PhysRevD.96.085012}
\bibinfo{author}{A.~Sachs}, \bibinfo{author}{R.~B. Mann}, \bibinfo{author}{E.~Mart\'{\i}n-Mart\'{\i}nez},
\newblock \bibinfo{title}{Entanglement harvesting and divergences in quadratic unruh-dewitt detector pairs},
\newblock \bibinfo{journal}{Phys. Rev. D} \bibinfo{volume}{96} (\bibinfo{year}{2017}) \bibinfo{pages}{085012}. \URLprefix \url{https://link.aps.org/doi/10.1103/PhysRevD.96.085012}. \DOIprefix\doi{10.1103/PhysRevD.96.085012}.
\bibitem[{Steeg and Menicucci(2009)}]{PhysRevD.79.044027}
\bibinfo{author}{G.~V. Steeg}, \bibinfo{author}{N.~C. Menicucci},
\newblock \bibinfo{title}{Entangling power of an expanding universe},
\newblock \bibinfo{journal}{Phys. Rev. D} \bibinfo{volume}{79} (\bibinfo{year}{2009}) \bibinfo{pages}{044027}. \URLprefix \url{https://link.aps.org/doi/10.1103/PhysRevD.79.044027}. \DOIprefix\doi{10.1103/PhysRevD.79.044027}.
\bibitem[{Mart\'{\i}n-Mart\'{\i}nez et~al.(2016)Mart\'{\i}n-Mart\'{\i}nez, Smith, and Terno}]{PhysRevD.93.044001}
\bibinfo{author}{E.~Mart\'{\i}n-Mart\'{\i}nez}, \bibinfo{author}{A.~R.~H. Smith}, \bibinfo{author}{D.~R. Terno},
\newblock \bibinfo{title}{Spacetime structure and vacuum entanglement},
\newblock \bibinfo{journal}{Phys. Rev. D} \bibinfo{volume}{93} (\bibinfo{year}{2016}) \bibinfo{pages}{044001}. \URLprefix \url{https://link.aps.org/doi/10.1103/PhysRevD.93.044001}. \DOIprefix\doi{10.1103/PhysRevD.93.044001}.
\bibitem[{Mart\'{\i}n-Mart\'{\i}nez et~al.(2013)Mart\'{\i}n-Mart\'{\i}nez, Montero, and del Rey}]{PhysRevD.87.064038}
\bibinfo{author}{E.~Mart\'{\i}n-Mart\'{\i}nez}, \bibinfo{author}{M.~Montero}, \bibinfo{author}{M.~del Rey},
\newblock \bibinfo{title}{Wavepacket detection with the unruh-dewitt model},
\newblock \bibinfo{journal}{Phys. Rev. D} \bibinfo{volume}{87} (\bibinfo{year}{2013}) \bibinfo{pages}{064038}. \URLprefix \url{https://link.aps.org/doi/10.1103/PhysRevD.87.064038}. \DOIprefix\doi{10.1103/PhysRevD.87.064038}.
\bibitem[{Mart\'{\i}n-Mart\'{\i}nez et~al.(2021)Mart\'{\i}n-Mart\'{\i}nez, Perche, and Torres}]{PhysRevD.103.025007}
\bibinfo{author}{E.~Mart\'{\i}n-Mart\'{\i}nez}, \bibinfo{author}{T.~R. Perche}, \bibinfo{author}{B.~d. S.~L. Torres},
\newblock \bibinfo{title}{Broken covariance of particle detector models in relativistic quantum information},
\newblock \bibinfo{journal}{Phys. Rev. D} \bibinfo{volume}{103} (\bibinfo{year}{2021}) \bibinfo{pages}{025007}. \URLprefix \url{https://link.aps.org/doi/10.1103/PhysRevD.103.025007}. \DOIprefix\doi{10.1103/PhysRevD.103.025007}.
\bibitem[{de~Ram\'on et~al.(2021)de~Ram\'on, Papageorgiou, and Mart\'{\i}n-Mart\'{\i}nez}]{PhysRevD.103.085002}
\bibinfo{author}{J.~de~Ram\'on}, \bibinfo{author}{M.~Papageorgiou}, \bibinfo{author}{E.~Mart\'{\i}n-Mart\'{\i}nez},
\newblock \bibinfo{title}{Relativistic causality in particle detector models: Faster-than-light signaling and impossible measurements},
\newblock \bibinfo{journal}{Phys. Rev. D} \bibinfo{volume}{103} (\bibinfo{year}{2021}) \bibinfo{pages}{085002}. \URLprefix \url{https://link.aps.org/doi/10.1103/PhysRevD.103.085002}. \DOIprefix\doi{10.1103/PhysRevD.103.085002}.
\bibitem[{de~Ram\'on et~al.(2023)de~Ram\'on, Papageorgiou, and Mart\'{\i}n-Mart\'{\i}nez}]{PhysRevD.108.045015}
\bibinfo{author}{J.~de~Ram\'on}, \bibinfo{author}{M.~Papageorgiou}, \bibinfo{author}{E.~Mart\'{\i}n-Mart\'{\i}nez},
\newblock \bibinfo{title}{Causality and signalling in noncompact detector-field interactions},
\newblock \bibinfo{journal}{Phys. Rev. D} \bibinfo{volume}{108} (\bibinfo{year}{2023}) \bibinfo{pages}{045015}. \URLprefix \url{https://link.aps.org/doi/10.1103/PhysRevD.108.045015}. \DOIprefix\doi{10.1103/PhysRevD.108.045015}.
\bibitem[{Ruep(2021)}]{Ruep:2021fjh}
\bibinfo{author}{M.~H. Ruep},
\newblock \bibinfo{title}{{Weakly coupled local particle detectors cannot harvest entanglement}},
\newblock \bibinfo{journal}{Class. Quant. Grav.} \bibinfo{volume}{38} (\bibinfo{year}{2021}) \bibinfo{pages}{195029}. \DOIprefix\doi{10.1088/1361-6382/ac1b08}. \href{http://arxiv.org/abs/2103.13400}{{\tt arXiv:2103.13400}}.
\bibitem[{Perche et~al.(2024)Perche, Polo-G\'omez, Torres, and Mart\'{\i}n-Mart\'{\i}nez}]{PhysRevD.109.045013}
\bibinfo{author}{T.~R. Perche}, \bibinfo{author}{J.~Polo-G\'omez}, \bibinfo{author}{B.~d. S.~L. Torres}, \bibinfo{author}{E.~Mart\'{\i}n-Mart\'{\i}nez},
\newblock \bibinfo{title}{Particle detectors from localized quantum field theories},
\newblock \bibinfo{journal}{Phys. Rev. D} \bibinfo{volume}{109} (\bibinfo{year}{2024}) \bibinfo{pages}{045013}. \URLprefix \url{https://link.aps.org/doi/10.1103/PhysRevD.109.045013}. \DOIprefix\doi{10.1103/PhysRevD.109.045013}.
\bibitem[{Torres(2024)}]{PhysRevD.109.065004}
\bibinfo{author}{B.~d. S.~L. Torres},
\newblock \bibinfo{title}{Particle detector models from path integrals of localized quantum fields},
\newblock \bibinfo{journal}{Phys. Rev. D} \bibinfo{volume}{109} (\bibinfo{year}{2024}) \bibinfo{pages}{065004}. \URLprefix \url{https://link.aps.org/doi/10.1103/PhysRevD.109.065004}. \DOIprefix\doi{10.1103/PhysRevD.109.065004}.
\bibitem[{Fewster and Verch(2020)}]{Fewster:2018qbm}
\bibinfo{author}{C.~J. Fewster}, \bibinfo{author}{R.~Verch},
\newblock \bibinfo{title}{{Quantum fields and local measurements}},
\newblock \bibinfo{journal}{Commun. Math. Phys.} \bibinfo{volume}{378} (\bibinfo{year}{2020}) \bibinfo{pages}{851--889}. \DOIprefix\doi{10.1007/s00220-020-03800-6}. \href{http://arxiv.org/abs/1810.06512}{{\tt arXiv:1810.06512}}.
\bibitem[{Fewster et~al.(2025)Fewster, Janssen, Loveridge, Rejzner, and Waldron}]{Fewster:2024pur}
\bibinfo{author}{J.~C. Fewster}, \bibinfo{author}{D.~W. Janssen}, \bibinfo{author}{L.~D. Loveridge}, \bibinfo{author}{K.~Rejzner}, \bibinfo{author}{J.~Waldron},
\newblock \bibinfo{title}{{Quantum Reference Frames, Measurement Schemes and the Type of Local Algebras in Quantum Field Theory}},
\newblock \bibinfo{journal}{Commun. Math. Phys.} \bibinfo{volume}{406} (\bibinfo{year}{2025}) \bibinfo{pages}{19}. \DOIprefix\doi{10.1007/s00220-024-05180-7}. \href{http://arxiv.org/abs/2403.11973}{{\tt arXiv:2403.11973}}.
\bibitem[{Polo-G\'omez et~al.(2022)Polo-G\'omez, Garay, and Mart\'{\i}n-Mart\'{\i}nez}]{PhysRevD.105.065003}
\bibinfo{author}{J.~Polo-G\'omez}, \bibinfo{author}{L.~J. Garay}, \bibinfo{author}{E.~Mart\'{\i}n-Mart\'{\i}nez},
\newblock \bibinfo{title}{A detector-based measurement theory for quantum field theory},
\newblock \bibinfo{journal}{Phys. Rev. D} \bibinfo{volume}{105} (\bibinfo{year}{2022}) \bibinfo{pages}{065003}. \URLprefix \url{https://link.aps.org/doi/10.1103/PhysRevD.105.065003}. \DOIprefix\doi{10.1103/PhysRevD.105.065003}.
\bibitem[{Papageorgiou and Fraser(2024)}]{Papageorgiou:2023nvf}
\bibinfo{author}{M.~Papageorgiou}, \bibinfo{author}{D.~Fraser},
\newblock \bibinfo{title}{{Eliminating the \textquoteleft{}Impossible\textquoteright{}: Recent Progress on Local Measurement Theory for Quantum Field Theory}},
\newblock \bibinfo{journal}{Found. Phys.} \bibinfo{volume}{54} (\bibinfo{year}{2024}) \bibinfo{pages}{26}. \DOIprefix\doi{10.1007/s10701-024-00756-8}. \href{http://arxiv.org/abs/2307.08524}{{\tt arXiv:2307.08524}}.
\bibitem[{Pranzini and Keski-Vakkuri(2025)}]{PhysRevD.111.045016}
\bibinfo{author}{N.~Pranzini}, \bibinfo{author}{E.~Keski-Vakkuri},
\newblock \bibinfo{title}{Detector-based measurements for qft: Two issues and an algebraic qft proposal},
\newblock \bibinfo{journal}{Phys. Rev. D} \bibinfo{volume}{111} (\bibinfo{year}{2025}) \bibinfo{pages}{045016}. \URLprefix \url{https://link.aps.org/doi/10.1103/PhysRevD.111.045016}. \DOIprefix\doi{10.1103/PhysRevD.111.045016}.
\bibitem[{Bombelli et~al.(1986)Bombelli, Koul, Lee, and Sorkin}]{PhysRevD.34.373}
\bibinfo{author}{L.~Bombelli}, \bibinfo{author}{R.~K. Koul}, \bibinfo{author}{J.~Lee}, \bibinfo{author}{R.~D. Sorkin},
\newblock \bibinfo{title}{Quantum source of entropy for black holes},
\newblock \bibinfo{journal}{Phys. Rev. D} \bibinfo{volume}{34} (\bibinfo{year}{1986}) \bibinfo{pages}{373--383}. \URLprefix \url{https://link.aps.org/doi/10.1103/PhysRevD.34.373}. \DOIprefix\doi{10.1103/PhysRevD.34.373}.
\bibitem[{Nishioka(2018)}]{RevModPhys.90.035007}
\bibinfo{author}{T.~Nishioka},
\newblock \bibinfo{title}{Entanglement entropy: Holography and renormalization group},
\newblock \bibinfo{journal}{Rev. Mod. Phys.} \bibinfo{volume}{90} (\bibinfo{year}{2018}) \bibinfo{pages}{035007}. \URLprefix \url{https://link.aps.org/doi/10.1103/RevModPhys.90.035007}. \DOIprefix\doi{10.1103/RevModPhys.90.035007}.
\bibitem[{Witten(2018)}]{RevModPhys.90.045003}
\bibinfo{author}{E.~Witten},
\newblock \bibinfo{title}{Aps medal for exceptional achievement in research: Invited article on entanglement properties of quantum field theory},
\newblock \bibinfo{journal}{Rev. Mod. Phys.} \bibinfo{volume}{90} (\bibinfo{year}{2018}) \bibinfo{pages}{045003}. \URLprefix \url{https://link.aps.org/doi/10.1103/RevModPhys.90.045003}. \DOIprefix\doi{10.1103/RevModPhys.90.045003}.
\bibitem[{Nishioka et~al.(2009)Nishioka, Ryu, and Takayanagi}]{Nishioka:2009un}
\bibinfo{author}{T.~Nishioka}, \bibinfo{author}{S.~Ryu}, \bibinfo{author}{T.~Takayanagi},
\newblock \bibinfo{title}{{Holographic Entanglement Entropy: An Overview}},
\newblock \bibinfo{journal}{J. Phys. A} \bibinfo{volume}{42} (\bibinfo{year}{2009}) \bibinfo{pages}{504008}. \DOIprefix\doi{10.1088/1751-8113/42/50/504008}. \href{http://arxiv.org/abs/0905.0932}{{\tt arXiv:0905.0932}}.
\bibitem[{Casini and Huerta(2009)}]{Casini:2009sr}
\bibinfo{author}{H.~Casini}, \bibinfo{author}{M.~Huerta},
\newblock \bibinfo{title}{{Entanglement entropy in free quantum field theory}},
\newblock \bibinfo{journal}{J. Phys. A} \bibinfo{volume}{42} (\bibinfo{year}{2009}) \bibinfo{pages}{504007}. \DOIprefix\doi{10.1088/1751-8113/42/50/504007}. \href{http://arxiv.org/abs/0905.2562}{{\tt arXiv:0905.2562}}.
\bibitem[{Riera and Latorre(2006)}]{Riera:2006vj}
\bibinfo{author}{A.~Riera}, \bibinfo{author}{J.~I. Latorre},
\newblock \bibinfo{title}{{Area law and vacuum reordering in harmonic networks}},
\newblock \bibinfo{journal}{Phys. Rev. A} \bibinfo{volume}{74} (\bibinfo{year}{2006}) \bibinfo{pages}{052326}. \DOIprefix\doi{10.1103/PhysRevA.74.052326}. \href{http://arxiv.org/abs/quant-ph/0605112}{{\tt arXiv:quant-ph/0605112}}.
\bibitem[{Das et~al.(2008)Das, Shankaranarayanan, and Sur}]{Das:2007mj}
\bibinfo{author}{S.~Das}, \bibinfo{author}{S.~Shankaranarayanan}, \bibinfo{author}{S.~Sur},
\newblock \bibinfo{title}{{Power-law corrections to entanglement entropy of black holes}},
\newblock \bibinfo{journal}{Phys. Rev. D} \bibinfo{volume}{77} (\bibinfo{year}{2008}) \bibinfo{pages}{064013}. \DOIprefix\doi{10.1103/PhysRevD.77.064013}. \href{http://arxiv.org/abs/0705.2070}{{\tt arXiv:0705.2070}}.
\bibitem[{Srednicki(1993)}]{Srednicki:1993im}
\bibinfo{author}{M.~Srednicki},
\newblock \bibinfo{title}{{Entropy and area}},
\newblock \bibinfo{journal}{Phys. Rev. Lett.} \bibinfo{volume}{71} (\bibinfo{year}{1993}) \bibinfo{pages}{666--669}. \DOIprefix\doi{10.1103/PhysRevLett.71.666}. \href{http://arxiv.org/abs/hep-th/9303048}{{\tt arXiv:hep-th/9303048}}.
\bibitem[{Katsinis and Pastras(2018)}]{Katsinis:2017qzh}
\bibinfo{author}{D.~Katsinis}, \bibinfo{author}{G.~Pastras},
\newblock \bibinfo{title}{{An Inverse Mass Expansion for Entanglement Entropy in Free Massive Scalar Field Theory}},
\newblock \bibinfo{journal}{Eur. Phys. J. C} \bibinfo{volume}{78} (\bibinfo{year}{2018}) \bibinfo{pages}{282}. \DOIprefix\doi{10.1140/epjc/s10052-018-5596-4}. \href{http://arxiv.org/abs/1711.02618}{{\tt arXiv:1711.02618}}.
\bibitem[{Belfiglio et~al.(2025{\natexlab{a}})Belfiglio, Luongo, Mancini, and Tomasi}]{Belfiglio:2024qsa}
\bibinfo{author}{A.~Belfiglio}, \bibinfo{author}{O.~Luongo}, \bibinfo{author}{S.~Mancini}, \bibinfo{author}{S.~Tomasi},
\newblock \bibinfo{title}{{Entanglement entropy in quantum black holes}},
\newblock \bibinfo{journal}{Class. Quant. Grav.} \bibinfo{volume}{42} (\bibinfo{year}{2025}{\natexlab{a}}) \bibinfo{pages}{035006}. \DOIprefix\doi{10.1088/1361-6382/ad9e66}. \href{http://arxiv.org/abs/2404.00715}{{\tt arXiv:2404.00715}}.
\bibitem[{Belfiglio et~al.(2025{\natexlab{b}})Belfiglio, Luongo, Mancini, and Tomasi}]{Belfiglio:2025hzo}
\bibinfo{author}{A.~Belfiglio}, \bibinfo{author}{O.~Luongo}, \bibinfo{author}{S.~Mancini}, \bibinfo{author}{S.~Tomasi},
\newblock \bibinfo{title}{{Entanglement entropy evolution during gravitational collapse}}  (\bibinfo{year}{2025}{\natexlab{b}}). \href{http://arxiv.org/abs/2502.14797}{{\tt arXiv:2502.14797}}.
\bibitem[{Bardeen et~al.(1973)Bardeen, Carter, and Hawking}]{Bardeen:1973gs}
\bibinfo{author}{J.~M. Bardeen}, \bibinfo{author}{B.~Carter}, \bibinfo{author}{S.~W. Hawking},
\newblock \bibinfo{title}{{The Four laws of black hole mechanics}},
\newblock \bibinfo{journal}{Commun. Math. Phys.} \bibinfo{volume}{31} (\bibinfo{year}{1973}) \bibinfo{pages}{161--170}. \DOIprefix\doi{10.1007/BF01645742}.
\bibitem[{Mukohyama et~al.(1997)Mukohyama, Seriu, and Kodama}]{PhysRevD.55.7666}
\bibinfo{author}{S.~Mukohyama}, \bibinfo{author}{M.~Seriu}, \bibinfo{author}{H.~Kodama},
\newblock \bibinfo{title}{Can the entanglement entropy be the origin of black-hole entropy?},
\newblock \bibinfo{journal}{Phys. Rev. D} \bibinfo{volume}{55} (\bibinfo{year}{1997}) \bibinfo{pages}{7666--7679}. \URLprefix \url{https://link.aps.org/doi/10.1103/PhysRevD.55.7666}. \DOIprefix\doi{10.1103/PhysRevD.55.7666}.
\bibitem[{Mukohyama et~al.(1998)Mukohyama, Seriu, and Kodama}]{PhysRevD.58.064001}
\bibinfo{author}{S.~Mukohyama}, \bibinfo{author}{M.~Seriu}, \bibinfo{author}{H.~Kodama},
\newblock \bibinfo{title}{Thermodynamics of entanglement in schwarzschild spacetime},
\newblock \bibinfo{journal}{Phys. Rev. D} \bibinfo{volume}{58} (\bibinfo{year}{1998}) \bibinfo{pages}{064001}. \URLprefix \url{https://link.aps.org/doi/10.1103/PhysRevD.58.064001}. \DOIprefix\doi{10.1103/PhysRevD.58.064001}.
\bibitem[{Wong et~al.(2013)Wong, Klich, Pando~Zayas, and Vaman}]{Wong:2013gua}
\bibinfo{author}{G.~Wong}, \bibinfo{author}{I.~Klich}, \bibinfo{author}{L.~A. Pando~Zayas}, \bibinfo{author}{D.~Vaman},
\newblock \bibinfo{title}{{Entanglement Temperature and Entanglement Entropy of Excited States}},
\newblock \bibinfo{journal}{JHEP} \bibinfo{volume}{12} (\bibinfo{year}{2013}) \bibinfo{pages}{020}. \DOIprefix\doi{10.1007/JHEP12(2013)020}. \href{http://arxiv.org/abs/1305.3291}{{\tt arXiv:1305.3291}}.
\bibitem[{Park(2016)}]{Park:2015hcz}
\bibinfo{author}{C.~Park},
\newblock \bibinfo{title}{{Thermodynamic law from the entanglement entropy bound}},
\newblock \bibinfo{journal}{Phys. Rev. D} \bibinfo{volume}{93} (\bibinfo{year}{2016}) \bibinfo{pages}{086003}. \DOIprefix\doi{10.1103/PhysRevD.93.086003}. \href{http://arxiv.org/abs/1511.02288}{{\tt arXiv:1511.02288}}.
\bibitem[{Belfiglio et~al.(2025)Belfiglio, Chandran, Luongo, and Mancini}]{PhysRevD.111.024013}
\bibinfo{author}{A.~Belfiglio}, \bibinfo{author}{S.~M. Chandran}, \bibinfo{author}{O.~Luongo}, \bibinfo{author}{S.~Mancini},
\newblock \bibinfo{title}{Horizon entanglement area law from regular black hole thermodynamics},
\newblock \bibinfo{journal}{Phys. Rev. D} \bibinfo{volume}{111} (\bibinfo{year}{2025}) \bibinfo{pages}{024013}. \URLprefix \url{https://link.aps.org/doi/10.1103/PhysRevD.111.024013}. \DOIprefix\doi{10.1103/PhysRevD.111.024013}.
\bibitem[{Komar(1959)}]{PhysRev.113.934}
\bibinfo{author}{A.~Komar},
\newblock \bibinfo{title}{Covariant conservation laws in general relativity},
\newblock \bibinfo{journal}{Phys. Rev.} \bibinfo{volume}{113} (\bibinfo{year}{1959}) \bibinfo{pages}{934--936}. \URLprefix \url{https://link.aps.org/doi/10.1103/PhysRev.113.934}. \DOIprefix\doi{10.1103/PhysRev.113.934}.
\bibitem[{Banerjee and Majhi(2010)}]{PhysRevD.81.124006}
\bibinfo{author}{R.~Banerjee}, \bibinfo{author}{B.~R. Majhi},
\newblock \bibinfo{title}{Statistical origin of gravity},
\newblock \bibinfo{journal}{Phys. Rev. D} \bibinfo{volume}{81} (\bibinfo{year}{2010}) \bibinfo{pages}{124006}. \URLprefix \url{https://link.aps.org/doi/10.1103/PhysRevD.81.124006}. \DOIprefix\doi{10.1103/PhysRevD.81.124006}.
\bibitem[{Raju(2022)}]{Raju:2020smc}
\bibinfo{author}{S.~Raju},
\newblock \bibinfo{title}{{Lessons from the information paradox}},
\newblock \bibinfo{journal}{Phys. Rept.} \bibinfo{volume}{943} (\bibinfo{year}{2022}) \bibinfo{pages}{1--80}. \DOIprefix\doi{10.1016/j.physrep.2021.10.001}. \href{http://arxiv.org/abs/2012.05770}{{\tt arXiv:2012.05770}}.
\bibitem[{Wall(2012)}]{PhysRevD.85.104049}
\bibinfo{author}{A.~C. Wall},
\newblock \bibinfo{title}{Proof of the generalized second law for rapidly changing fields and arbitrary horizon slices},
\newblock \bibinfo{journal}{Phys. Rev. D} \bibinfo{volume}{85} (\bibinfo{year}{2012}) \bibinfo{pages}{104049}. \URLprefix \url{https://link.aps.org/doi/10.1103/PhysRevD.85.104049}. \DOIprefix\doi{10.1103/PhysRevD.85.104049}.
\bibitem[{Bisognano and Wichmann(1976)}]{Bisognano:1976za}
\bibinfo{author}{J.~J. Bisognano}, \bibinfo{author}{E.~H. Wichmann},
\newblock \bibinfo{title}{{On the Duality Condition for Quantum Fields}},
\newblock \bibinfo{journal}{J. Math. Phys.} \bibinfo{volume}{17} (\bibinfo{year}{1976}) \bibinfo{pages}{303--321}. \DOIprefix\doi{10.1063/1.522898}.
\bibitem[{Gibbons and Hawking(1977)}]{PhysRevD.15.2752}
\bibinfo{author}{G.~W. Gibbons}, \bibinfo{author}{S.~W. Hawking},
\newblock \bibinfo{title}{Action integrals and partition functions in quantum gravity},
\newblock \bibinfo{journal}{Phys. Rev. D} \bibinfo{volume}{15} (\bibinfo{year}{1977}) \bibinfo{pages}{2752--2756}. \URLprefix \url{https://link.aps.org/doi/10.1103/PhysRevD.15.2752}. \DOIprefix\doi{10.1103/PhysRevD.15.2752}.
\bibitem[{Dowker et~al.(1992)Dowker, Gregory, and Traschen}]{PhysRevD.45.2762}
\bibinfo{author}{F.~Dowker}, \bibinfo{author}{R.~Gregory}, \bibinfo{author}{J.~Traschen},
\newblock \bibinfo{title}{Euclidean black-hole vortices},
\newblock \bibinfo{journal}{Phys. Rev. D} \bibinfo{volume}{45} (\bibinfo{year}{1992}) \bibinfo{pages}{2762--2771}. \URLprefix \url{https://link.aps.org/doi/10.1103/PhysRevD.45.2762}. \DOIprefix\doi{10.1103/PhysRevD.45.2762}.
\bibitem[{Christodoulou(2008)}]{Christodoulou:2008nj}
\bibinfo{author}{D.~Christodoulou},
\newblock \bibinfo{title}{{The Formation of Black Holes in General Relativity}},
\newblock in: \bibinfo{booktitle}{{12th Marcel Grossmann Meeting on General Relativity}}, \bibinfo{year}{2008}, pp. \bibinfo{pages}{24--34}. \href{http://arxiv.org/abs/0805.3880}{{\tt arXiv:0805.3880}}.
\bibitem[{Mathur(2009)}]{Mathur:2009hf}
\bibinfo{author}{S.~D. Mathur},
\newblock \bibinfo{title}{{The Information paradox: A Pedagogical introduction}},
\newblock \bibinfo{journal}{Class. Quant. Grav.} \bibinfo{volume}{26} (\bibinfo{year}{2009}) \bibinfo{pages}{224001}. \DOIprefix\doi{10.1088/0264-9381/26/22/224001}. \href{http://arxiv.org/abs/0909.1038}{{\tt arXiv:0909.1038}}.
\bibitem[{Almheiri et~al.(2013{\natexlab{a}})Almheiri, Marolf, Polchinski, and Sully}]{Almheiri:2012rt}
\bibinfo{author}{A.~Almheiri}, \bibinfo{author}{D.~Marolf}, \bibinfo{author}{J.~Polchinski}, \bibinfo{author}{J.~Sully},
\newblock \bibinfo{title}{{Black Holes: Complementarity or Firewalls?}},
\newblock \bibinfo{journal}{JHEP} \bibinfo{volume}{02} (\bibinfo{year}{2013}{\natexlab{a}}) \bibinfo{pages}{062}. \DOIprefix\doi{10.1007/JHEP02(2013)062}. \href{http://arxiv.org/abs/1207.3123}{{\tt arXiv:1207.3123}}.
\bibitem[{Almheiri et~al.(2013{\natexlab{b}})Almheiri, Marolf, Polchinski, Stanford, and Sully}]{Almheiri:2013hfa}
\bibinfo{author}{A.~Almheiri}, \bibinfo{author}{D.~Marolf}, \bibinfo{author}{J.~Polchinski}, \bibinfo{author}{D.~Stanford}, \bibinfo{author}{J.~Sully},
\newblock \bibinfo{title}{{An Apologia for Firewalls}},
\newblock \bibinfo{journal}{JHEP} \bibinfo{volume}{09} (\bibinfo{year}{2013}{\natexlab{b}}) \bibinfo{pages}{018}. \DOIprefix\doi{10.1007/JHEP09(2013)018}. \href{http://arxiv.org/abs/1304.6483}{{\tt arXiv:1304.6483}}.
\bibitem[{Hubeny et~al.(2007)Hubeny, Rangamani, and Takayanagi}]{Hubeny:2007xt}
\bibinfo{author}{V.~E. Hubeny}, \bibinfo{author}{M.~Rangamani}, \bibinfo{author}{T.~Takayanagi},
\newblock \bibinfo{title}{{A Covariant holographic entanglement entropy proposal}},
\newblock \bibinfo{journal}{JHEP} \bibinfo{volume}{07} (\bibinfo{year}{2007}) \bibinfo{pages}{062}. \DOIprefix\doi{10.1088/1126-6708/2007/07/062}. \href{http://arxiv.org/abs/0705.0016}{{\tt arXiv:0705.0016}}.
\bibitem[{Faulkner et~al.(2013)Faulkner, Lewkowycz, and Maldacena}]{Faulkner:2013ana}
\bibinfo{author}{T.~Faulkner}, \bibinfo{author}{A.~Lewkowycz}, \bibinfo{author}{J.~Maldacena},
\newblock \bibinfo{title}{{Quantum corrections to holographic entanglement entropy}},
\newblock \bibinfo{journal}{JHEP} \bibinfo{volume}{11} (\bibinfo{year}{2013}) \bibinfo{pages}{074}. \DOIprefix\doi{10.1007/JHEP11(2013)074}. \href{http://arxiv.org/abs/1307.2892}{{\tt arXiv:1307.2892}}.
\bibitem[{Engelhardt and Wall(2015)}]{Engelhardt:2014gca}
\bibinfo{author}{N.~Engelhardt}, \bibinfo{author}{A.~C. Wall},
\newblock \bibinfo{title}{{Quantum Extremal Surfaces: Holographic Entanglement Entropy beyond the Classical Regime}},
\newblock \bibinfo{journal}{JHEP} \bibinfo{volume}{01} (\bibinfo{year}{2015}) \bibinfo{pages}{073}. \DOIprefix\doi{10.1007/JHEP01(2015)073}. \href{http://arxiv.org/abs/1408.3203}{{\tt arXiv:1408.3203}}.
\bibitem[{Wall(2014)}]{Wall:2012uf}
\bibinfo{author}{A.~C. Wall},
\newblock \bibinfo{title}{{Maximin Surfaces, and the Strong Subadditivity of the Covariant Holographic Entanglement Entropy}},
\newblock \bibinfo{journal}{Class. Quant. Grav.} \bibinfo{volume}{31} (\bibinfo{year}{2014}) \bibinfo{pages}{225007}. \DOIprefix\doi{10.1088/0264-9381/31/22/225007}. \href{http://arxiv.org/abs/1211.3494}{{\tt arXiv:1211.3494}}.
\bibitem[{Akers et~al.(2020)Akers, Engelhardt, Penington, and Usatyuk}]{Akers:2019lzs}
\bibinfo{author}{C.~Akers}, \bibinfo{author}{N.~Engelhardt}, \bibinfo{author}{G.~Penington}, \bibinfo{author}{M.~Usatyuk},
\newblock \bibinfo{title}{{Quantum Maximin Surfaces}},
\newblock \bibinfo{journal}{JHEP} \bibinfo{volume}{08} (\bibinfo{year}{2020}) \bibinfo{pages}{140}. \DOIprefix\doi{10.1007/JHEP08(2020)140}. \href{http://arxiv.org/abs/1912.02799}{{\tt arXiv:1912.02799}}.
\bibitem[{Hubeny(2015)}]{Hubeny:2014bla}
\bibinfo{author}{V.~E. Hubeny},
\newblock \bibinfo{title}{{The AdS/CFT Correspondence}},
\newblock \bibinfo{journal}{Class. Quant. Grav.} \bibinfo{volume}{32} (\bibinfo{year}{2015}) \bibinfo{pages}{124010}. \DOIprefix\doi{10.1088/0264-9381/32/12/124010}. \href{http://arxiv.org/abs/1501.00007}{{\tt arXiv:1501.00007}}.
\bibitem[{Bousso(2002)}]{RevModPhys.74.825}
\bibinfo{author}{R.~Bousso},
\newblock \bibinfo{title}{The holographic principle},
\newblock \bibinfo{journal}{Rev. Mod. Phys.} \bibinfo{volume}{74} (\bibinfo{year}{2002}) \bibinfo{pages}{825--874}. \URLprefix \url{https://link.aps.org/doi/10.1103/RevModPhys.74.825}. \DOIprefix\doi{10.1103/RevModPhys.74.825}.
\bibitem[{Hayden and Preskill(2007)}]{Hayden:2007cs}
\bibinfo{author}{P.~Hayden}, \bibinfo{author}{J.~Preskill},
\newblock \bibinfo{title}{{Black holes as mirrors: Quantum information in random subsystems}},
\newblock \bibinfo{journal}{JHEP} \bibinfo{volume}{09} (\bibinfo{year}{2007}) \bibinfo{pages}{120}. \DOIprefix\doi{10.1088/1126-6708/2007/09/120}. \href{http://arxiv.org/abs/0708.4025}{{\tt arXiv:0708.4025}}.
\bibitem[{Sekino and Susskind(2008)}]{Sekino:2008he}
\bibinfo{author}{Y.~Sekino}, \bibinfo{author}{L.~Susskind},
\newblock \bibinfo{title}{{Fast Scramblers}},
\newblock \bibinfo{journal}{JHEP} \bibinfo{volume}{10} (\bibinfo{year}{2008}) \bibinfo{pages}{065}. \DOIprefix\doi{10.1088/1126-6708/2008/10/065}. \href{http://arxiv.org/abs/0808.2096}{{\tt arXiv:0808.2096}}.
\bibitem[{Dvali et~al.(2013)Dvali, Flassig, Gomez, Pritzel, and Wintergerst}]{PhysRevD.88.124041}
\bibinfo{author}{G.~Dvali}, \bibinfo{author}{D.~Flassig}, \bibinfo{author}{C.~Gomez}, \bibinfo{author}{A.~Pritzel}, \bibinfo{author}{N.~Wintergerst},
\newblock \bibinfo{title}{Scrambling in the black hole portrait},
\newblock \bibinfo{journal}{Phys. Rev. D} \bibinfo{volume}{88} (\bibinfo{year}{2013}) \bibinfo{pages}{124041}. \URLprefix \url{https://link.aps.org/doi/10.1103/PhysRevD.88.124041}. \DOIprefix\doi{10.1103/PhysRevD.88.124041}.
\bibitem[{Penington et~al.(2022)Penington, Shenker, Stanford, and Yang}]{Penington:2019kki}
\bibinfo{author}{G.~Penington}, \bibinfo{author}{S.~H. Shenker}, \bibinfo{author}{D.~Stanford}, \bibinfo{author}{Z.~Yang},
\newblock \bibinfo{title}{{Replica wormholes and the black hole interior}},
\newblock \bibinfo{journal}{JHEP} \bibinfo{volume}{03} (\bibinfo{year}{2022}) \bibinfo{pages}{205}. \DOIprefix\doi{10.1007/JHEP03(2022)205}. \href{http://arxiv.org/abs/1911.11977}{{\tt arXiv:1911.11977}}.
\bibitem[{Almheiri et~al.(2020)Almheiri, Hartman, Maldacena, Shaghoulian, and Tajdini}]{Almheiri:2019qdq}
\bibinfo{author}{A.~Almheiri}, \bibinfo{author}{T.~Hartman}, \bibinfo{author}{J.~Maldacena}, \bibinfo{author}{E.~Shaghoulian}, \bibinfo{author}{A.~Tajdini},
\newblock \bibinfo{title}{{Replica Wormholes and the Entropy of Hawking Radiation}},
\newblock \bibinfo{journal}{JHEP} \bibinfo{volume}{05} (\bibinfo{year}{2020}) \bibinfo{pages}{013}. \DOIprefix\doi{10.1007/JHEP05(2020)013}. \href{http://arxiv.org/abs/1911.12333}{{\tt arXiv:1911.12333}}.
\bibitem[{Goswami and Narayan(2022)}]{Goswami:2022ylc}
\bibinfo{author}{K.~Goswami}, \bibinfo{author}{K.~Narayan},
\newblock \bibinfo{title}{{Small Schwarzschild de Sitter black holes, quantum extremal surfaces and islands}},
\newblock \bibinfo{journal}{JHEP} \bibinfo{volume}{10} (\bibinfo{year}{2022}) \bibinfo{pages}{031}. \DOIprefix\doi{10.1007/JHEP10(2022)031}. \href{http://arxiv.org/abs/2207.10724}{{\tt arXiv:2207.10724}}.
\bibitem[{Wang et~al.(2021)Wang, Li, and Wang}]{Wang:2021woy}
\bibinfo{author}{X.~Wang}, \bibinfo{author}{R.~Li}, \bibinfo{author}{J.~Wang},
\newblock \bibinfo{title}{{Islands and Page curves of Reissner-Nordstr\"om black holes}},
\newblock \bibinfo{journal}{JHEP} \bibinfo{volume}{04} (\bibinfo{year}{2021}) \bibinfo{pages}{103}. \DOIprefix\doi{10.1007/JHEP04(2021)103}. \href{http://arxiv.org/abs/2101.06867}{{\tt arXiv:2101.06867}}.
\bibitem[{Lin et~al.(2024)Lin, Yu, Ge, and Tian}]{PhysRevD.110.046008}
\bibinfo{author}{S.-Y. Lin}, \bibinfo{author}{M.-H. Yu}, \bibinfo{author}{X.-H. Ge}, \bibinfo{author}{L.-J. Tian},
\newblock \bibinfo{title}{Entanglement entropy, phase transition, and island rule for reissner-nordstr\"om-ads black holes},
\newblock \bibinfo{journal}{Phys. Rev. D} \bibinfo{volume}{110} (\bibinfo{year}{2024}) \bibinfo{pages}{046008}. \URLprefix \url{https://link.aps.org/doi/10.1103/PhysRevD.110.046008}. \DOIprefix\doi{10.1103/PhysRevD.110.046008}.
\bibitem[{Wang and Li(2024)}]{PhysRevD.110.066012}
\bibinfo{author}{L.~Wang}, \bibinfo{author}{R.~Li},
\newblock \bibinfo{title}{Entanglement islands and the page curve of hawking radiation for rotating kerr black holes},
\newblock \bibinfo{journal}{Phys. Rev. D} \bibinfo{volume}{110} (\bibinfo{year}{2024}) \bibinfo{pages}{066012}. \URLprefix \url{https://link.aps.org/doi/10.1103/PhysRevD.110.066012}. \DOIprefix\doi{10.1103/PhysRevD.110.066012}.
\bibitem[{Luongo et~al.(2023)Luongo, Mancini, and Pierosara}]{PhysRevD.108.104059}
\bibinfo{author}{O.~Luongo}, \bibinfo{author}{S.~Mancini}, \bibinfo{author}{P.~Pierosara},
\newblock \bibinfo{title}{Entanglement entropy for spherically symmetric regular black holes},
\newblock \bibinfo{journal}{Phys. Rev. D} \bibinfo{volume}{108} (\bibinfo{year}{2023}) \bibinfo{pages}{104059}. \URLprefix \url{https://link.aps.org/doi/10.1103/PhysRevD.108.104059}. \DOIprefix\doi{10.1103/PhysRevD.108.104059}.
\bibitem[{Perez-Gonzalez(2025)}]{PhysRevD.111.083015}
\bibinfo{author}{Y.~F. Perez-Gonzalez},
\newblock \bibinfo{title}{Page time of primordial black holes in the standard model and beyond},
\newblock \bibinfo{journal}{Phys. Rev. D} \bibinfo{volume}{111} (\bibinfo{year}{2025}) \bibinfo{pages}{083015}. \URLprefix \url{https://link.aps.org/doi/10.1103/PhysRevD.111.083015}. \DOIprefix\doi{10.1103/PhysRevD.111.083015}.
\bibitem[{Balasubramanian et~al.(2021)Balasubramanian, Kar, and Ugajin}]{Balasubramanian:2020xqf}
\bibinfo{author}{V.~Balasubramanian}, \bibinfo{author}{A.~Kar}, \bibinfo{author}{T.~Ugajin},
\newblock \bibinfo{title}{{Islands in de Sitter space}},
\newblock \bibinfo{journal}{JHEP} \bibinfo{volume}{02} (\bibinfo{year}{2021}) \bibinfo{pages}{072}. \DOIprefix\doi{10.1007/JHEP02(2021)072}. \href{http://arxiv.org/abs/2008.05275}{{\tt arXiv:2008.05275}}.
\bibitem[{Seo(2022)}]{Seo:2022ezk}
\bibinfo{author}{M.-S. Seo},
\newblock \bibinfo{title}{{Information paradox and island in quasi-de Sitter space}},
\newblock \bibinfo{journal}{Eur. Phys. J. C} \bibinfo{volume}{82} (\bibinfo{year}{2022}) \bibinfo{pages}{1082}. \DOIprefix\doi{10.1140/epjc/s10052-022-11068-4}. \href{http://arxiv.org/abs/2204.04585}{{\tt arXiv:2204.04585}}.
\bibitem[{Piao(2023)}]{PhysRevD.107.123509}
\bibinfo{author}{Y.-S. Piao},
\newblock \bibinfo{title}{Implication of the island rule for inflation and primordial perturbations},
\newblock \bibinfo{journal}{Phys. Rev. D} \bibinfo{volume}{107} (\bibinfo{year}{2023}) \bibinfo{pages}{123509}. \URLprefix \url{https://link.aps.org/doi/10.1103/PhysRevD.107.123509}. \DOIprefix\doi{10.1103/PhysRevD.107.123509}.
\bibitem[{Faulkner et~al.(2014)Faulkner, Guica, Hartman, Myers, and Van~Raamsdonk}]{Faulkner:2013ica}
\bibinfo{author}{T.~Faulkner}, \bibinfo{author}{M.~Guica}, \bibinfo{author}{T.~Hartman}, \bibinfo{author}{R.~C. Myers}, \bibinfo{author}{M.~Van~Raamsdonk},
\newblock \bibinfo{title}{{Gravitation from Entanglement in Holographic CFTs}},
\newblock \bibinfo{journal}{JHEP} \bibinfo{volume}{03} (\bibinfo{year}{2014}) \bibinfo{pages}{051}. \DOIprefix\doi{10.1007/JHEP03(2014)051}. \href{http://arxiv.org/abs/1312.7856}{{\tt arXiv:1312.7856}}.
\bibitem[{Lin et~al.(2015)Lin, Marcolli, Ooguri, and Stoica}]{PhysRevLett.114.221601}
\bibinfo{author}{J.~Lin}, \bibinfo{author}{M.~Marcolli}, \bibinfo{author}{H.~Ooguri}, \bibinfo{author}{B.~Stoica},
\newblock \bibinfo{title}{Locality of gravitational systems from entanglement of conformal field theories},
\newblock \bibinfo{journal}{Phys. Rev. Lett.} \bibinfo{volume}{114} (\bibinfo{year}{2015}) \bibinfo{pages}{221601}. \URLprefix \url{https://link.aps.org/doi/10.1103/PhysRevLett.114.221601}. \DOIprefix\doi{10.1103/PhysRevLett.114.221601}.
\bibitem[{Czech et~al.(2016)Czech, Lamprou, McCandlish, and Sully}]{Czech:2015kbp}
\bibinfo{author}{B.~Czech}, \bibinfo{author}{L.~Lamprou}, \bibinfo{author}{S.~McCandlish}, \bibinfo{author}{J.~Sully},
\newblock \bibinfo{title}{{Tensor Networks from Kinematic Space}},
\newblock \bibinfo{journal}{JHEP} \bibinfo{volume}{07} (\bibinfo{year}{2016}) \bibinfo{pages}{100}. \DOIprefix\doi{10.1007/JHEP07(2016)100}. \href{http://arxiv.org/abs/1512.01548}{{\tt arXiv:1512.01548}}.
\bibitem[{Faulkner et~al.(2017)Faulkner, Haehl, Hijano, Parrikar, Rabideau, and Van~Raamsdonk}]{Faulkner:2017tkh}
\bibinfo{author}{T.~Faulkner}, \bibinfo{author}{F.~M. Haehl}, \bibinfo{author}{E.~Hijano}, \bibinfo{author}{O.~Parrikar}, \bibinfo{author}{C.~Rabideau}, \bibinfo{author}{M.~Van~Raamsdonk},
\newblock \bibinfo{title}{{Nonlinear Gravity from Entanglement in Conformal Field Theories}},
\newblock \bibinfo{journal}{JHEP} \bibinfo{volume}{08} (\bibinfo{year}{2017}) \bibinfo{pages}{057}. \DOIprefix\doi{10.1007/JHEP08(2017)057}. \href{http://arxiv.org/abs/1705.03026}{{\tt arXiv:1705.03026}}.
\bibitem[{Czech(2018)}]{Czech:2017ryf}
\bibinfo{author}{B.~Czech},
\newblock \bibinfo{title}{{Einstein Equations from Varying Complexity}},
\newblock \bibinfo{journal}{Phys. Rev. Lett.} \bibinfo{volume}{120} (\bibinfo{year}{2018}) \bibinfo{pages}{031601}. \DOIprefix\doi{10.1103/PhysRevLett.120.031601}. \href{http://arxiv.org/abs/1706.00965}{{\tt arXiv:1706.00965}}.
\bibitem[{Barcelo et~al.(2005)Barcelo, Liberati, and Visser}]{Barcelo:2005fc}
\bibinfo{author}{C.~Barcelo}, \bibinfo{author}{S.~Liberati}, \bibinfo{author}{M.~Visser},
\newblock \bibinfo{title}{{Analogue gravity}},
\newblock \bibinfo{journal}{Living Rev. Rel.} \bibinfo{volume}{8} (\bibinfo{year}{2005}) \bibinfo{pages}{12}. \DOIprefix\doi{10.12942/lrr-2005-12}. \href{http://arxiv.org/abs/gr-qc/0505065}{{\tt arXiv:gr-qc/0505065}}.
\bibitem[{Hung et~al.(2013)Hung, Gurarie, and Chin}]{Hung:2012nc}
\bibinfo{author}{C.-L. Hung}, \bibinfo{author}{V.~Gurarie}, \bibinfo{author}{C.~Chin},
\newblock \bibinfo{title}{{From Cosmology to Cold Atoms: Observation of Sakharov Oscillations in Quenched Atomic Superfluids}},
\newblock \bibinfo{journal}{Science} \bibinfo{volume}{341} (\bibinfo{year}{2013}) \bibinfo{pages}{1213--1215}. \DOIprefix\doi{10.1126/science.1237557}. \href{http://arxiv.org/abs/1209.0011}{{\tt arXiv:1209.0011}}.
\bibitem[{Wittemer et~al.(2019)Wittemer, Hakelberg, Kiefer, Schr\"oder, Fey, Sch\"utzhold, Warring, and Schaetz}]{PhysRevLett.123.180502}
\bibinfo{author}{M.~Wittemer}, \bibinfo{author}{F.~Hakelberg}, \bibinfo{author}{P.~Kiefer}, \bibinfo{author}{J.-P. Schr\"oder}, \bibinfo{author}{C.~Fey}, \bibinfo{author}{R.~Sch\"utzhold}, \bibinfo{author}{U.~Warring}, \bibinfo{author}{T.~Schaetz},
\newblock \bibinfo{title}{Phonon pair creation by inflating quantum fluctuations in an ion trap},
\newblock \bibinfo{journal}{Phys. Rev. Lett.} \bibinfo{volume}{123} (\bibinfo{year}{2019}) \bibinfo{pages}{180502}. \URLprefix \url{https://link.aps.org/doi/10.1103/PhysRevLett.123.180502}. \DOIprefix\doi{10.1103/PhysRevLett.123.180502}.
\bibitem[{Wilson et~al.(2011)Wilson, Johansson, Pourkabirian, Simoen, Johansson, Duty, Nori, and Delsing}]{Wilson:2011rsw}
\bibinfo{author}{C.~M. Wilson}, \bibinfo{author}{G.~Johansson}, \bibinfo{author}{A.~Pourkabirian}, \bibinfo{author}{M.~Simoen}, \bibinfo{author}{J.~R. Johansson}, \bibinfo{author}{T.~Duty}, \bibinfo{author}{F.~Nori}, \bibinfo{author}{P.~Delsing},
\newblock \bibinfo{title}{{Observation of the dynamical Casimir effect in a superconducting circuit}},
\newblock \bibinfo{journal}{Nature} \bibinfo{volume}{479} (\bibinfo{year}{2011}) \bibinfo{pages}{376--379}. \DOIprefix\doi{10.1038/nature10561}.
\bibitem[{Vezzoli et~al.(2019)Vezzoli, Mussot, Westerberg, Kudlinski, Dinparasti~Saleh, Prain, Biancalana, Lantz, and Faccio}]{Vezzoli2019}
\bibinfo{author}{S.~Vezzoli}, \bibinfo{author}{A.~Mussot}, \bibinfo{author}{N.~Westerberg}, \bibinfo{author}{A.~Kudlinski}, \bibinfo{author}{H.~Dinparasti~Saleh}, \bibinfo{author}{A.~Prain}, \bibinfo{author}{F.~Biancalana}, \bibinfo{author}{E.~Lantz}, \bibinfo{author}{D.~Faccio},
\newblock \bibinfo{title}{Optical analogue of the dynamical casimir effect in a dispersion-oscillating fibre},
\newblock \bibinfo{journal}{Communications Physics} \bibinfo{volume}{2} (\bibinfo{year}{2019}) \bibinfo{pages}{84}. \URLprefix \url{https://doi.org/10.1038/s42005-019-0183-z}. \DOIprefix\doi{10.1038/s42005-019-0183-z}.
\bibitem[{Steinhauer et~al.(2022)Steinhauer, Abuzarli, Aladjidi, Bienaim\'e, Piekarski, Liu, Giacobino, Bramati, and Glorieux}]{Steinhauer:2021fhb}
\bibinfo{author}{J.~Steinhauer}, \bibinfo{author}{M.~Abuzarli}, \bibinfo{author}{T.~Aladjidi}, \bibinfo{author}{T.~Bienaim\'e}, \bibinfo{author}{C.~Piekarski}, \bibinfo{author}{W.~Liu}, \bibinfo{author}{E.~Giacobino}, \bibinfo{author}{A.~Bramati}, \bibinfo{author}{Q.~Glorieux},
\newblock \bibinfo{title}{{Analogue cosmological particle creation in an ultracold quantum fluid of light}},
\newblock \bibinfo{journal}{Nature Commun.} \bibinfo{volume}{13} (\bibinfo{year}{2022}) \bibinfo{pages}{2890}. \DOIprefix\doi{10.1038/s41467-022-30603-1}. \href{http://arxiv.org/abs/2102.08279}{{\tt arXiv:2102.08279}}.
\bibitem[{Steinhauer(2016)}]{Steinhauer:2015saa}
\bibinfo{author}{J.~Steinhauer},
\newblock \bibinfo{title}{{Observation of quantum Hawking radiation and its entanglement in an analogue black hole}},
\newblock \bibinfo{journal}{Nature Phys.} \bibinfo{volume}{12} (\bibinfo{year}{2016}) \bibinfo{pages}{959}. \DOIprefix\doi{10.1038/nphys3863}. \href{http://arxiv.org/abs/1510.00621}{{\tt arXiv:1510.00621}}.
\bibitem[{Berti et~al.(2024)Berti, Fernandes, Butera, Recati, Wouters, and Carusotto}]{Bec_hawk24}
\bibinfo{author}{A.~Berti}, \bibinfo{author}{L.~Fernandes}, \bibinfo{author}{S.~Butera}, \bibinfo{author}{A.~Recati}, \bibinfo{author}{M.~Wouters}, \bibinfo{author}{I.~Carusotto},
\newblock \bibinfo{title}{Analog {Hawking} radiation from a spin-sonic horizon in a two-component {Bose{\textendash}Einstein} condensate},
\newblock \bibinfo{journal}{Comptes Rendus. Physique}  (\bibinfo{year}{2024}).
\bibitem[{Agullo et~al.(2022)Agullo, Brady, and Kranas}]{PhysRevLett.128.091301}
\bibinfo{author}{I.~Agullo}, \bibinfo{author}{A.~J. Brady}, \bibinfo{author}{D.~Kranas},
\newblock \bibinfo{title}{Quantum aspects of stimulated hawking radiation in an optical analog white-black hole pair},
\newblock \bibinfo{journal}{Phys. Rev. Lett.} \bibinfo{volume}{128} (\bibinfo{year}{2022}) \bibinfo{pages}{091301}. \URLprefix \url{https://link.aps.org/doi/10.1103/PhysRevLett.128.091301}. \DOIprefix\doi{10.1103/PhysRevLett.128.091301}.
\bibitem[{Brady et~al.(2022)Brady, Agullo, and Kranas}]{PhysRevD.106.105021}
\bibinfo{author}{A.~J. Brady}, \bibinfo{author}{I.~Agullo}, \bibinfo{author}{D.~Kranas},
\newblock \bibinfo{title}{Symplectic circuits, entanglement, and stimulated hawking radiation in analogue gravity},
\newblock \bibinfo{journal}{Phys. Rev. D} \bibinfo{volume}{106} (\bibinfo{year}{2022}) \bibinfo{pages}{105021}. \URLprefix \url{https://link.aps.org/doi/10.1103/PhysRevD.106.105021}. \DOIprefix\doi{10.1103/PhysRevD.106.105021}.

\end{thebibliography}

\end{document}